\documentclass[
reprint, showkeys, amsmath, amssymb, aps, prd]
{revtex4-2} 

\usepackage{graphicx}
\usepackage{dcolumn}
\usepackage{bm}

\def\MXB{MXB $1659\text{-}29$}
\def\SAX{SAX J$1808.4\text{-}3658$}

\begin{document}

\preprint{APS/123-QED}

\title{Constraining twin stars with cold neutron star cooling data}

\author{Melissa Mendes}
\email{melissa.mendes@tu-darmstadt.de}
\affiliation{McGill University, Department of Physics and Trottier Space Institute at McGill, H3A2T8 Montreal, Canada,\\
Technische Universit\"at Darmstadt, Department of Physics, 64289 Darmstadt, Germany,\\
ExtreMe Matter Institute EMMI, GSI Helmholtzzentrum f\"ur Schwerionenforschung GmbH, 64291 Darmstadt, Germany,\\
Max-Planck-Institut f\"ur Kernphysik, Saupfercheckweg 1, 69117 Heidelberg, Germany}

\author{Jan-Erik Christian}%
\email{jan-erik.christian@uni-hamburg.de}
\affiliation{Hamburger Sternwarte, University of Hamburg, Gojenbergsweg 112, 21029 Hamburg, Germany}

\author{Farrukh J. Fattoyev}
\email{ffattoyev01@manhattan.edu} 
\affiliation{Department of Mathematics and Physics, Manhattan University, Riverdale, NY 10471, USA}

\author{J{\"u}rgen Schaffner-Bielich}
\email{schaffner@astro.uni-frankfurt.de}
\affiliation{Institut für Theoretische Physik, Goethe Universität,\\
Max von Laue-Str.~1, D-60438 Frankfurt am Main, Germany}%

\date{\today}

\begin{abstract}
We investigate the influence of a phase transition from hadronic matter to a deconfined quark phase inside a neutron star on its cooling behaviour including the appearance of twin star solutions in the mass-radius diagram. We find that while the inferred neutrino luminosity of cold transiently-accreting star in \MXB\ is reproduced in all of the constructed twin star models, the luminosity of a colder source, the neutron star in \SAX, cannot be described by equations of state with quark-hadron transition densities below $1.7$ saturation density, suggesting that twin stars with such low density transitions to the quark phase are not realized in nature. We also discuss how constraints to the quark-hadron phase transition density are strongly dependent on the cooling effectiveness of neutrino reactions in the quark phase.
\end{abstract}


\maketitle

\section{\label{sec:intro} Introduction}

The unique high-density regime of neutron stars has long been established as an ideal laboratory for the investigation of nuclear bulk matter properties, specifically the equation of state (EOS). Nonetheless, there are great technical difficulties in both theoretically determining the EOS by performing a first-principles calculation from QCD, and experimentally constraining the EOS with nuclear or astrophysical measurements. 

Currently, the low-density region of equations of state (EOSs), up to at least saturation density, can be constrained with experimentally determined ground state properties of finite nuclei, such as binding energy per nucleon, charge radii and neutron skin thickness (see, for example, \cite{Lattimer:2023rpe, MUSES:2023hyz}). A viable way to construct an EOS able to reproduce these properties is the relativistic mean field (RMF) approach  \cite{PhysRev.98.783, Duerr56, Walecka74, Boguta:1977xi, Serot:1984ey, Mueller:1996pm, Todd:2005, Chen:2014, Fattoyev:2017}. It generates EOSs that are highly parameterized by the strengths of various meson interactions, which are sensitive to different baryon densities of the EOS. In addition, in constant-coupling RMF parametrizations, its EOSs are covariant, so when extrapolated to the neutron-star regime, causality is naturally preserved, as demonstrated by Mueller and Serot \cite{Mueller:1996pm}. Thus, this approach is particularly suited for the incorporation of the known nuclear experimental constraints, at densities below and around saturation density, and observational astrophysical constraints, at larger densities. The high-density neutron star EOS constraints include heavy ion collision experiments \cite{Danielewicz:2002pu, LeFevre:2015paj, Morfouace:2019jky} and neutron stars' observables such as maximum mass \cite{Demorest:2010bx, Antoniadis:2013pzd, Fonseca:2016tux, Cromartie:2019, Romani:2022jhd}, tidal deformability ($\Lambda$) \cite{Abbott:2017, Abbott:2019} and radius measurements. Specifically, the NICER X-ray mission has provided particularly constraining simultaneous measurements of mass and radius of selected pulsars \cite{Riley:2019, Miller:2019, Riley:2021, Miller:2021,Vinciguerra:2023qxq,Choudhury:2024xbk}, setting strong constraints on the EOS \cite{Raaijmakers:2019a, Raaijmakers:2019b, Raaijmakers:2021,Rutherford:2024}. However, by combining all these measurements, one notices that there is an apparent tension between neutron star masses, which reach at least values of $2\,M_\odot$ \cite{Demorest:2010bx, Antoniadis:2013pzd, Fonseca:2016tux, Cromartie:2019}, therefore requiring a stiff EOS, and the tidal deformability reported for GW170817, which suggests a soft EOS \cite{Abbott:2019}. Nonetheless, EOSs usually considered too stiff to reproduce the tidal deformability constraint can be brought into agreement with it if a quark-hadron phase transition is present \cite{Paschalidis:2017qmb, Christian:2018jyd, Montana:2018bkb, Sieniawska:2018zzj, Christian:2019}. This phase transition generates so-called hybrid stars, formed by an inner quark core with nucleonic outer core and crust \cite{Ivanenko:1965dg, Itoh:1970uw, Alford:2004pf, Coelho:2010fv, Chen:2011my, Masuda:2012kf, Yasutake:2014oxa, Zacchi:2015oma}, which feature smaller values of the tidal deformability, more in line with naturally soft EOSs. 

A subcategory of hybrid stars with some significance are twin stars, which are two stars with the same mass and different radii. Such configurations can only be generated by an EOS with a strong phase transition or a rapid cross-over\cite{Kampfer:1981yr, Glendenning:1998ag, Schertler:2000xq, SchaffnerBielich:2002ki, Zdunik:2012dj, Alford:2015dpa, Blaschke:2015uva, Alford:2017qgh, Christian:2017jni}. Thus, observing twin stars would be a smoking gun signal for the presence of a transition to a new phase in the core of at least some neutron stars. In this context, it is interesting to investigate whether cooling data could further constrain the existence of twin stars or set limits on the density for which such a phase transition would take place.

Previous studies have investigated the cooling behavior of quark-hadron hybrid stars with different EOSs \cite{deCarvalho:2015, Lyra:2023}, suggesting that they are able to reproduce current stars' temperature evolution data. Another study \cite{Mendes:2022} has performed estimates on the effectiveness of quark direct Urca (dUrca) cooling reactions in hybrid stars that are consistent with cold stars' luminosities, but they lacked a consistent EOS for the quark phase. To our knowledge, the present work is the first study focusing on twin stars without strangeness in the hadronic phase. Furthermore, we focus on very cold transiently-accreting neutron stars, whose low temperatures strongly indicate the presence of fast-cooling processes in the stars' core \cite{Potekhin:2023}. Reproducing the inferred neutrino luminosity of cold sources can set an upper limit on the star core volume governed by fast-cooling processes, thus potentially constrain the quark-hadron phase transition density, particularly for EOSs with dUrca neutrino emission reactions in the quark phase only. 

In section \ref{sec:framework}, we describe the hadronic and quark EOSs used in this investigation, as well as the calculation of dUrca emissivities and the estimation of the neutrino luminosity of the sources. The corresponding luminosities are confronted with astrophysical data from cooling neutron stars in section \ref{sec:results}, where we also show the possible hybrid star masses of each source as a function of the quark-hadron phase transition density. The relationship between the EOSs parameters and the reproduction of neutron stars' luminosities is discussed and summarized in section \ref{sec:conclusion}.

\section{\label{sec:framework}Theoretical Framework}

\subsection{\label{sec:EOS}Equation of state}

To provide a detailed understanding of the EOS in the context of the relativistic mean field theory \cite{PhysRev.98.783,Duerr56,Walecka74,Boguta:1977xi,Serot:1984ey,Mueller:1996pm,Todd:2005,Chen:2014,Fattoyev:2017}, we first discuss the Lagrangian density  given by $\mathcal{L}=\mathcal{L}_{0}+\mathcal{L}_{i n t}$, where the first term is the free Lagrangian density, in natural units, 
\begin{widetext}
\begin{equation}
\label{L0}
\mathcal{L}_{0}=\bar{\psi}\left(i \gamma^{\mu} \partial_{\mu}-M\right) \psi+\frac{1}{2}\left(\partial_{\mu} \phi \partial^{\mu} \phi-m_{\mathrm{s}}^{2} \phi^{2}\right)+\frac{1}{2} m_{\mathrm{v}}^{2} V_{\mu} V^{\mu}-\frac{1}{4} F_{\mu \nu} F^{\mu \nu}+\frac{1}{2} m_{\rho}^{2} \boldsymbol{b_{\mu}} \cdot \boldsymbol{b^{\mu}}-\frac{1}{4} V_{\mu \nu} V^{\mu \nu}-\frac{1}{4} \boldsymbol{b_{\mu \nu}} \cdot \boldsymbol{b^{\mu \nu}}, 
\end{equation}
and the second term is the interacting Lagrangian density,
\begin{eqnarray}
\label{Lint}
\mathcal{L}_{i n t} &=& \bar{\psi}\left[g_{s} \phi-\left(g_{\mathrm{v}} V_{\mu}+\frac{g_{\rho}}{2} \tau \cdot \boldsymbol{b_{\mu}}+\frac{e}{2}\left(1+\tau_{3}\right) A_{\mu}\right) \gamma^{\mu}\right] \psi-\frac{\kappa}{3 !}\left(g_{\mathrm{s}} \phi\right)^{3}-\frac{\lambda}{4 !}\left(g_{\mathrm{s}} \phi\right)^{4}+\frac{\zeta}{4 !}\left(g_{\mathrm{v}}^{2} V_{\mu} V^{\mu}\right)^{2}
\nonumber\\
&& +\Lambda_{\mathrm{v}}\left(g_{\mathrm{v}}^{2} V_{\mu} V^{\mu}\right)\left(g_{\rho}^{2} \boldsymbol{b_{\mu}} \cdot \boldsymbol{b^{\mu}}\right),
\end{eqnarray}
\end{widetext} 
\noindent
where 
\begin{equation}
    \begin{split}
    F_{\mu \nu} & =\partial_\mu A_\nu-\partial_\nu A_\mu, \\
    V_{\mu \nu} & =\partial_\mu V_\nu-\partial_\nu V_\mu, \\
    \mathbf{b}_{\mu \nu} & =\partial_\mu \mathbf{b}_\nu-\partial_\nu \mathbf{b}_\mu,
    \end{split}
\end{equation}
and $M$ is the nucleon mass. In this Lagrangian, the nucleon interactions are modeled by exchanging photons, represented by the field $A_{\mu}$, and various mesons. The field $\phi$ represents the interactions of the scalar-isoscalar $\sigma$-meson responsible for intermediate attractive interactions, $V^{\mu}$ represents the vector-isoscalar $\omega$-meson responsible for short-range repulsive interactions and $\boldsymbol{b}_{\mu}$ represents the vector-isovector $\rho$-meson. The latter introduces the isospin dependence in the interactions and is crucial for describing the differences between neutron-rich and isospin-symmetric nuclear matter. In addition to meson-nucleon interactions, the Lagrangian density includes scalar and vector self-interactions. In particular, the $\omega$-meson self-interactions, as described by the parameter $\zeta$, soften the EOS at high density and can be tuned to reproduce the maximum mass of a neutron star~\cite{Mueller:1996pm}. Hence, we use $\zeta$ as one of our free parameters ranging between $0.00 \leq \zeta \leq 0.02$ to make sure that the EOSs are consistent with the latest pulsar mass measurements \cite{Cromartie:2019, Demorest:2010bx, Antoniadis:2013pzd}.

It is customary to characterize the behavior of both symmetric nuclear matter (SNM) and the symmetry energy in terms of a few bulk parameters. To introduce these parameters in terms of the EOS, the total binding energy per nucleon can be expanded as
\begin{equation}
    E(n, \alpha) = E(n, 0) +  E_{\rm sym}(n) \alpha^2 + \cdots
\end{equation}
where $E(n, 0)$ is the energy per nucleon of symmetric matter, $\alpha=1-2y$ is the asymmetry parameter with $y=n_P/(n_N+n_P)$, where $n_N$ is the neutron and $n_P$ the proton number density, respectively, and $E_{\rm sym}(n)$ is the symmetry energy. By performing a Taylor series expansion around saturation density, these energies can be expressed as
\begin{equation}
\label{eq:esym1}
    E(n, 0) = \epsilon_0 + \frac{1}{2} K x^2  + \cdots,
\end{equation}
and
\begin{equation}
\label{eq:esym2}
    E_{\rm sym}(n) = J + L x + \frac{1}{2} K_{\rm sym} x^2 + \cdots,
\end{equation}
where $x=(n-n_{\rm sat})/3$ is a dimensionless expansion parameter that quantifies the deviations of the density from its value at the
saturation density $n_\text{sat}$.  Here $\epsilon_0$ and $K$ represent the energy per nucleon and the incompressibility coefficient of SNM, while $J$, $L$ and $K_{\rm sym}$ are the value, slope and curvature of the symmetry energy at saturation density, respectively. These coefficients can be constrained by nuclear experiments, but, particularly for the density slope of the symmetry energy, there are conflicting constraints (see, for example \cite{Reed:2023cap} and references within). While the measurement of the electric dipole polarizability suggests a soft value of $L=47\pm8$ MeV~\cite{Zhang:2014yfa} in agreement with the CREX result of the neutron skin measurement of Ca$^{48}$~\cite{CREX:2022kgg}, the PREX measurements of the neutron skin in Pb$^{208}$~\cite{Reed:2021nqk} prefer a stiffer value for the slope of the symmetry energy, of $L=106\pm37$ MeV. A broader range of estimations for the slope parameter $L$ from neutron skin experiments place it between either $L = 40-60\,$MeV \cite{CREX:2022kgg, Zhang:2022bni, Lattimer:2023rpe} or much higher $L= 121\pm47\,$MeV \cite{PREX:2021umo, Zhang:2022bni, Lattimer:2023rpe}. To cover a broad range of the slope parameter $L$ consistent with these experimental constraints, we use EOSs with $L=50$ MeV and $L=90$ MeV.

Additionally, we take the Dirac effective mass, $M^* = M - g_s \phi$, as our third free parameter. This parameter affects the pressure of SNM at intermediate densities \cite{Cai:2014kya}, and for an optimal parametrization of the relativistic mean field model, $m^{*} = M^*/M$ should be in the range of $0.55 \lesssim m^* \lesssim 0.65$, \cite{Rufa:1988zz, Fattoyev:2012uu}, although larger values of $m^{*}$, up to $0.75$ may also be allowed \cite{Ghosh:2022lam}. We use two values of $m^*$ in the EOSs, of $0.55$ and $0.60$. Thus, we have three parameters that control the density dependence of the EOS at various densities: $m^*$ at intermediate densities, $\zeta$ at high densities, and $L$ around saturation density for the nuclear symmetry energy. We combine those parameters within 4 EOSs: $\mathrm{L}=50$ MeV, $m^{*}=0.60$, $\zeta=0$ (EOS I); $\mathrm{L}=50$ MeV, $m^{*}=0.60$, $\zeta=0.02$ (EOS II); $\mathrm{L}=90$ MeV, $m^{*}=0.55$, $\zeta=0.02$ (EOS III) and $\mathrm{L}=90$ MeV, $m^{*}=0.60$, $\zeta=0.02$ (EOS IV). The remaining parameters of the Lagrangian are then optimized to predict the binding energies and charged radii of closed shell nuclei $^{40}$Ca, $^{48}$Ca, $^{90}$Zr, $^{132}$Sn and $^{208}$Pb within less than 2\% of the experimental data. We also ensure that the incompressibility of neutron-rich matter \cite{Piekarewicz:2008nh} is consistent with the centroid energies of the giant monopole resonance in $^{208}${Pb}. In Table~\ref{tab:couplings} we provide the full set of coupling constants for the four EOSs, and in Table~\ref{tab:BE_RC}, we present a representative set of finite nuclei predictions for each of these EOS configurations. Finally, in Table \ref{tab:bulk_props} we display some of the bulk parameters of infinite nuclear matter for these EOSs.

\begin{table*}
\caption{\label{tab:couplings} Coupling constants of the four EOSs discussed in the text. For each of the four models, we fix $m_{\rm v}=782.5$ MeV and $m_{\rho}=763$ MeV. The scalar meson masses and the nonlinear $\kappa$ values are given in MeV.}
\begin{ruledtabular}
\begin{tabular}{ccccccccc}
       Model & $m_{\rm s}$ &  $g_s^2$ &  $g_v^2$ & $g_{\rho}^2$ & $\kappa$ & $\lambda$ & $\zeta$ & $\Lambda_{\rm v}$ \\
    \hline
       EOS I & $501.366$ & $99.880356$ &  $159.122616$ & $144.434050$ & $4.701218$ & $-0.0200808$ & $0.000$ & $0.03608328$ \\
       EOS II & $499.693$ & $104.831132$ &  $173.492123$ & $164.004957$ & $3.535450$ & $-0.00539112$ & $0.020$ & $0.03915850$ \\
       EOS III & $509.767$ & $128.601266$ &  $215.590598$ & $92.415742$ & $2.519619$ & $-0.00218582$ & $0.020$ & $0.009667709$ \\
       EOS IV & $503.057$ & $107.170473$ &  $176.982544$ & $92.079086$ & $3.129633$ & $-0.00319217$ & $0.020$ & $0.009050690$ \\
    \end{tabular}
\end{ruledtabular}    
\end{table*}

\begin{table*}
\caption{\label{tab:BE_RC} Experimental data for the binding energy per nucleon (MeV)\,\cite{Wang:2012}, charge radii (fm)\,\cite{Angeli:2013}, and neutron skins (fm)\,for $^{48}$Ca and $^{208}$Pb representing doubly closed shell nuclei, are shown alongside theoretical predictions from IUFSU \cite{Fattoyev:2010mx}, FSUGold2 \cite{Chen:2014sca}, and the four EOSs discussed in the text. Predictions for other closed shell nuclei properties also lie within less than 2\% of the experimental data.}
\begin{ruledtabular}
\begin{tabular}{ccccccccc}
       Nucleus & Observable & Experiment & IUFSU & FSUGold2 & EOS I & EOS II & EOS III & EOS IV \\
       \hline
       & $B/A$ & 8.667 & 8.549 & 8.621 & 8.594 & 8.595 & 8.753 & 8.628 \\
       $^{48}$Ca & $R_{\rm ch}$ & $3.477\pm0.002$ & 3.416 & 3.413 & 3.442 & 3.428 & 3.400 & 3.425 \\
       & $R_{\rm skin}$ & $0.121 \pm 0.035$ & 0.173 & 0.232 & 0.182 & 0.178 & 0.211 & 0.219 \\
       \hline
       & $B/A$  & 7.867 & 7.896 & 7.872 & 7.867 & 7.867 & 7.867 & 7.868 \\
       $^{208}$Pb & $R_{\rm ch}$ & $5.501\pm0.001$ & 5.476 & 5.489 & 5.502 & 5.502 & 5.502 & 5.502 \\
       & $R_{\rm skin}$ & $0.283 \pm 0.071$ & 0.162 & 0.286 & 0.183 & 0.176 & 0.269 & 0.260 \\
    \end{tabular}
\end{ruledtabular}
\end{table*}

\begin{table*}
\caption{\label{tab:bulk_props} Bulk properties of nuclear matter for the four sets of EOS discussed in the text. Saturation density $(n_{\rm{sat}})$ given in (fm$^{-3}$), energy per nucleon of SNM $(\epsilon_0)$, incompressibility coefficient of SNM $(K_0)$, value $(J)$, slope $(L)$ and curvature $(K_{\rm sym})$ of the symmetry energy at saturation density given in MeV.}
\begin{ruledtabular}
\begin{tabular}{ccccccc}
       Model & $n_{\rm{sat}}$  &  $\epsilon_0$ & $K_0$  & $J$ & $L$ & $K_{\rm sym}$ \\
    \hline
       EOS I & $0.1517$ & $-16.19$ & $234.0$ & $31.63$ & $50.00$ & $26.76$  \\
       EOS II & $0.1513$ & $-16.21$ & $233.0$ & $31.45$ & $50.00$ & $21.88$  \\
       EOS III & $0.1463$ & $-15.98$ & $254.0$ & $35.15$ & $90.00$ & $-36.13$  \\
       EOS IV & $0.1489$ & $-16.21$ & $246.0$ & $35.64$ & $90.00$ & $-56.56$  \\                 
    \end{tabular}
\end{ruledtabular}
\end{table*}

Comparing the results from each of the EOSs in pairs, one can check the effect of a given parameter on hybrid star cooling. In particular, an important quantity to examine is the direct Urca threshold, which refers to the proton fraction $Y_{\rm p} = Z/A$ in neutron star matter above which direct Urca reactions are allowed by energy-momentum conservation. In the simplest case, when muons are not present in the EOS, the dUrca threshold is found to be $Y_{\mathrm{p\, dUrca}} = 1/9$, but in more realistic EOSs including muons \cite{Lattimer:1991}, it can be found by solving the charge neutrality and momentum conservation equations that lead to the following dynamical equation~\cite{Klahn:2006, Fattoyev:2011}, 
\begin{equation}
\left(\frac{1-Y_{\rm p}}{Y_{\rm p}}\right)^{1/3} = 1 + (1-\mathcal{Y})^{1/3} \ ,
\end{equation}
where $\mathcal{Y} = Z_{\mu}/Z$ is the muon charge fraction. Therefore, the dUrca threshold is strongly dependent on the EOS and it has been shown \cite{Lattimer:2014, Mendes:2021} that EOSs with a larger value of 
$L$ have a lower dUrca threshold density. This trend can be understood by observing that EOSs with a larger value of $L$ have a larger symmetry energy at high densities, which makes the neutron star matter more symmetric and allows the proton fraction to more easily exceed the required threshold proton fraction at lower densities. In other words, $Y_{\rm p}$ is directly proportional to $L$, according to the relation \cite{Mendes:2021}
\begin{equation}
\label{eq:Yp}
    Y_{\rm p} \simeq \frac{64}{3 \pi^{2} n_{\rm sat}}\left(J + L x + \frac{1}{2} K_{\rm sym} x^2+\cdots\right)^{3},
\end{equation}
for matter in beta equilibrium. Increasing the slope of the symmetry energy independently of the other expansion parameters directly increases the proton fraction, thus reducing the density at which the dUrca threshold $Y_{\rm p}$ is reached. There is no similar one-to-one correspondence between the remaining parameters $m^{*}$ and $\zeta$ and the expansion parameters in Eq. \ref{eq:Yp} with the proton fraction, so a simple relation between the dUrca threshold and $m^{*}$ or $\zeta$ cannot be constructed. Nonetheless, EOSs with larger $m^{*}$ or $\zeta$ present slightly increased dUrca threshold densities, as seen in Table \ref{tab:EOS}, but we note in passing that the influence of those quantities in hybrid star cooling will be more clearly seen in the dUrca emissivity rather than the dUrca threshold.

\begin{table}
\caption{\label{tab:EOS}Slope of symmetry energy at saturation $(L)$ in MeV and the dimensionless quantities of Dirac effective mass $(m^{*})$, $\omega$-meson self-interactions $(\zeta)$ and dUrca threshold proton fraction $(Y_{p \ \mathrm{dUrca}})$ for purely hadronic EOS I, EOS II, EOS III and EOS IV. The dUrca threshold density $(n_{\mathrm{dUrca}})$ is given in terms of the saturation density and the dUrca threshold mass $(m_{\mathrm{dUrca}})$ and maximum mass $(m_{\mathrm{max}})$, in terms of $M_{\odot}$.}
\begin{ruledtabular}
\begin{tabular}{cccccccc}
 & $L$ & $m^{*}$ & $\zeta$ & $Y_{p \ \mathrm{dUrca}}$ & $n_{\mathrm{dUrca}}$ & $m_{\mathrm{dUrca}}$ & $m_{\mathrm{max}}$\\
\hline
EOS I & $50$ & $0.60$ & $0.0$ & $0.1375$ & $3.51$ & $2.58$ & $2.70$\\
EOS II & $50$ & $0.60$ & $0.02$ & $0.1376$ & $3.55$ & $1.87$ & $2.09$\\
EOS III & $90$ & $0.55$ & $0.02$ & $0.1319$ & $1.77$ & $1.01$ & $2.19$\\
EOS IV & $90$ & $0.60$ & $0.02$ & $0.1321$ & $1.78$ & $0.92$ & $2.11$\\
\end{tabular}
\end{ruledtabular}
\end{table}

\begin{table*}
\caption{\label{tab:phasetrans} Parameters of the hybrid EOSs built with EOS I (upper left), EOS II (upper right), EOS III (lower left) and EOS IV (lower right). Transition density $(n_{\rm trans})$ given in terms of saturation density, transition pressure $(p_{\rm trans})$ and energy density jump $(\Delta \epsilon)$ given in $\mathrm{MeV/fm^3}$. The cases where nucleonic dUrca processes are allowed are marked with a star.}
\begin{ruledtabular}
\begin{tabular}{@{\hspace{4em}} l @{\hspace{4em}} | @{\hspace{4em}} l @{\hspace{7em}}}
    \multicolumn{1}{l}{{\hspace{10em}} EOS I} &\multicolumn{1}{l}{{\hspace{7em}} EOS II}\\ \hline
    \colrule
    $n_{\rm {trans}} = 1$, $p_{\rm {trans}}= 3$, \, $\Delta \epsilon = 170$, \,  & $n_{\rm {trans}} = 1.2$, $p_{\rm {trans}}= 5$, \, $\Delta \epsilon = 180$\\
    $n_{\rm {trans}} = 1.29$, $p_{\rm {trans}}= 6$, \, $\Delta \epsilon = 260$, \,  & $n_{\rm {trans}} = 1.87$, $p_{\rm {trans}}= 20$, \, $\Delta \epsilon = 223$\\
    $n_{\rm {trans}} = 1.84$, $p_{\rm {trans}}= 25$, \, $\Delta \epsilon = 220$, \,  & $n_{\rm {trans}} = 2.2$, $p_{\rm {trans}}= 35$, \, $\Delta \epsilon = 241$\\
    $n_{\rm {trans}} = 2$, $p_{\rm {trans}}= 38$, \, $\Delta \epsilon = 236$, \,  & $ n_{\rm {trans}}^{*} = 4$, $p_{\rm {trans}}= 153$, \, $\Delta \epsilon = 250$\\
    $n_{\rm {trans}} = 2$, $p_{\rm {trans}}= 38$, \, $\Delta \epsilon = 280$ \,  & $ n_{\rm {trans}}^{*} = 4.07$, $p_{\rm {trans}}= 160$, \, $\Delta \epsilon = 245$\\
    & $n_{\rm {trans}}^{*} = 4.3$, $p_{\rm {trans}}= 180$, \, $\Delta \epsilon = 229$\\
    & $n_{\rm {trans}}^{*} = 4.6$, $p_{\rm {trans}}= 205$, \, $\Delta \epsilon = 200$\\[0.5em]
    \hline
     \multicolumn{1}{l}{{\hspace{10em}} EOS III} &\multicolumn{1}{l}{{\hspace{7em}} EOS IV}\\ \hline
    \colrule
    $n_{\rm {trans}} = 1$, $p_{\rm {trans}}= 4$, \, $\Delta \epsilon = 130$, \,  & $n_{\rm {trans}} = 0.9$, $p_{\rm {trans}}= 3.2$, \, $\Delta \epsilon = 132$\\
    $n_{\rm {trans}} = 1.5$, $p_{\rm {trans}}= 12$, \, $\Delta \epsilon = 172$, \,  & $n_{\rm {trans}} = 1.28$, $p_{\rm {trans}}= 7$, \, $\Delta \epsilon = 285$\\
    $n_{\rm {trans}}^{*} = 2$, $p_{\rm {trans}}= 27$, \, $\Delta \epsilon = 206$, \,  & $n_{\rm {trans}} = 1.67$, $p_{\rm {trans}}= 15$, \, $\Delta \epsilon = 182$\\
    $n_{\rm {trans}}^{*} = 2.18$, $p_{\rm {trans}}= 36$, \, $\Delta \epsilon = 220$ \,  & $n_{\rm {trans}}^{*} = 1.93$, $p_{\rm {trans}}= 23$, \, $\Delta \epsilon = 211$\\
    $n_{\rm {trans}}^{*} = 2.2$, $p_{\rm {trans}}= 38$, \, $\Delta \epsilon = 296$ \, & $n_{\rm {trans}}^{*} = 2.16$, $p_{\rm {trans}}= 32$, \, $\Delta \epsilon = 226$\\
     & $n_{\rm {trans}}^{*} = 2.37$, $p_{\rm {trans}}= 42$, \, $\Delta \epsilon = 255$\\    
\end{tabular}
\end{ruledtabular}
\end{table*}

To build quark-hadron phase transitions in the hadronic EOSs, we assume that at a sufficiently high pressure, $p_{\rm trans}$, the hadronic EOS is replaced by a quark phase. We model this by using a Maxwell construction, meaning that the phase transition is realized as a discontinuity in energy density at a constant pressure of $p_{\rm trans}$. The quark phase is described using a constant speed of sound EOS. This ansatz has been well established \cite{Zdunik:2012dj,Alford:2013} and takes the following mathematical form:
	\begin{equation}
		\epsilon(p) =
		\begin{cases} 
			\epsilon_{\rm RMF}(p)	&  p < p_{\rm trans}\\
			\epsilon_{\rm trans}+\Delta\epsilon + c_{QM}^{-2}(p-p_{\rm trans})	& p > p_{\rm trans},\\
		\end{cases}
	\end{equation} 
where $c_{\rm QM}$ is the constant sound speed of the quark phase. We set the speed of sound in the quark phase to $c_{QM} = c = 1$ in natural units, since higher sound speeds increase the possibility for twin star configurations. The transitional pressure $p_{\rm trans}$ and the jump in energy density $\Delta\epsilon$ are free parameters. The transition pressure can be used interchangeably with the energy density at the point of transition $\epsilon_{\rm trans}$ in the nucleonic phase through the relation 
$\epsilon_{\rm trans} = \epsilon_{\rm RMF}(p_{\rm trans})$. 

Following the works of \cite{Christian:2017jni,Christian:2018jyd,Christian:2023hez}, we build quark-hadron phase transitions that generate twin stars. This criterion is a strong constraint mostly for EOS II, forbidding phase transitions between $2.3 \leq n \leq 4 \, n_{\rm sat}$ for that EOS. Thus, hybrid stars in this density region are not investigated in this work, even though they could be compatible with observational data. We vary the phase transition parameters $p_{\rm trans}$ and $\Delta \epsilon$ to allow for the smallest and largest possible $n_{\rm trans}$ generating twin stars that fit the astrophysical constraints. Different combinations of the parameters $L$ and $\zeta$ affect the stiffness of the hadronic part of the EOSs, which restricts the quark phase transitions resulting in hybrid stars compatible with astrophysical data constraints. In particular, EOSs with larger $\zeta$ accommodate larger quark phase transition densities, but the tidal deformability observation restricts EOSs with larger values of $L$ to relatively low-density phase transitions, around $2 \, n_{\rm sat}$. The parameters of all phase transitions built in this work are shown in Table \ref{tab:phasetrans}. Finally, we combine the quark-hadron hybrid core EOS with the Baym-Pethick-Sutherland EOS \cite{Baym:1971} for the outer crust and the Negele-Vautherin EOS \cite{Negele:1973} for the inner crust. Using accreted crust EOSs would not significantly change the calculated masses or tidal deformabilities \cite{Piekarewicz:2018sgy, Kalaitzis:2019dqc, Biswas:2019, Ji:2019hxe, Perot:2020gux, Gittins:2020mll}, even if it may alter radii predictions \cite{Piekarewicz:2014, Ji:2019hxe, Perot:2020gux}. For each EOS, we use the random-phase approximation (RPA) analysis\,\cite{Carriere:2002bx} to calculate the crust-core transition density, with the inner crust EOS ending at approximately $0.5\,n_{\rm sat}$ and the remaining small gap in densities smoothly connected using a cubic spline interpolation.

All EOSs constructed in this work are compatible with the most recent astrophysical observations of mass and radius of neutron stars. However, simultaneously reproducing the large mass of PSR J$0740$+$6620$ and the tidal deformability detected in the merger event GW$170817$ can be challenging. The smallest radius at two solar masses featured by any of our hadronic EOSs is $R_{2\,M_\odot} =  11.85\,$km, placing it comfortably within the radius range of $R = 12.76^{+1.49}_{-1.02}\,$km for the 68\% confidence interval of PSR J$0740$+$6620$ \cite{Dittmann:2024mbo}, also in agreement with this source's recent data re-analysis of \cite{Salmi:2024}. The largest radius at two solar masses is generated by the stiffest EOS, at $R_{2\,M_\odot} = 13.7\,$km, which is within this pulsar data constraint as well. All hybrid EOSs also respect this limit. On the other hand, all of our purely hadronic EOSs have a tidal deformability outside the bounds of GW$170817$. However, for most of our hybrid stars, we are able to choose appropriately small transition points to the quark phase such that compatibility with GW$170817$ was regained. Exceptions to this case are EOS II hybrid stars with phase transition above $4 n_{sat}$, that presented a tidal deformability of $\Lambda_{1.4\,M_\odot}= 596$, which is marginally outside GW$170817$'s upper limit of $\Lambda_{1.4\,M_\odot}= 580$ \cite{LIGO:2018}, but close enough to warrant investigation. The tidal deformabilities of all other hybrid EOSs listed in Table \ref{tab:phasetrans} for $1.4 M_{\odot}$ stars range between 100 and 550. The other mass and radius contours reported by NICER, for sources PSR J$0437$-$4715$ \cite{Choudhury:2024xbk} and PSR J$0030$+$0451$ \cite{Miller:2019,Riley:2019}, do not impose stricter constraints to the hybrid EOSs, which are also mostly consistent with the data re-analysis of PSR J$0030$+$0451$ of Ref.~\cite{Vinciguerra:2023qxq}. Model \texttt{PDT-U} for this source is the exception, as it tends to exclude hybrid EOSs with low density phase transitions. Nonetheless, we chose to proceed analyzing those EOSs, mostly because they remain consistent with PSR J$0030$+$0451$'s favored mass-radius contours, \texttt{ST+PDT} \footnote{Mass-radius posteriors for all models of \cite{Vinciguerra:2023qxq} can be found at the Zenodo repository: https://zenodo.org/records/7646352}.

While solving the TOV equations, we have often observed the presence of a single stable hybrid star in the first mass-radius branch, even when usual stability criteria, like the so-called Seidov limit \cite{Seidov:1971}, would suggest otherwise. This phenomenon is a function of the step-size of the radial grid used, or equivalently, the error tolerance chosen in solving the TOV equations. For an Euler ODE solver with fixed step-size, we have observed its critical value to be around $50$-$100$m, depending on the size of the quark core in the hybrid EOSs. To our knowledge, this phenomenon has been observed before, but not documented \cite{Alford:2024}. It does not affect the calculations performed in this work, and investigating it was out of the scope of this paper.

\subsection{\label{sec:NScooling}Hybrid star cooling}

The neutron stars under study have high inferred neutrino luminosities, thus we focus on neutrino emitting fast-cooling processes, which take place at the neutron star core, and ignore slow-cooling, as well as crust-cooling processes, because their cooling rates are subdominant in cold neutron stars \cite{Yakovlev:2001}. The determination of neutron stars surface temperature and envelope composition in this simplified scenario is discussed in section \ref{sec:MXBandSAX}. We use a 1D general relativity routine to calculate the hybrid stars' core cooling.

For the EOSs used in this work, the relevant reactions are nucleonic neutrino direct Urca reactions for the nucleonic part of the EOS, $n \rightarrow p + l + \bar{\nu}_{l}$ and $\quad p + l \rightarrow n + \nu_{l}$, where $l$ stands for leptons, and quark direct Urca reactions for the quark part of the EOS, $d \rightarrow u+e^{-}+\bar{\nu}_{e}$ and $u+e^{-} \rightarrow d+\nu_{e}$, with corresponding neutrino emissivities in natural units, respectively derived in \cite{Yakovlev:2001} and \cite{Iwamoto:1980},
\begin{eqnarray}
\label{eq:dUrca}
\epsilon_{\mathrm{Urca}}=&&\frac{457 \pi}{10080} G_{\mathrm{F}}^{2} \cos ^{2} \theta_{\mathrm{C}}\left(1+3 g_{\mathrm{A}}^{*2}\right) m_{n}^{L \, *} m_{p}^{L \, *}\times \nonumber\\
&&\times \left(3 \pi^{2} Y_{e} n\right)^{1 / 3}T^{6}
\end{eqnarray}
\begin{equation}
\label{eq:qdUrca}
    \epsilon_{\mathrm{q} \ \mathrm{Urca}}= \frac{914}{315} G_{\mathrm{F}}^2 \cos^{2}\theta_{\mathrm{C}} (3 Y_e)^{1/3} \alpha \pi^2 n  T^6.
\end{equation}
Corrections due to in-medium interactions are included in the effective Landau masses $m_{n}^{L \, *}, m_{p}^{L \, *}$, given by
\begin{equation}
    \mathrm{m}_{i}^{L \, *}=\sqrt{{M_{\rm i}^{*}}^2+k_{\rm F \, i}^2}
\end{equation}
with $M_{\rm i}^{*}$ being the Dirac effective mass. Here $g_{A}^{*}\simeq g_{A}\left(1-\frac{n}{4.15\left(n_{\rm sat}+n\right)}\right)$ is the axial vector coupling that depends on the baryon density~\cite{Carter:2002}, $G_{\mathrm{F}}$ is the weak interaction constant, $\alpha$ is the strong coupling constant and $\theta_{\mathrm{C}}$ is the Cabibbo angle.  

Considering up and down quarks cooling processes only, we fix the electron fraction in the quark phase to $Y_e=10^{-5}$ and the strong coupling constant $\alpha=0.12$, unchanging with density. The effect of these approximations on the estimated masses of the sources and phase transition density is discussed in section \ref{sec:conclusion}. The quark EOS we use is agnostic about the presence of strange quarks in the quark phase, but for simplicity we assume they do not participate in the cooling reactions here. Possible color superconducting phases are also ignored for the same reason. 

Given the nature of first-order transitions, no quark-hadron mixed phase is present for a Maxwell construction in neutron stars. Therefore, if the phase transition happens before the nucleonic dUrca threshold is reached, quark dUrca processes will be the only fast-cooling neutrino reactions of the hybrid star. The hybrid EOSs where nucleonic dUrca cooling takes place are indicated by star symbols $(*)$ in Table \ref{tab:phasetrans}.

Some gap models describing superfluidity and superconductivity in the nucleonic core are included, following the parametrization originally described in \cite{Ho:2015}: 
\begin{equation}
\label{eq:gapmodel}
    \Delta\left(k_{\mathrm{Fx}}\right)= \Delta_{0} \frac{\left(k_{\mathrm{Fx}}-k_{0}\right)^{2}}{\left(k_{\mathrm{Fx}}-k_{0}\right)^{2}+k_{1}} \frac{\left(k_{\mathrm{Fx}}-k_{2}\right)^{2}}{\left(k_{\mathrm{Fx}}-k_{2}\right)^{2}+k_{3}} \quad ,
\end{equation}
where $k_{\mathrm{Fx}}$ is the Fermi momentum of the nucleons $x$ and the quantities $\Delta_{0}$, $k_{0}$, $k_{1}$, $k_{2}$ and $k_{3}$ are free parameters. We work with combinations of neutron triplet superfluid gap models NT SYHHP \cite{Shternin:2011}, NT AO \cite{Amundsen_2:1985}, NT EEHO \cite{Elgaroy_3:1996} and proton singlet superconducting gap models PS CCDK \cite{Chen:1993}, PS BCLL \cite{Baldo:1992} and PS CCYps \cite{Chao:1972}, which form a representative sample of nuclear pairing models, including examples where both most and very few of the nucleons in the core are paired. When nuclear pairing is included, the dUrca emissivity (Eq. \ref{eq:dUrca}) is reduced by a factor $R_{\mathrm{L}}$, such that 
\begin{equation}
    \epsilon_{\mathrm{Urca}}^{\mathrm{reduced}}=\epsilon_{\mathrm{Urca}} R_L,    
\end{equation}
\noindent
where \cite{Yakovlev:2001}
\begin{equation} \label{eq:reduced1}
\begin{split}
    R_{\mathrm{L}} &=\left[0.2312+\sqrt{(0.7688)^{2}+(0.1438 ~ v_{\mathrm{S}})^{2}}\right]^{5.5}\\
    & \qquad \times \exp \left(3.427-\sqrt{(3.427)^{2}+v_{\mathrm{S}}^{2}}\right),
    \\
    v_{\mathrm{S}} &=\sqrt{1-\tau}\left(1.456-\frac{0.157}{\sqrt{\tau}}+\frac{1.764}{\tau}\right),
\end{split}
\end{equation}
for proton singlets and
\begin{equation} \label{eq:reduced2}
\begin{split}
    R_{\mathrm{L}} &=\left[0.2546+\sqrt{(0.7454)^{2}+(0.1284 \, v_{\mathrm{T}})^{2}}\right]^{5}\\
    & \qquad \times \exp \left(2.701-\sqrt{(2.701)^{2}+v_{\mathrm{T}}^{2}}\right),\\
    v_{\mathrm{T}} &=\sqrt{1-\tau}\left(0.7893+\frac{1.188}{\tau}\right).
\end{split}
\end{equation}
for neutron triplets, such that when both species are paired, 
\begin{equation} \label{eq:bothpairing}
R_{L} \sim \min \left(R_{L,\mathrm{singlet}}, R_{L,\mathrm{triplet}}\right).
\end{equation}
In these expressions, $\tau = T/T_c$, where $T_c$ is the critical temperature of each gap model parametrization and $T=T(r)$ is the local temperature such that
\begin{equation}\label{eq:tempinf}
    T(r) = \tilde{T} e^{-\phi(r)},   
\end{equation}
with $\tilde{T}$ being the temperature of the core at infinity. Eq.~(\ref{eq:bothpairing}) is valid for strong superfluidity \citep{Yakovlev:1994}, but since $T_c \gg T$ for most of the neutron star's core density, weak superfluidity corrections to these factors would only be minor.

\subsection{\label{sec:MXBandSAX} Transiently-accreting neutron stars envelope and temperature}

We reproduce the inferred neutrino luminosity of two transiently-accreting neutron stars in the low-mass X-ray binaries \MXB\ and \SAX. These sources are chosen for their low luminosity, which suggests that fast-cooling processes must be active in their core \cite{Potekhin:2023}. In particular, the neutron star in \SAX\ is one of the coldest ever observed, thus its cooling pattern might strongly constrain the core EOS. \MXB\ was selected because three cycles of accretion-quiescence have been observed for this neutron star, which has improved the accuracy of its crust heat transport models and temperature estimation, allowing for a more precise determination of its neutrino luminosity \cite{Brown:2018, Mendes:2022}. 

The estimated temperature of the neutron star in \MXB\ at quiescence as measured by an observer at infinity, $T_{e}^{\infty}$, is taken from \cite{Cumming:2017, Brown:2018} to be $55$ eV. Therefore, we assume for this source a helium envelope composition, following their detailed modelling of crust cooling, consistent with this source's latest data \cite{Parikh:2019}. We then use relation \cite{Gudmundsson:1983}, in natural units,
\begin{eqnarray}\label{light_el}
    \tilde{T} =&& 0.552 e^{\phi(R)} \times \nonumber\\
    &&\times 10^8 K \left[\left(\frac{T_e^{\infty}}{10^{24}\, K}\right)^{4} e^{-3\phi(R)} \frac{R^2}{9 \times 10^{4} GM} \right]^{0.413},
\end{eqnarray}
valid when a neutron stars' envelope consists of mostly light elements, to describe how the temperature varies from the surface to the bottom of the envelope. The inferred neutrino luminosity of the neutron star in \MXB\ is estimated from the amount of deposited energy from accretion, considering its inferred average mass accretion rate $\langle \dot{M} \rangle \approx  1.5 \times 10^{-9}\, \mathrm{M}_{\odot}/\mathrm{yr}$ \cite{Cumming:2016}, following the procedure described in \cite{Potekhin:2019}, which leads to an expected value of neutrino luminosity observed at infinity $L_{\nu} = (3.91 \pm 2) \times 10^{34}$ erg/s with $1 \sigma$ standard deviation. The major sources of uncertainty to this neutrino luminosity estimation are the estimated source's distance and the envelope composition. Other possible sources of uncertainty in the crust cooling modelling, such as the uniformity of the outburst accretion rate or deep crustal heating rates, were already included in the analysis of \cite{Brown:2018}, leading to the previously mentioned error bars in the luminosity. 

Detailed models for the crust cooling of the neutron star in \SAX\ are not available. Thus, we work with an estimated upper limit to its neutrino luminosity of $L_{\nu} = 1 \times 10^{32} - 1 \times 10^{33}$ erg/s, obtained considering its average accreted mass $\langle \dot{M} \rangle = 9 \times 10^{-12} \, \mathrm{M}_{\odot}/\mathrm{yr}$ \cite{Heinke:2009}, such that the approximate neutrino luminosity is the mass accretion rate times the energy released by deep crustal heating reactions, $Q\approx 0.5-2$ MeV/nucleon mass. We take $T_{e}^{\infty}= 36^{+4}_{-8}$ eV for this source \cite{Heinke:2009}, and calculate its neutrino luminosity considering two envelope composition scenarios: a ``light elements'' one, where we assume Eq.~(\ref{light_el}) can be used and a ``heavy elements'' one, where the expression \cite{Potekhin:1997}
\begin{eqnarray}
    \tilde{T} =&& 1.288 e^{\phi(R)} \times \nonumber\\
    && \times 10^8 K \left[\left(\frac{T_e^{\infty}}{10^{24}\, K}\right)^{4} e^{-3\phi(R)} \frac{R^2}{9 \times 10^{4} GM} \right]^{0.455}
\end{eqnarray}
valid for iron envelopes, is used.

We assume that the inferred neutrino luminosity of each source is released entirely by core cooling processes. Since the masses of these sources are unconstrained, we calculate hybrid star masses consistent with the estimated luminosities and their corresponding quark core volumes.

\section{\label{sec:results}Results}

In this section, we calculate the range of neutron star masses corresponding to the estimated lower and upper limits of neutrino luminosity for each source. This provides a check to the feasibility of the phase transitions built-in and allows us to delineate the effects of twin stars in the luminosity relations. Since core collapse supernova simulations suggest that neutron stars lighter than $\approx 1\,M_\odot$ are not formed \cite{Burrows:2019, Suwa2018, Radice2017, Janka2008, Fischer2010}, we do not include such stars in our analysis, but still display them in some plots for a more complete visualization of the results.

\subsection{\label{sec:L50}Hybrid stars with hadronic $L=50$ EOSs}

Depending on the EOS, nucleonic superfluidity model, and the onset of the quark phase, the inferred neutrino luminosities of the neutron stars in \MXB\ and \SAX\ can be reached with quark dUrca processes alone, nucleonic dUrca processes alone or a combination of them. For EOS I, all quark phase transitions take place before the onset of the dUrca threshold, thus the neutrino luminosities of cold sources are reached with quark dUrca processes only. Increasing the value of the $\zeta$ parameter to $0.02$ (EOS II) considerably softens the EOS and allows for larger phase transition densities. In that case, all hybrid stars with phase transition densities smaller than the dUrca threshold, $3.55 \, n_{sat}$, cool predominantly through quark dUrca processes, while the ones with larger phase transition densities cool mainly through nucleonic dUrca processes.

\begin{figure}
    \centering
    \includegraphics[scale=0.6]{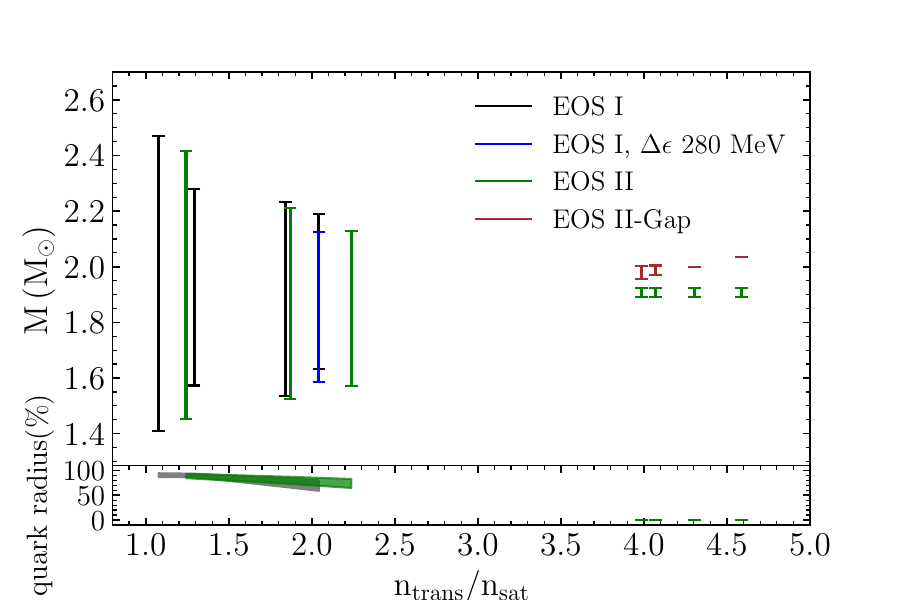}
    \caption{Total masses of the hybrid stars with \MXB's neutron star lower and upper inferred neutrino luminosities, built with EOS I (in black) and EOS II (in green). The density transitions $\mathrm{n}_\mathrm{trans} \, (\mathrm{n}_\mathrm{sat})$ are represented on the x-axis. The blue line marks the special case of EOS I, $\mathrm{p}_\mathrm{trans}= 38$ MeV, $\mathrm{n}_\mathrm{trans}= 2 \, \mathrm{n}_\mathrm{sat}$, $\Delta \epsilon = 280$ MeV to differentiate it from $\Delta \epsilon = 236$ MeV with same transition density, in black. In brown, the results for EOS II with nuclear pairing combination NT AO+PS CCYps are shown. In the lower plot, the hybrid stars corresponding quark radius fraction in relation to total radius is shown.}
    \label{fig:T55_Minf_50}
\end{figure}

Figure \ref{fig:T55_Minf_50} displays the predicted hybrid star masses for \MXB's neutron star, considering also nuclear pairing in the nucleonic phase for EOS II stars. The data points in blue correspond to EOS I with a phase transition at $\mathrm{n}_\mathrm{trans}=2 \, \mathrm{n}_\mathrm{sat}$, but $\Delta \epsilon=280 \, \rm{MeV}$, to differentiate it from the point in black with phase transition at $\mathrm{n}_\mathrm{trans}=2 \, \mathrm{n}_\mathrm{sat}$ and $\Delta \epsilon=236 \, \rm{MeV}$. Increasing $\Delta \epsilon$ while keeping the other phase transition parameters unaltered slightly reduces the mass range, but the most important parameter for determining the range of hybrid masses is still the transition density $n_{trans}$, that determines the quark volume fraction in the hybrid star. For EOS II, the gap model combination NT AO+PS CCYps, shown in Figure \ref{fig:T55_Minf_50}, provides a strong dUrca suppression, resulting in a large difference to the non-paired case, which eventually fails to reproduce the lower luminosity limit of the source in \MXB\ for large phase transition densities, $\mathrm{n}_\mathrm{trans} \geq 4.3 \, \mathrm{n}_\mathrm{sat}$. That is an example situation where including nuclear pairing in the nucleonic phase does not necessarily increase the calculated mass range, because then, the quark dUrca cooling again becomes dominant and the decisive factor becomes the quark volume, which might not be large enough to reproduce a source's luminosity. This situation is particularly clear for combinations of strong superfluidity and superconductivity gap models such as PS AO+NT SYHHP or PS CCDK+NT AO.

\begin{figure}
    \centering
    \includegraphics[scale=0.6]{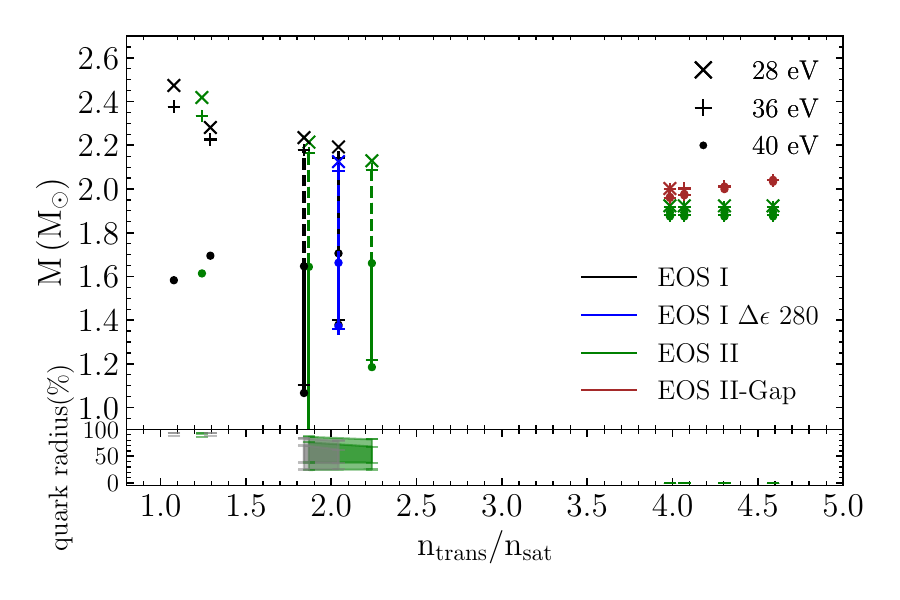}
    \caption{Total masses of the hybrid stars with \SAX's neutron star lower and upper inferred neutrino luminosities, built with EOS I (in black) and EOS II (in green), for T$^{\infty}=28$ eV (crosses and dotted lines), T$^{\infty}=36$ eV (plus signs and dashed lines) and T$^{\infty}=40$ eV (circles and solid lines), with light element envelope. The blue line marks the special case of EOS I, $\mathrm{p}_\mathrm{trans}= 38$ MeV, $\mathrm{n}_\mathrm{trans}= 2 \, \mathrm{n}_\mathrm{sat}$, $\Delta \epsilon = 280$ MeV. In brown, the results for EOS II with nuclear pairing combination NT AO+PS CCYps are shown. In the lower plot, the hybrid stars corresponding quark radius fraction in relation to total radius is shown.}
    \label{fig:allT_Minf_50}
\end{figure}

\begin{figure}
    \centering
    \includegraphics[scale=0.6]{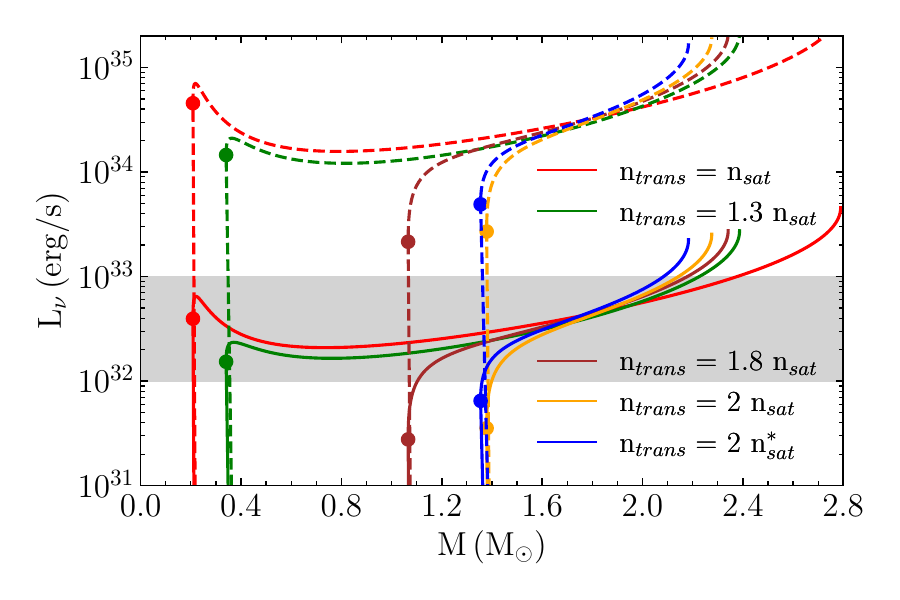}
    \caption{Total luminosity and mass of EOS I hybrid stars for T$^{\infty}=36$ eV with light elements envelope (solid lines) and heavy elements envelope (dashed lines). \SAX's neutron star predicted luminosity is highlighted by the grey region. Different phase transitions are indicated by different colors according to the label. Circles mark the first stable star for each EOS. The blue line with label $\mathrm{n}_\mathrm{trans}= 2 \, \mathrm{n}^{*}_\mathrm{sat}$ is the case $\mathrm{p}_\mathrm{trans}= 38$ MeV, $\mathrm{n}_\mathrm{trans}= 2 \, \mathrm{n}_\mathrm{sat}$ and $\Delta \epsilon = 280$ MeV.}
    \label{fig:LxM_T36_50}
\end{figure}

For both EOSs, the mass ranges predicted for hybrid stars with phase transitions at low densities ($\mathrm{n}_\mathrm{trans} \lesssim 2.5 \, \mathrm{n}_\mathrm{sat}$) in Figure~\ref{fig:T55_Minf_50} are larger because these stars have a larger quark phase volume fraction (close to $100 \%$), thus  present a larger volume undergoing quark dUrca cooling. This suggests that, for those EOSs and phase transitions, the full range of inferred luminosities of colder sources is potentially not achieved in the quark volume available, that is, these twin stars would not reproduce the neutrino luminosity of colder sources. This is indeed the case for EOS I and EOS II, which fail to reproduce \SAX's neutron star inferred luminosity (between $10^{32}$ to $10^{33}$ erg/s) completely for phase transitions with $\mathrm{n}_\mathrm{trans}< 1.8 \, \mathrm{n}_\mathrm{sat}$, for all envelope compositions tested. This scenario is examined in more detail in Figure~\ref{fig:allT_Minf_50} and Figure~\ref{fig:LxM_T36_50}. In Figure~\ref{fig:allT_Minf_50}, the calculated masses for the neutron star in \SAX\ are shown, considering a light element envelope and $28 \leq $ T$^{\infty} \leq 40$ eV. The predicted masses for each temperature are indicated with different symbols. Despite the general trend of lower effective temperatures resulting in larger masses for \SAX's neutron star, the fraction of quark radius relative to the total radius necessary to reach the inferred luminosity is always between $90\%$ and $100\%$ for $\mathrm{n}_\mathrm{trans}< 1.8 \, \mathrm{n}_\mathrm{sat}$. In Figure~\ref{fig:LxM_T36_50}, we examine in more detail the luminosity curves of EOS~I with T$^{\infty}=36$ eV, where the inferred luminosity of the neutron star in \SAX\ is indicated by the horizontal grey band and the first stable star for each phase transition is represented by a dot, such that lower masses correspond to unstable stars. Hence, it is clear that the lower limit of \SAX's neutron star luminosities is reached only in unstable stars for phase transitions $\mathrm{n}_\mathrm{trans}< 1.8 \, \mathrm{n}_\mathrm{sat}$, with a light element envelope, thus these phase transitions are inconsistent with the full error band of cooling data. 

Including nuclear pairing in EOS II with \SAX's neutron star effective temperatures produces a similar effect to that seen in Figure~\ref{fig:T55_Minf_50} for \MXB. Strong pairing such as NT AO+PS CCYps reduces the predicted mass range because it suppresses the nucleonic dUrca reactions, forcing most of the luminosity to be generated by quark dUrca processes. In this scenario, the cases with low temperatures in particular have problems in reproducing the lower limit of \SAX's neutron star luminosity. We note that, for both EOSs, assuming a heavy element envelope composition for the neutron star in \SAX\ results in luminosities inconsistent with the source's inferred cooling data for most phase transitions and temperatures.

\subsection{\label{sec:L90}Hybrid stars with hadronic $L=90$ EOSs}

The dUrca threshold of hadronic $L=90$ EOSs is lower than for the $L=50$ EOSs (see Table \ref{tab:EOS}), thus there are more cases of hybrid stars constructed involving both nucleonic and quark dUrca cooling. However, the increased stiffness of the nucleonic EOSs only allows for relatively low density phase transitions, of at most $n_{trans} \lesssim 2.4 \, n_{sat}$, to be compatible with astrophysical data \cite{Christian:2019,Christian:2021,Christian:2023hez}, hence only a relatively small range of possible phase transition densities can be probed. The different Dirac effective masses $M^{*}$ of EOS III and EOS IV only marginally affect the allowed quark-hadron phase transition densities and luminosity curves. 

\begin{figure}
    \centering
    \includegraphics[scale=0.6]{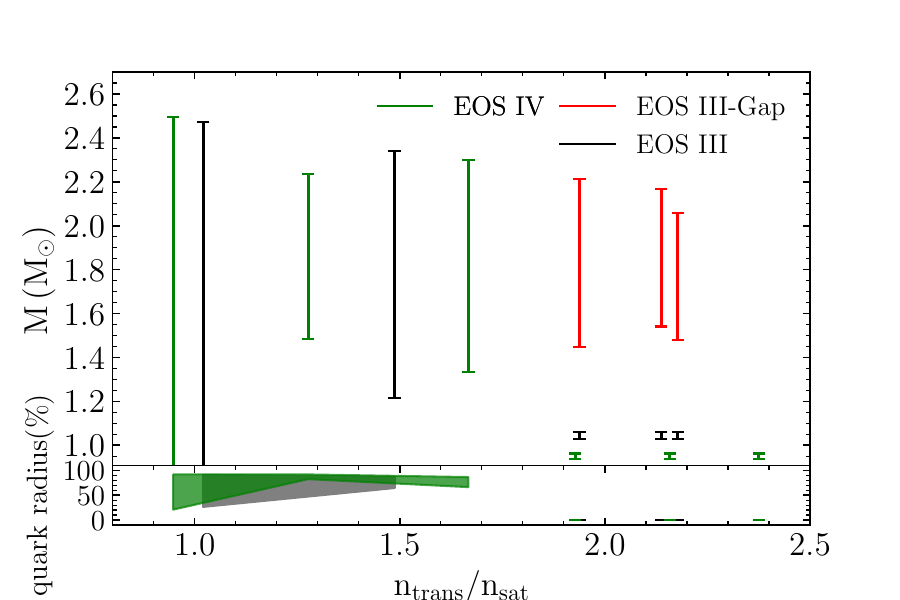}
    \caption{Total masses of the hybrid stars with \MXB's neutron star lower and upper inferred neutrino luminosities, including superfluidity in EOS III (PS CCDK+NT EEHOr gap model combination in red), without superfluidity (black), and EOS IV without superfluidity (green). The density transitions $\mathrm{n}_\mathrm{trans} \, (\mathrm{n}_\mathrm{sat})$ are represented on the x-axis. In the lower plot, the hybrid stars corresponding quark radius fraction in relation to total radius is shown.}
    \label{fig:T55_Minf_90}
\end{figure}

\begin{figure}
    \centering
    \includegraphics[scale=0.6]{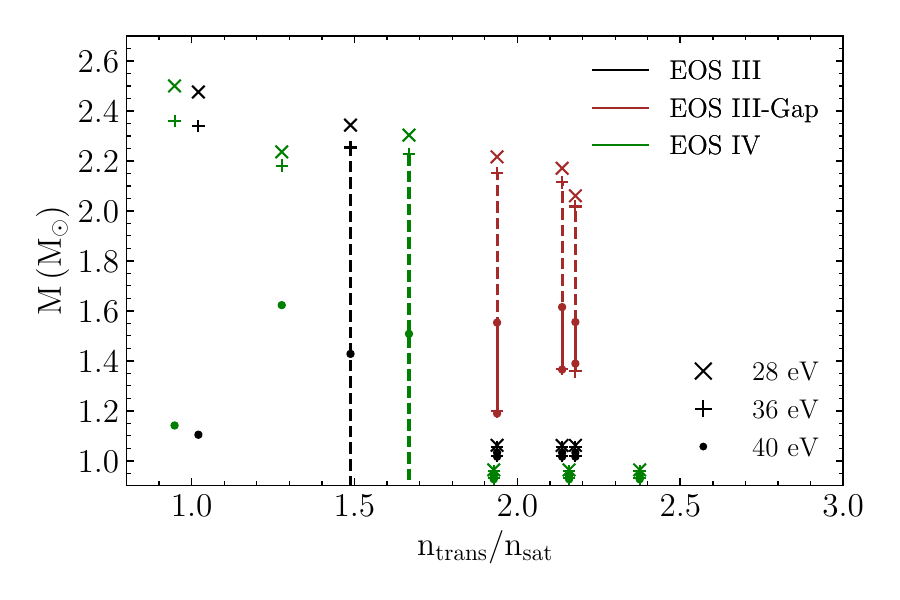}
    \caption{Total masses of the hybrid stars with \SAX's lower and upper inferred neutrino luminosities, including superfluidity in EOS III (PS CCYps+NT AO gap model combination in brown), without superfluidity (black), and EOS IV without superfluidity (green), for T$^{\infty}=28$ eV (crosses and dotted-dashed lines), T$^{\infty}=36$ eV (plus signs and dashed lines) and T$^{\infty}=40$ eV (circles and solid lines), with a light element envelope. The density transitions $\mathrm{n}_\mathrm{trans} \, (\mathrm{n}_\mathrm{sat})$ are represented on the x-axis.}
    \label{fig:allT_Minf_90}
\end{figure}

The general pattern of larger mass ranges for lower phase transition densities is repeated for the $L=90$ EOSs, as seen in Figure~\ref{fig:T55_Minf_90} and \ref{fig:allT_Minf_90}. For these EOSs, large quark volumes are still needed to reproduce the upper limit of inferred luminosities for the neutron star in \MXB, thus colder sources tend to be incompatible with small phase transition densities, around $n_{trans} < 1.8 \, n_{sat}$. This trend is also observed for reproducing \SAX's neutron star data, whose full inferred luminosity band is only achieved assuming a light element envelope composition and $n_{trans} \gtrsim 1.9 \, n_{sat}$, for both EOS III and IV and $28 <$ T$^{\infty} \leq 40$~eV. A less likely but still possible scenario for this source is a heavy element envelope composition with T$^{\infty}=28$~eV and $n_{trans} \gtrsim 1.9 \, n_{sat}$, where the luminosities are reached by nucleonic dUrca, thus the mass ranges are very small, or a light element envelope with T$^{\infty}=36$~eV and $n_{trans}= 1.5 \, n_{sat}$ for EOS III or $n_{trans}= 1.7 \, n_{sat}$ for EOS IV, where the source luminosity is reached by quark dUrca processes only, as shown in Figure~\ref{fig:allT_Minf_90} for the light element envelope composition. However, in these cases, \SAX's neutron star lower luminosity limit is reached for sources with masses $M < 1 M_{\odot}$ .

Including nuclear pairing results in quark dUrca processes being dominant for densities $n_{trans} > 1.9 \, n_{sat}$, which leads to a visible increasing of the mass ranges from around one solar mass, for the non-superfluid case, to values centered around $1.8 \, \mathrm{M}_{\odot}$ for \MXB\ and nuclear pairing model PS CCDK+NT EEHOr, as seen in Figure~\ref{fig:T55_Minf_90}. For \SAX, including nuclear pairing also significantly increases the mass ranges above one solar mass for both EOSs, as seen in Figure~\ref{fig:allT_Minf_90} for EOS III. Both EOSs favour strong proton superconductivity gap models, such as PS CCDK and PS CCYps, and are agnostic about the neutron superfluidity ones, given the relatively low densities of onset of the quark phase, which starts before neutron pairing can significantly affect the luminosity curves.

\section{\label{sec:conclusion}Discussion and conclusions}

Reproducing the luminosity of fast cooling sources with twin stars is challenging. Firstly, one must build quark-hadron EOSs consistent with astrophysical data, for which the biggest limiting factor is the tidal deformability constraint provided by GW170817 \citep{Abbott:2019}. Since our hadronic EOSs feature comparatively large radii with large values of $\Lambda$, a small transition density is needed to achieve compatibility with this constraint, thus the range of allowed phase transition densities for EOS III and IV, with L$=90$ MeV, is limited to n $< \, 2.5 \, \mathrm{n}_\mathrm{sat}$ and for EOS I, with L$=50$ MeV, to n $< \, 2 \, \mathrm{n}_\mathrm{sat}$.  
The quark-hadron phase transition density for EOS II can be as high as $4.6 \, \mathrm{n}_\mathrm{sat}$, but the requirement that mass-radius contours from NICER \citep{Riley:2021, Miller:2021} are reproduced leaves a gap between the allowed phase transition densities such that $1.2 \, \mathrm{n}_\mathrm{sat} \leq \mathrm{n}_\mathrm{trans}\leq \, 2.2 \, \mathrm{n}_\mathrm{sat}$ or $4 \, \mathrm{n}_\mathrm{sat} \leq \mathrm{n}_\mathrm{trans}\leq \, 4.6 \, \mathrm{n}_\mathrm{sat}$. This EOS allows for a larger range of transition densities generating twin stars, opening up the high density regime. 

However, due to the smaller efficiency of quark dUrca processes compared to nucleonic dUrca processes, stable twin stars with low density phase transitions tend to struggle reproducing the low luminosities of cold neutron stars. In particular, quark-hadron phase transitions densities below $1.7 \, \mathrm{n}_\mathrm{sat}$ in twin stars of all EOSs under study fail to reproduce the full range of \SAX's neutron star inferred luminosities, usually only succeeding to reproduce their upper limit. Therefore, these low density quark-hadron phase transitions, up to $1.7$ $n_{\rm{sat}}$, are disfavoured over the ones able to reproduce the full luminosity range, suggesting that twin stars with low density transitions to the quark phase are not realized in nature.

\begin{figure}
    \centering
    \includegraphics[scale=0.6]{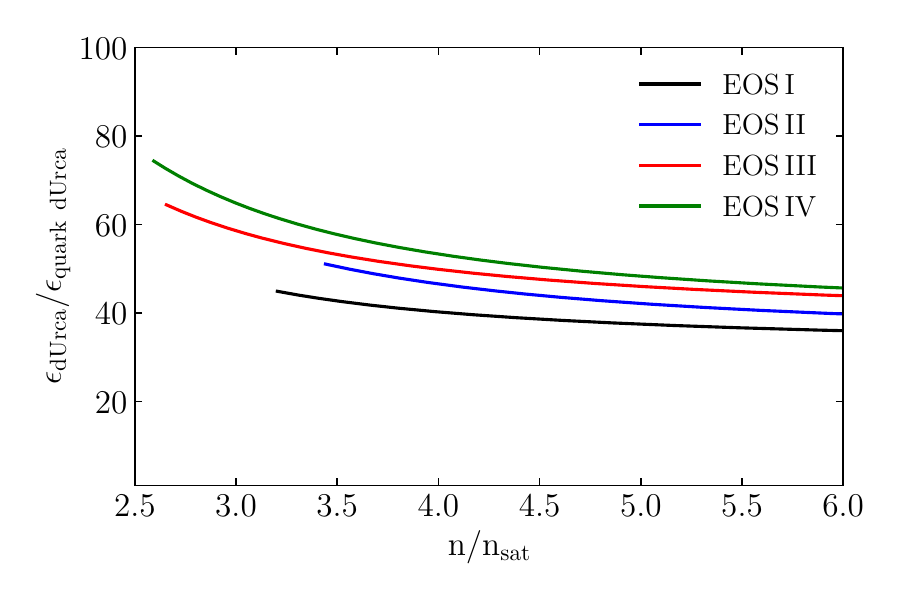}
    \caption{Ratio of nucleonic dUrca emissivity $(\epsilon_{\mathrm{dUrca}})$ to quark dUrca emissivity $(\epsilon_{\mathrm{quark} \ \mathrm{dUrca}})$, assuming these reactions are simultaneously active in the quark phase, for all EOSs.}
    \label{fig:ehadxeq}
\end{figure}

EOSs with larger phase transition densities successfully reproduce \SAX's neutron star luminosity, often with a light element envelope composition and $28 \leq$ T$^{\infty} \leq 36$ eV.
For phase transition densities above the nucleonic dUrca threshold, that luminosity is obtained with nucleonic processes only and the quark cooling contribution is very minor and hardly detectable. In these cases, including nuclear pairing in the nucleonic phase suppresses nucleonic dUrca reactions, turning quark dUrca into the leading cooling process. However, because the hybrid star volume in the quark phase is small, these stars often fail to reach low luminosities consistent with cold neutron stars. In this scenario, the full range of \SAX's neutron star inferred luminosity has been successfully reproduced for EOS II containing weak nuclear pairing models both for protons and neutrons, and for EOS III and IV with strong proton superconductivity models.
All phase transitions of EOS I are below the nucleonic dUrca threshold. This result could be an indication that twin stars with low $L$ hadronic EOSs, with transition density to the quark phase $4 \, \mathrm{n}_\mathrm{sat} \leq \mathrm{n}_\mathrm{trans}$, favor weak nuclear pairing whereas large $L$ ones, which can only hold transition densities of $2.5 \, \mathrm{n}_\mathrm{sat} < \mathrm{n}_\mathrm{trans}$, favor strong proton pairing and very weak neutron triplet pairing in the nucleonic phase.

Hybrid stars that are not twin stars allow for more phase transitions, particularly above $2 \, \mathrm{n}_\mathrm{sat}$, consistent with astrophysical constraints, including cooling data. Thus, while twin stars can still describe observations within a certain range of phase transition densities for some EOSs, this could be an indication that cooling data favours non-twin star solutions. Overall, hadronic EOSs with lower values of the slope of symmetry density $L$ tend to allow for twin stars with larger quark-hadron phase transition densities. Furthermore, their hybrid stars tend to have lower radii, so they more easily match NICER constraints than their large $L$ counterparts.  Therefore, it is more likely for EOSs with small values of $L$ to feature twin stars which respect to astrophysical constraints, including cooling.

The cooling calculations of the quark phase have been performed with an estimate of the electron fraction of $Y_e=10^{-5}$, and strong coupling constant, $\alpha=0.12$. These factors determine the difference in effectiveness of nucleonic and quark dUrca emissivities. Their ratio, which is independent of the source's temperature, is shown in Figure~\ref{fig:ehadxeq} for all EOSs studied in this work. Nucleonic dUrca reactions are just under $100$ times more effective than quark dUrca reactions for these values of $Y_e$ and $\alpha$, a ratio very similar to all EOSs, peaking at low densities. Therefore, when both processes are active, quark dUrca cooling is subdominant and hardly noticeable. Our estimate for the electron fraction of the quark phase was based on the assumption that it consists of up, down and strange quarks in weak equilibrium. Then, assuming the quark phase is charge neutral and that up and down quarks are approximately massless and non-interacting fermions, one can estimate that
\begin{equation}
    \mu_e \approx \frac{m_s^2}{4\mu_q},   
\end{equation}
hence, 
\begin{equation}
    Y_e \approx \frac{1}{3}\left(\frac{m_s}{2\mu_q}\right)^6 \approx 10^{-5}-10^{-7},  
\end{equation}
for typical quark chemical potentials of $\mu_q = 300-500$ MeV \cite{Schaffner-Bielich:2020}. This estimate is consistent with the calculations of more realistic NJL quark EOSs. NJL models without color superconductivity estimate that the electron fraction for two or three flavour quark matter is, at most, of the order of $Y_e \approx 10^{-3}$, reducing with increasing density \cite{Buballa:2003}. Performing the cooling calculations with this value of electron fraction for the quark phase, we obtain qualitatively similar results to the $Y_e =10^{-5}$ case discussed in this paper: low density phase transitions for all EOSs struggle to reproduce the full range of cold neutron stars' luminosities. Furthermore, the calculated masses in the $Y_e =10^{-3}$ scenario are smaller and the mass ranges are reduced, as expected from a more efficient quark cooling. Future work will study the effects of consistently calculated neutrino emissivities for a specific quark EOS, including color superconductivity phases, to investigate the sensitivity of these results. 

\begin{acknowledgments}
The authors thank Charles Gale and Andrew Cumming for insightful discussions as well as the anonymous referee for suggestions that greatly improved the text. MM and JEC especially thank the late Stephan Wystub for suggestions and comments and dedicate this work to his memory.

MM was supported by the MITACS Globalink Research Award IT32929, the Schlumberger Foundation with a Faculty for the Future Fellowship, the European Research Council (ERC) under the European Union's Horizon 2020 research and innovation programme (Grant Agreement No.~101020842), and by the Deutsche Forschungsgemeinschaft (DFG, German Research Foundation)---Project-ID 279384907---SFB 1245.
JEC was funded by the European Research Council (ERC) Advanced Grant INSPIRATION under the European Union's Horizon 2020 research and innovation programme (Grant agreement No. 101053985). 
JSB acknowledges support by the Deutsche Forschungsgemeinschaft (DFG, German Research Foundation) through the CRC-TR 211 `Strong-interaction matter under extreme conditions'– project number 315477589 – TRR 211.
\end{acknowledgments}

\bibliography{Bibliography}

\begin{thebibliography}{122}%
\makeatletter
\providecommand \@ifxundefined [1]{%
 \@ifx{#1\undefined}
}%
\providecommand \@ifnum [1]{%
 \ifnum #1\expandafter \@firstoftwo
 \else \expandafter \@secondoftwo
 \fi
}%
\providecommand \@ifx [1]{%
 \ifx #1\expandafter \@firstoftwo
 \else \expandafter \@secondoftwo
 \fi
}%
\providecommand \natexlab [1]{#1}%
\providecommand \enquote  [1]{``#1''}%
\providecommand \bibnamefont  [1]{#1}%
\providecommand \bibfnamefont [1]{#1}%
\providecommand \citenamefont [1]{#1}%
\providecommand \href@noop [0]{\@secondoftwo}%
\providecommand \href [0]{\begingroup \@sanitize@url \@href}%
\providecommand \@href[1]{\@@startlink{#1}\@@href}%
\providecommand \@@href[1]{\endgroup#1\@@endlink}%
\providecommand \@sanitize@url [0]{\catcode `\\12\catcode `\$12\catcode `\&12\catcode `\#12\catcode `\^12\catcode `\_12\catcode `\%12\relax}%
\providecommand \@@startlink[1]{}%
\providecommand \@@endlink[0]{}%
\providecommand \url  [0]{\begingroup\@sanitize@url \@url }%
\providecommand \@url [1]{\endgroup\@href {#1}{\urlprefix }}%
\providecommand \urlprefix  [0]{URL }%
\providecommand \Eprint [0]{\href }%
\providecommand \doibase [0]{http://dx.doi.org/}%
\providecommand \selectlanguage [0]{\@gobble}%
\providecommand \bibinfo  [0]{\@secondoftwo}%
\providecommand \bibfield  [0]{\@secondoftwo}%
\providecommand \translation [1]{[#1]}%
\providecommand \BibitemOpen [0]{}%
\providecommand \bibitemStop [0]{}%
\providecommand \bibitemNoStop [0]{.\EOS\space}%
\providecommand \EOS [0]{\spacefactor3000\relax}%
\providecommand \BibitemShut  [1]{\csname bibitem#1\endcsname}%
\let\auto@bib@innerbib\@empty
\bibitem [{\citenamefont {Lattimer}(2023)}]{Lattimer:2023rpe}%
  \BibitemOpen
  \bibfield  {author} {\bibinfo {author} {\bibfnamefont {J.~M.}\ \bibnamefont {Lattimer}},\ }\href {\doibase 10.3390/particles6010003} {\bibfield  {journal} {\bibinfo  {journal} {Particles}\ }\textbf {\bibinfo {volume} {6}},\ \bibinfo {pages} {30} (\bibinfo {year} {2023})},\ \Eprint {http://arxiv.org/abs/2301.03666} {arXiv:2301.03666 [nucl-th]} \BibitemShut {NoStop}%
\bibitem [{\citenamefont {Kumar}\ \emph {et~al.}(2024)\citenamefont {Kumar} \emph {et~al.}}]{MUSES:2023hyz}%
  \BibitemOpen
  \bibfield  {author} {\bibinfo {author} {\bibfnamefont {R.}~\bibnamefont {Kumar}} \emph {et~al.} (\bibinfo {collaboration} {MUSES}),\ }\href {\doibase 10.1007/s41114-024-00049-6} {\bibfield  {journal} {\bibinfo  {journal} {Living Rev. Rel.}\ }\textbf {\bibinfo {volume} {27}},\ \bibinfo {pages} {3} (\bibinfo {year} {2024})},\ \Eprint {http://arxiv.org/abs/2303.17021} {arXiv:2303.17021 [nucl-th]} \BibitemShut {NoStop}%
\bibitem [{\citenamefont {Johnson}\ and\ \citenamefont {Teller}(1955)}]{PhysRev.98.783}%
  \BibitemOpen
  \bibfield  {author} {\bibinfo {author} {\bibfnamefont {M.~H.}\ \bibnamefont {Johnson}}\ and\ \bibinfo {author} {\bibfnamefont {E.}~\bibnamefont {Teller}},\ }\href {\doibase 10.1103/PhysRev.98.783} {\bibfield  {journal} {\bibinfo  {journal} {Phys. Rev.}\ }\textbf {\bibinfo {volume} {98}},\ \bibinfo {pages} {783} (\bibinfo {year} {1955})}\BibitemShut {NoStop}%
\bibitem [{\citenamefont {Duerr}(1956)}]{Duerr56}%
  \BibitemOpen
  \bibfield  {author} {\bibinfo {author} {\bibfnamefont {H.-P.}\ \bibnamefont {Duerr}},\ }\href@noop {} {\bibfield  {journal} {\bibinfo  {journal} {Phys. Rev.}\ }\textbf {\bibinfo {volume} {103}},\ \bibinfo {pages} {469} (\bibinfo {year} {1956})}\BibitemShut {NoStop}%
\bibitem [{\citenamefont {{Walecka}}(1974)}]{Walecka74}%
  \BibitemOpen
  \bibfield  {author} {\bibinfo {author} {\bibfnamefont {J.~D.}\ \bibnamefont {{Walecka}}},\ }\href {\doibase 10.1016/0003-4916(74)90208-5} {\bibfield  {journal} {\bibinfo  {journal} {Annals of Physics}\ }\textbf {\bibinfo {volume} {83}},\ \bibinfo {pages} {491} (\bibinfo {year} {1974})}\BibitemShut {NoStop}%
\bibitem [{\citenamefont {Boguta}\ and\ \citenamefont {Bodmer}(1977)}]{Boguta:1977xi}%
  \BibitemOpen
  \bibfield  {author} {\bibinfo {author} {\bibfnamefont {J.}~\bibnamefont {Boguta}}\ and\ \bibinfo {author} {\bibfnamefont {A.~R.}\ \bibnamefont {Bodmer}},\ }\href {\doibase 10.1016/0375-9474(77)90626-1} {\bibfield  {journal} {\bibinfo  {journal} {Nucl. Phys. A}\ }\textbf {\bibinfo {volume} {292}},\ \bibinfo {pages} {413} (\bibinfo {year} {1977})}\BibitemShut {NoStop}%
\bibitem [{\citenamefont {Serot}\ and\ \citenamefont {Walecka}(1968)}]{Serot:1984ey}%
  \BibitemOpen
  \bibfield  {author} {\bibinfo {author} {\bibfnamefont {B.~D.}\ \bibnamefont {Serot}}\ and\ \bibinfo {author} {\bibfnamefont {J.~D.}\ \bibnamefont {Walecka}},\ }\href {https://inspirehep.net/literature/207866} {\emph {\bibinfo {title} {The Relativistic Nuclear Many-body Problem}}},\ edited by\ \bibinfo {editor} {\bibfnamefont {M.}~\bibnamefont {Baranger}}\ and\ \bibinfo {editor} {\bibfnamefont {E.}~\bibnamefont {Vogt.}}\ (\bibinfo  {publisher} {PLENUM PRESS · NEW YORK-LONDON},\ \bibinfo {year} {1968})\BibitemShut {NoStop}%
\bibitem [{\citenamefont {Mueller}\ and\ \citenamefont {Serot}(1996)}]{Mueller:1996pm}%
  \BibitemOpen
  \bibfield  {author} {\bibinfo {author} {\bibfnamefont {H.}~\bibnamefont {Mueller}}\ and\ \bibinfo {author} {\bibfnamefont {B.~D.}\ \bibnamefont {Serot}},\ }\href {\doibase 10.1016/0375-9474(96)00187-X} {\bibfield  {journal} {\bibinfo  {journal} {Nucl. Phys.}\ }\textbf {\bibinfo {volume} {A606}},\ \bibinfo {pages} {508} (\bibinfo {year} {1996})},\ \Eprint {http://arxiv.org/abs/nucl-th/9603037} {arXiv:nucl-th/9603037 [nucl-th]} \BibitemShut {NoStop}%
\bibitem [{\citenamefont {Todd-Rutel}\ and\ \citenamefont {Piekarewicz}(2005)}]{Todd:2005}%
  \BibitemOpen
  \bibfield  {author} {\bibinfo {author} {\bibfnamefont {B.~G.}\ \bibnamefont {Todd-Rutel}}\ and\ \bibinfo {author} {\bibfnamefont {J.}~\bibnamefont {Piekarewicz}},\ }\href {\doibase 10.1103/PhysRevLett.95.122501} {\bibfield  {journal} {\bibinfo  {journal} {Phys. Rev. Lett.}\ }\textbf {\bibinfo {volume} {95}},\ \bibinfo {pages} {122501} (\bibinfo {year} {2005})}\BibitemShut {NoStop}%
\bibitem [{\citenamefont {Chen}\ and\ \citenamefont {Piekarewicz}(2014{\natexlab{a}})}]{Chen:2014}%
  \BibitemOpen
  \bibfield  {author} {\bibinfo {author} {\bibfnamefont {W.-C.}\ \bibnamefont {Chen}}\ and\ \bibinfo {author} {\bibfnamefont {J.}~\bibnamefont {Piekarewicz}},\ }\href {\doibase https://doi.org/10.1103/PhysRevC.90.044305} {\bibfield  {journal} {\bibinfo  {journal} {Phys. Rev. C}\ }\textbf {\bibinfo {volume} {90}},\ \bibinfo {pages} {044305} (\bibinfo {year} {2014}{\natexlab{a}})}\BibitemShut {NoStop}%
\bibitem [{\citenamefont {Fattoyev}\ \emph {et~al.}(2018)\citenamefont {Fattoyev}, \citenamefont {Piekarewicz},\ and\ \citenamefont {Horowitz}}]{Fattoyev:2017}%
  \BibitemOpen
  \bibfield  {author} {\bibinfo {author} {\bibfnamefont {F.~J.}\ \bibnamefont {Fattoyev}}, \bibinfo {author} {\bibfnamefont {J.}~\bibnamefont {Piekarewicz}}, \ and\ \bibinfo {author} {\bibfnamefont {C.~J.}\ \bibnamefont {Horowitz}},\ }\href {\doibase 10.1103/PhysRevLett.120.172702} {\bibfield  {journal} {\bibinfo  {journal} {Phys. Rev. Lett.}\ }\textbf {\bibinfo {volume} {120}},\ \bibinfo {pages} {172702} (\bibinfo {year} {2018})},\ \Eprint {http://arxiv.org/abs/1711.06615} {arXiv:1711.06615 [nucl-th]} \BibitemShut {NoStop}%
\bibitem [{\citenamefont {Danielewicz}\ \emph {et~al.}(2002)\citenamefont {Danielewicz}, \citenamefont {Lacey},\ and\ \citenamefont {Lynch}}]{Danielewicz:2002pu}%
  \BibitemOpen
  \bibfield  {author} {\bibinfo {author} {\bibfnamefont {P.}~\bibnamefont {Danielewicz}}, \bibinfo {author} {\bibfnamefont {R.}~\bibnamefont {Lacey}}, \ and\ \bibinfo {author} {\bibfnamefont {W.~G.}\ \bibnamefont {Lynch}},\ }\href {\doibase 10.1126/science.1078070} {\bibfield  {journal} {\bibinfo  {journal} {Science}\ }\textbf {\bibinfo {volume} {298}},\ \bibinfo {pages} {1592} (\bibinfo {year} {2002})},\ \Eprint {http://arxiv.org/abs/nucl-th/0208016} {arXiv:nucl-th/0208016} \BibitemShut {NoStop}%
\bibitem [{\citenamefont {Le~F\`evre}\ \emph {et~al.}(2016)\citenamefont {Le~F\`evre}, \citenamefont {Leifels}, \citenamefont {Reisdorf}, \citenamefont {Aichelin},\ and\ \citenamefont {Hartnack}}]{LeFevre:2015paj}%
  \BibitemOpen
  \bibfield  {author} {\bibinfo {author} {\bibfnamefont {A.}~\bibnamefont {Le~F\`evre}}, \bibinfo {author} {\bibfnamefont {Y.}~\bibnamefont {Leifels}}, \bibinfo {author} {\bibfnamefont {W.}~\bibnamefont {Reisdorf}}, \bibinfo {author} {\bibfnamefont {J.}~\bibnamefont {Aichelin}}, \ and\ \bibinfo {author} {\bibfnamefont {C.}~\bibnamefont {Hartnack}},\ }\href {\doibase 10.1016/j.nuclphysa.2015.09.015} {\bibfield  {journal} {\bibinfo  {journal} {Nucl. Phys. A}\ }\textbf {\bibinfo {volume} {945}},\ \bibinfo {pages} {112} (\bibinfo {year} {2016})},\ \Eprint {http://arxiv.org/abs/1501.05246} {arXiv:1501.05246 [nucl-ex]} \BibitemShut {NoStop}%
\bibitem [{\citenamefont {Morfouace}\ \emph {et~al.}(2019)\citenamefont {Morfouace} \emph {et~al.}}]{Morfouace:2019jky}%
  \BibitemOpen
  \bibfield  {author} {\bibinfo {author} {\bibfnamefont {P.}~\bibnamefont {Morfouace}} \emph {et~al.},\ }\href {\doibase 10.1016/j.physletb.2019.135045} {\bibfield  {journal} {\bibinfo  {journal} {Phys. Lett. B}\ }\textbf {\bibinfo {volume} {799}},\ \bibinfo {pages} {135045} (\bibinfo {year} {2019})},\ \Eprint {http://arxiv.org/abs/1904.12471} {arXiv:1904.12471 [nucl-ex]} \BibitemShut {NoStop}%
\bibitem [{\citenamefont {Demorest}\ \emph {et~al.}(2010)\citenamefont {Demorest}, \citenamefont {Pennucci}, \citenamefont {Ransom}, \citenamefont {Roberts},\ and\ \citenamefont {Hessels}}]{Demorest:2010bx}%
  \BibitemOpen
  \bibfield  {author} {\bibinfo {author} {\bibfnamefont {P.}~\bibnamefont {Demorest}}, \bibinfo {author} {\bibfnamefont {T.}~\bibnamefont {Pennucci}}, \bibinfo {author} {\bibfnamefont {S.}~\bibnamefont {Ransom}}, \bibinfo {author} {\bibfnamefont {M.}~\bibnamefont {Roberts}}, \ and\ \bibinfo {author} {\bibfnamefont {J.}~\bibnamefont {Hessels}},\ }\href {\doibase 10.1038/nature09466} {\bibfield  {journal} {\bibinfo  {journal} {Nature}\ }\textbf {\bibinfo {volume} {467}},\ \bibinfo {pages} {1081} (\bibinfo {year} {2010})},\ \Eprint {http://arxiv.org/abs/1010.5788} {arXiv:1010.5788 [astro-ph.HE]} \BibitemShut {NoStop}%
\bibitem [{\citenamefont {Antoniadis}\ \emph {et~al.}(2013)\citenamefont {Antoniadis}, \citenamefont {Freire}, \citenamefont {Wex}, \citenamefont {Tauris}, \citenamefont {Lynch}, \citenamefont {van Kerkwijk}, \citenamefont {Kramer}, \citenamefont {Bassa}, \citenamefont {Dhillon}, \citenamefont {Driebe}, \citenamefont {Hessels}, \citenamefont {Kaspi}, \citenamefont {Kondratiev}, \citenamefont {Langer}, \citenamefont {Marsh}, \citenamefont {McLaughlin}, \citenamefont {Pennucci}, \citenamefont {Ransom}, \citenamefont {Stairs}, \citenamefont {van Leeuwen}, \citenamefont {Verbiest},\ and\ \citenamefont {Whelan}}]{Antoniadis:2013pzd}%
  \BibitemOpen
  \bibfield  {author} {\bibinfo {author} {\bibfnamefont {J.}~\bibnamefont {Antoniadis}}, \bibinfo {author} {\bibfnamefont {P.~C.}\ \bibnamefont {Freire}}, \bibinfo {author} {\bibfnamefont {N.}~\bibnamefont {Wex}}, \bibinfo {author} {\bibfnamefont {T.~M.}\ \bibnamefont {Tauris}}, \bibinfo {author} {\bibfnamefont {R.~S.}\ \bibnamefont {Lynch}}, \bibinfo {author} {\bibfnamefont {M.~H.}\ \bibnamefont {van Kerkwijk}}, \bibinfo {author} {\bibfnamefont {M.}~\bibnamefont {Kramer}}, \bibinfo {author} {\bibfnamefont {C.}~\bibnamefont {Bassa}}, \bibinfo {author} {\bibfnamefont {V.~S.}\ \bibnamefont {Dhillon}}, \bibinfo {author} {\bibfnamefont {T.}~\bibnamefont {Driebe}}, \bibinfo {author} {\bibfnamefont {J.~W.~T.}\ \bibnamefont {Hessels}}, \bibinfo {author} {\bibfnamefont {V.~M.}\ \bibnamefont {Kaspi}}, \bibinfo {author} {\bibfnamefont {V.~I.}\ \bibnamefont {Kondratiev}}, \bibinfo {author} {\bibfnamefont {N.}~\bibnamefont {Langer}}, \bibinfo {author} {\bibfnamefont {T.~R.}\ \bibnamefont {Marsh}}, \bibinfo {author}
  {\bibfnamefont {M.~A.}\ \bibnamefont {McLaughlin}}, \bibinfo {author} {\bibfnamefont {T.~T.}\ \bibnamefont {Pennucci}}, \bibinfo {author} {\bibfnamefont {S.~M.}\ \bibnamefont {Ransom}}, \bibinfo {author} {\bibfnamefont {I.~H.}\ \bibnamefont {Stairs}}, \bibinfo {author} {\bibfnamefont {J.}~\bibnamefont {van Leeuwen}}, \bibinfo {author} {\bibfnamefont {J.~P.~W.}\ \bibnamefont {Verbiest}}, \ and\ \bibinfo {author} {\bibfnamefont {D.~G.}\ \bibnamefont {Whelan}},\ }\href {\doibase 10.1126/science.1233232} {\bibfield  {journal} {\bibinfo  {journal} {Science}\ }\textbf {\bibinfo {volume} {340}},\ \bibinfo {pages} {6131} (\bibinfo {year} {2013})},\ \Eprint {http://arxiv.org/abs/1304.6875} {arXiv:1304.6875 [astro-ph.HE]} \BibitemShut {NoStop}%
\bibitem [{\citenamefont {Fonseca}\ \emph {et~al.}(2016)\citenamefont {Fonseca}, \citenamefont {Pennucci}, \citenamefont {Ellis},\ and\ \citenamefont {other}}]{Fonseca:2016tux}%
  \BibitemOpen
  \bibfield  {author} {\bibinfo {author} {\bibfnamefont {E.}~\bibnamefont {Fonseca}}, \bibinfo {author} {\bibfnamefont {T.~T.}\ \bibnamefont {Pennucci}}, \bibinfo {author} {\bibfnamefont {J.~A.}\ \bibnamefont {Ellis}}, \ and\ \bibinfo {author} {\bibnamefont {other}},\ }\href {\doibase 10.3847/0004-637X/832/2/167} {\bibfield  {journal} {\bibinfo  {journal} {Astrophys. J.}\ }\textbf {\bibinfo {volume} {832}},\ \bibinfo {pages} {167} (\bibinfo {year} {2016})},\ \Eprint {http://arxiv.org/abs/1603.00545} {arXiv:1603.00545 [astro-ph.HE]} \BibitemShut {NoStop}%
\bibitem [{\citenamefont {Cromartie}\ \emph {et~al.}(2019)\citenamefont {Cromartie} \emph {et~al.}}]{Cromartie:2019}%
  \BibitemOpen
  \bibfield  {author} {\bibinfo {author} {\bibfnamefont {H.~T.}\ \bibnamefont {Cromartie}} \emph {et~al.} (\bibinfo {collaboration} {NANOGrav}),\ }\href {\doibase 10.1038/s41550-019-0880-2} {\bibfield  {journal} {\bibinfo  {journal} {Nature Astron.}\ }\textbf {\bibinfo {volume} {4}},\ \bibinfo {pages} {72} (\bibinfo {year} {2019})},\ \Eprint {http://arxiv.org/abs/1904.06759} {arXiv:1904.06759 [astro-ph.HE]} \BibitemShut {NoStop}%
\bibitem [{\citenamefont {Romani}\ \emph {et~al.}(2022)\citenamefont {Romani}, \citenamefont {Kandel}, \citenamefont {Filippenko}, \citenamefont {Brink},\ and\ \citenamefont {Zheng}}]{Romani:2022jhd}%
  \BibitemOpen
  \bibfield  {author} {\bibinfo {author} {\bibfnamefont {R.~W.}\ \bibnamefont {Romani}}, \bibinfo {author} {\bibfnamefont {D.}~\bibnamefont {Kandel}}, \bibinfo {author} {\bibfnamefont {A.~V.}\ \bibnamefont {Filippenko}}, \bibinfo {author} {\bibfnamefont {T.~G.}\ \bibnamefont {Brink}}, \ and\ \bibinfo {author} {\bibfnamefont {W.}~\bibnamefont {Zheng}},\ }\href {\doibase 10.3847/2041-8213/ac8007} {\bibfield  {journal} {\bibinfo  {journal} {Astrophys. J. Lett.}\ }\textbf {\bibinfo {volume} {934}},\ \bibinfo {pages} {L18} (\bibinfo {year} {2022})},\ \Eprint {http://arxiv.org/abs/2207.05124} {arXiv:2207.05124 [astro-ph.HE]} \BibitemShut {NoStop}%
\bibitem [{\citenamefont {Abbott}\ \emph {et~al.}(2017)\citenamefont {Abbott}, \citenamefont {Abbott}, \citenamefont {Abbott}, \citenamefont {Acernese}, \citenamefont {Ackley}, \citenamefont {Adams}, \citenamefont {Adams}, \citenamefont {Addesso}, \citenamefont {Adhikari}, \citenamefont {Adya}, \citenamefont {Affeldt}, \citenamefont {Afrough}, \citenamefont {Agarwal}, \citenamefont {Agathos}, \citenamefont {Agatsuma}, \citenamefont {Aggarwal}, \citenamefont {Aguiar}, \citenamefont {Aiello}, \citenamefont {Ain}, \citenamefont {Ajith}, \citenamefont {Allen}, \citenamefont {Allen}, \citenamefont {Allocca}, \citenamefont {Altin}, \citenamefont {Amato}, \citenamefont {Ananyeva}, \citenamefont {Anderson}, \citenamefont {Anderson}, \citenamefont {Angelova}, \citenamefont {Antier}, \citenamefont {Appert}, \citenamefont {Arai}, \citenamefont {Araya}, \citenamefont {Areeda}, \citenamefont {Arnaud}, \citenamefont {Arun}, \citenamefont {Ascenzi}, \citenamefont {Ashton}, \citenamefont {Ast}, \citenamefont {Aston},
  \citenamefont {Astone}, \citenamefont {Atallah}, \citenamefont {Aufmuth}, \citenamefont {Aulbert}, \citenamefont {AultONeal}, \citenamefont {Austin}, \citenamefont {Avila-Alvarez}, \citenamefont {Babak}, \citenamefont {Bacon}, \citenamefont {Bader}, \citenamefont {Bae}, \citenamefont {Bailes}, \citenamefont {Baker}, \citenamefont {Baldaccini}, \citenamefont {Ballardin}, \citenamefont {Ballmer}, \citenamefont {Banagiri}, \citenamefont {Barayoga}, \citenamefont {Barclay}, \citenamefont {Barish}, \citenamefont {Barker},\ and\ \citenamefont {et~al}}]{Abbott:2017}%
  \BibitemOpen
  \bibfield  {author} {\bibinfo {author} {\bibfnamefont {B.~P.}\ \bibnamefont {Abbott}}, \bibinfo {author} {\bibfnamefont {R.}~\bibnamefont {Abbott}}, \bibinfo {author} {\bibfnamefont {T.~D.}\ \bibnamefont {Abbott}}, \bibinfo {author} {\bibfnamefont {F.}~\bibnamefont {Acernese}}, \bibinfo {author} {\bibfnamefont {K.}~\bibnamefont {Ackley}}, \bibinfo {author} {\bibfnamefont {C.}~\bibnamefont {Adams}}, \bibinfo {author} {\bibfnamefont {T.}~\bibnamefont {Adams}}, \bibinfo {author} {\bibfnamefont {P.}~\bibnamefont {Addesso}}, \bibinfo {author} {\bibfnamefont {R.~X.}\ \bibnamefont {Adhikari}}, \bibinfo {author} {\bibfnamefont {V.~B.}\ \bibnamefont {Adya}}, \bibinfo {author} {\bibfnamefont {C.}~\bibnamefont {Affeldt}}, \bibinfo {author} {\bibfnamefont {M.}~\bibnamefont {Afrough}}, \bibinfo {author} {\bibfnamefont {B.}~\bibnamefont {Agarwal}}, \bibinfo {author} {\bibfnamefont {M.}~\bibnamefont {Agathos}}, \bibinfo {author} {\bibfnamefont {K.}~\bibnamefont {Agatsuma}}, \bibinfo {author} {\bibfnamefont
  {N.}~\bibnamefont {Aggarwal}}, \bibinfo {author} {\bibfnamefont {O.~D.}\ \bibnamefont {Aguiar}}, \bibinfo {author} {\bibfnamefont {L.}~\bibnamefont {Aiello}}, \bibinfo {author} {\bibfnamefont {A.}~\bibnamefont {Ain}}, \bibinfo {author} {\bibfnamefont {P.}~\bibnamefont {Ajith}}, \bibinfo {author} {\bibfnamefont {B.}~\bibnamefont {Allen}}, \bibinfo {author} {\bibfnamefont {G.}~\bibnamefont {Allen}}, \bibinfo {author} {\bibfnamefont {A.}~\bibnamefont {Allocca}}, \bibinfo {author} {\bibfnamefont {P.~A.}\ \bibnamefont {Altin}}, \bibinfo {author} {\bibfnamefont {A.}~\bibnamefont {Amato}}, \bibinfo {author} {\bibfnamefont {A.}~\bibnamefont {Ananyeva}}, \bibinfo {author} {\bibfnamefont {S.~B.}\ \bibnamefont {Anderson}}, \bibinfo {author} {\bibfnamefont {W.~G.}\ \bibnamefont {Anderson}}, \bibinfo {author} {\bibfnamefont {S.~V.}\ \bibnamefont {Angelova}}, \bibinfo {author} {\bibfnamefont {S.}~\bibnamefont {Antier}}, \bibinfo {author} {\bibfnamefont {S.}~\bibnamefont {Appert}}, \bibinfo {author} {\bibfnamefont
  {K.}~\bibnamefont {Arai}}, \bibinfo {author} {\bibfnamefont {M.~C.}\ \bibnamefont {Araya}}, \bibinfo {author} {\bibfnamefont {J.~S.}\ \bibnamefont {Areeda}}, \bibinfo {author} {\bibfnamefont {N.}~\bibnamefont {Arnaud}}, \bibinfo {author} {\bibfnamefont {K.~G.}\ \bibnamefont {Arun}}, \bibinfo {author} {\bibfnamefont {S.}~\bibnamefont {Ascenzi}}, \bibinfo {author} {\bibfnamefont {G.}~\bibnamefont {Ashton}}, \bibinfo {author} {\bibfnamefont {M.}~\bibnamefont {Ast}}, \bibinfo {author} {\bibfnamefont {S.~M.}\ \bibnamefont {Aston}}, \bibinfo {author} {\bibfnamefont {P.}~\bibnamefont {Astone}}, \bibinfo {author} {\bibfnamefont {D.~V.}\ \bibnamefont {Atallah}}, \bibinfo {author} {\bibfnamefont {P.}~\bibnamefont {Aufmuth}}, \bibinfo {author} {\bibfnamefont {C.}~\bibnamefont {Aulbert}}, \bibinfo {author} {\bibfnamefont {K.}~\bibnamefont {AultONeal}}, \bibinfo {author} {\bibfnamefont {C.}~\bibnamefont {Austin}}, \bibinfo {author} {\bibfnamefont {A.}~\bibnamefont {Avila-Alvarez}}, \bibinfo {author} {\bibfnamefont
  {S.}~\bibnamefont {Babak}}, \bibinfo {author} {\bibfnamefont {P.}~\bibnamefont {Bacon}}, \bibinfo {author} {\bibfnamefont {M.~K.~M.}\ \bibnamefont {Bader}}, \bibinfo {author} {\bibfnamefont {S.}~\bibnamefont {Bae}}, \bibinfo {author} {\bibfnamefont {M.}~\bibnamefont {Bailes}}, \bibinfo {author} {\bibfnamefont {P.~T.}\ \bibnamefont {Baker}}, \bibinfo {author} {\bibfnamefont {F.}~\bibnamefont {Baldaccini}}, \bibinfo {author} {\bibfnamefont {G.}~\bibnamefont {Ballardin}}, \bibinfo {author} {\bibfnamefont {S.~W.}\ \bibnamefont {Ballmer}}, \bibinfo {author} {\bibfnamefont {S.}~\bibnamefont {Banagiri}}, \bibinfo {author} {\bibfnamefont {J.~C.}\ \bibnamefont {Barayoga}}, \bibinfo {author} {\bibfnamefont {S.~E.}\ \bibnamefont {Barclay}}, \bibinfo {author} {\bibfnamefont {B.~C.}\ \bibnamefont {Barish}}, \bibinfo {author} {\bibfnamefont {D.}~\bibnamefont {Barker}}, \ and\ \bibinfo {author} {\bibnamefont {et~al}} (\bibinfo {collaboration} {LIGO Scientific Collaboration and Virgo Collaboration}),\ }\href {\doibase
  10.1103/PhysRevLett.119.161101} {\bibfield  {journal} {\bibinfo  {journal} {Phys. Rev. Lett.}\ }\textbf {\bibinfo {volume} {119}},\ \bibinfo {pages} {161101} (\bibinfo {year} {2017})}\BibitemShut {NoStop}%
\bibitem [{\citenamefont {Abbott}\ \emph {et~al.}(2019)\citenamefont {Abbott}, \citenamefont {Abbott}, \citenamefont {Abbott}, \citenamefont {Acernese}, \citenamefont {Ackley}, \citenamefont {Adams}, \citenamefont {Adams}, \citenamefont {Addesso}, \citenamefont {Adhikari}, \citenamefont {Adya},\ and\ \citenamefont {et~al}}]{Abbott:2019}%
  \BibitemOpen
  \bibfield  {author} {\bibinfo {author} {\bibfnamefont {B.~P.}\ \bibnamefont {Abbott}}, \bibinfo {author} {\bibfnamefont {R.}~\bibnamefont {Abbott}}, \bibinfo {author} {\bibfnamefont {T.~D.}\ \bibnamefont {Abbott}}, \bibinfo {author} {\bibfnamefont {F.}~\bibnamefont {Acernese}}, \bibinfo {author} {\bibfnamefont {K.}~\bibnamefont {Ackley}}, \bibinfo {author} {\bibfnamefont {C.}~\bibnamefont {Adams}}, \bibinfo {author} {\bibfnamefont {T.}~\bibnamefont {Adams}}, \bibinfo {author} {\bibfnamefont {P.}~\bibnamefont {Addesso}}, \bibinfo {author} {\bibfnamefont {R.~X.}\ \bibnamefont {Adhikari}}, \bibinfo {author} {\bibfnamefont {V.~B.}\ \bibnamefont {Adya}}, \ and\ \bibinfo {author} {\bibnamefont {et~al}} (\bibinfo {collaboration} {LIGO Scientific Collaboration and Virgo Collaboration}),\ }\href {\doibase 10.1103/PhysRevX.9.011001} {\bibfield  {journal} {\bibinfo  {journal} {Phys. Rev. X}\ }\textbf {\bibinfo {volume} {9}},\ \bibinfo {pages} {011001} (\bibinfo {year} {2019})}\BibitemShut {NoStop}%
\bibitem [{\citenamefont {Riley}\ \emph {et~al.}(2019)\citenamefont {Riley} \emph {et~al.}}]{Riley:2019}%
  \BibitemOpen
  \bibfield  {author} {\bibinfo {author} {\bibfnamefont {T.~E.}\ \bibnamefont {Riley}} \emph {et~al.},\ }\href {\doibase 10.3847/2041-8213/ab481c} {\bibfield  {journal} {\bibinfo  {journal} {Astrophys. J. Lett.}\ }\textbf {\bibinfo {volume} {887}},\ \bibinfo {pages} {L21} (\bibinfo {year} {2019})},\ \Eprint {http://arxiv.org/abs/1912.05702} {arXiv:1912.05702 [astro-ph.HE]} \BibitemShut {NoStop}%
\bibitem [{\citenamefont {Miller}\ \emph {et~al.}(2019)\citenamefont {Miller} \emph {et~al.}}]{Miller:2019}%
  \BibitemOpen
  \bibfield  {author} {\bibinfo {author} {\bibfnamefont {M.~C.}\ \bibnamefont {Miller}} \emph {et~al.},\ }\href {\doibase 10.3847/2041-8213/ab50c5} {\bibfield  {journal} {\bibinfo  {journal} {Astrophys. J. Lett.}\ }\textbf {\bibinfo {volume} {887}},\ \bibinfo {pages} {L24} (\bibinfo {year} {2019})},\ \Eprint {http://arxiv.org/abs/1912.05705} {arXiv:1912.05705 [astro-ph.HE]} \BibitemShut {NoStop}%
\bibitem [{\citenamefont {Riley}\ \emph {et~al.}(2021)\citenamefont {Riley} \emph {et~al.}}]{Riley:2021}%
  \BibitemOpen
  \bibfield  {author} {\bibinfo {author} {\bibfnamefont {T.~E.}\ \bibnamefont {Riley}} \emph {et~al.},\ }\href {\doibase 10.3847/2041-8213/ac0a81} {\bibfield  {journal} {\bibinfo  {journal} {Astrophys. J. Lett.}\ }\textbf {\bibinfo {volume} {918}},\ \bibinfo {pages} {L27} (\bibinfo {year} {2021})},\ \Eprint {http://arxiv.org/abs/2105.06980} {arXiv:2105.06980 [astro-ph.HE]} \BibitemShut {NoStop}%
\bibitem [{\citenamefont {{Miller}}\ \emph {et~al.}(2021)\citenamefont {{Miller}}, \citenamefont {{Lamb}}, \citenamefont {{Dittmann}}, \citenamefont {{Bogdanov}}, \citenamefont {{Arzoumanian}}, \citenamefont {{Gendreau}}, \citenamefont {{Guillot}}, \citenamefont {{Ho}}, \citenamefont {{Lattimer}}, \citenamefont {{Loewenstein}}, \citenamefont {{Morsink}}, \citenamefont {{Ray}}, \citenamefont {{Wolff}}, \citenamefont {{Baker}}, \citenamefont {{Cazeau}}, \citenamefont {{Manthripragada}}, \citenamefont {{Markwardt}}, \citenamefont {{Okajima}}, \citenamefont {{Pollard}}, \citenamefont {{Cognard}}, \citenamefont {{Cromartie}}, \citenamefont {{Fonseca}}, \citenamefont {{Guillemot}}, \citenamefont {{Kerr}}, \citenamefont {{Parthasarathy}}, \citenamefont {{Pennucci}}, \citenamefont {{Ransom}},\ and\ \citenamefont {{Stairs}}}]{Miller:2021}%
  \BibitemOpen
  \bibfield  {author} {\bibinfo {author} {\bibfnamefont {M.~C.}\ \bibnamefont {{Miller}}}, \bibinfo {author} {\bibfnamefont {F.~K.}\ \bibnamefont {{Lamb}}}, \bibinfo {author} {\bibfnamefont {A.~J.}\ \bibnamefont {{Dittmann}}}, \bibinfo {author} {\bibfnamefont {S.}~\bibnamefont {{Bogdanov}}}, \bibinfo {author} {\bibfnamefont {Z.}~\bibnamefont {{Arzoumanian}}}, \bibinfo {author} {\bibfnamefont {K.~C.}\ \bibnamefont {{Gendreau}}}, \bibinfo {author} {\bibfnamefont {S.}~\bibnamefont {{Guillot}}}, \bibinfo {author} {\bibfnamefont {W.~C.~G.}\ \bibnamefont {{Ho}}}, \bibinfo {author} {\bibfnamefont {J.~M.}\ \bibnamefont {{Lattimer}}}, \bibinfo {author} {\bibfnamefont {M.}~\bibnamefont {{Loewenstein}}}, \bibinfo {author} {\bibfnamefont {S.~M.}\ \bibnamefont {{Morsink}}}, \bibinfo {author} {\bibfnamefont {P.~S.}\ \bibnamefont {{Ray}}}, \bibinfo {author} {\bibfnamefont {M.~T.}\ \bibnamefont {{Wolff}}}, \bibinfo {author} {\bibfnamefont {C.~L.}\ \bibnamefont {{Baker}}}, \bibinfo {author} {\bibfnamefont {T.}~\bibnamefont
  {{Cazeau}}}, \bibinfo {author} {\bibfnamefont {S.}~\bibnamefont {{Manthripragada}}}, \bibinfo {author} {\bibfnamefont {C.~B.}\ \bibnamefont {{Markwardt}}}, \bibinfo {author} {\bibfnamefont {T.}~\bibnamefont {{Okajima}}}, \bibinfo {author} {\bibfnamefont {S.}~\bibnamefont {{Pollard}}}, \bibinfo {author} {\bibfnamefont {I.}~\bibnamefont {{Cognard}}}, \bibinfo {author} {\bibfnamefont {H.~T.}\ \bibnamefont {{Cromartie}}}, \bibinfo {author} {\bibfnamefont {E.}~\bibnamefont {{Fonseca}}}, \bibinfo {author} {\bibfnamefont {L.}~\bibnamefont {{Guillemot}}}, \bibinfo {author} {\bibfnamefont {M.}~\bibnamefont {{Kerr}}}, \bibinfo {author} {\bibfnamefont {A.}~\bibnamefont {{Parthasarathy}}}, \bibinfo {author} {\bibfnamefont {T.~T.}\ \bibnamefont {{Pennucci}}}, \bibinfo {author} {\bibfnamefont {S.}~\bibnamefont {{Ransom}}}, \ and\ \bibinfo {author} {\bibfnamefont {I.}~\bibnamefont {{Stairs}}},\ }\href {\doibase 10.3847/2041-8213/ac089b} {\bibfield  {journal} {\bibinfo  {journal} {The Astrophysical Journal}\ }\textbf
  {\bibinfo {volume} {918}},\ \bibinfo {eid} {L28} (\bibinfo {year} {2021})},\ \Eprint {http://arxiv.org/abs/2105.06979} {arXiv:2105.06979 [astro-ph.HE]} \BibitemShut {NoStop}%
\bibitem [{\citenamefont {Vinciguerra}\ \emph {et~al.}(2024)\citenamefont {Vinciguerra} \emph {et~al.}}]{Vinciguerra:2023qxq}%
  \BibitemOpen
  \bibfield  {author} {\bibinfo {author} {\bibfnamefont {S.}~\bibnamefont {Vinciguerra}} \emph {et~al.},\ }\href {\doibase 10.3847/1538-4357/acfb83} {\bibfield  {journal} {\bibinfo  {journal} {Astrophys. J.}\ }\textbf {\bibinfo {volume} {961}},\ \bibinfo {pages} {62} (\bibinfo {year} {2024})},\ \Eprint {http://arxiv.org/abs/2308.09469} {arXiv:2308.09469 [astro-ph.HE]} \BibitemShut {NoStop}%
\bibitem [{\citenamefont {Choudhury}\ \emph {et~al.}(2024)\citenamefont {Choudhury} \emph {et~al.}}]{Choudhury:2024xbk}%
  \BibitemOpen
  \bibfield  {author} {\bibinfo {author} {\bibfnamefont {D.}~\bibnamefont {Choudhury}} \emph {et~al.},\ }\href {\doibase 10.3847/2041-8213/ad5a6f} {\bibfield  {journal} {\bibinfo  {journal} {Astrophys. J. Lett.}\ }\textbf {\bibinfo {volume} {971}},\ \bibinfo {pages} {L20} (\bibinfo {year} {2024})},\ \Eprint {http://arxiv.org/abs/2407.06789} {arXiv:2407.06789 [astro-ph.HE]} \BibitemShut {NoStop}%
\bibitem [{\citenamefont {Raaijmakers}\ \emph {et~al.}(2019)\citenamefont {Raaijmakers} \emph {et~al.}}]{Raaijmakers:2019a}%
  \BibitemOpen
  \bibfield  {author} {\bibinfo {author} {\bibfnamefont {G.}~\bibnamefont {Raaijmakers}} \emph {et~al.},\ }\href {\doibase 10.3847/2041-8213/ab451a} {\bibfield  {journal} {\bibinfo  {journal} {Astrophys. J. Lett.}\ }\textbf {\bibinfo {volume} {887}},\ \bibinfo {pages} {L22} (\bibinfo {year} {2019})},\ \Eprint {http://arxiv.org/abs/1912.05703} {arXiv:1912.05703 [astro-ph.HE]} \BibitemShut {NoStop}%
\bibitem [{\citenamefont {Raaijmakers}\ \emph {et~al.}(2020)\citenamefont {Raaijmakers} \emph {et~al.}}]{Raaijmakers:2019b}%
  \BibitemOpen
  \bibfield  {author} {\bibinfo {author} {\bibfnamefont {G.}~\bibnamefont {Raaijmakers}} \emph {et~al.},\ }\href {\doibase 10.3847/2041-8213/ab822f} {\bibfield  {journal} {\bibinfo  {journal} {Astrophys. J. Lett.}\ }\textbf {\bibinfo {volume} {893}},\ \bibinfo {pages} {L21} (\bibinfo {year} {2020})},\ \Eprint {http://arxiv.org/abs/1912.11031} {arXiv:1912.11031 [astro-ph.HE]} \BibitemShut {NoStop}%
\bibitem [{\citenamefont {Raaijmakers}\ \emph {et~al.}(2021)\citenamefont {Raaijmakers}, \citenamefont {Greif}, \citenamefont {Hebeler}, \citenamefont {Hinderer}, \citenamefont {Nissanke}, \citenamefont {Schwenk}, \citenamefont {Riley}, \citenamefont {Watts}, \citenamefont {Lattimer},\ and\ \citenamefont {Ho}}]{Raaijmakers:2021}%
  \BibitemOpen
  \bibfield  {author} {\bibinfo {author} {\bibfnamefont {G.}~\bibnamefont {Raaijmakers}}, \bibinfo {author} {\bibfnamefont {S.~K.}\ \bibnamefont {Greif}}, \bibinfo {author} {\bibfnamefont {K.}~\bibnamefont {Hebeler}}, \bibinfo {author} {\bibfnamefont {T.}~\bibnamefont {Hinderer}}, \bibinfo {author} {\bibfnamefont {S.}~\bibnamefont {Nissanke}}, \bibinfo {author} {\bibfnamefont {A.}~\bibnamefont {Schwenk}}, \bibinfo {author} {\bibfnamefont {T.~E.}\ \bibnamefont {Riley}}, \bibinfo {author} {\bibfnamefont {A.~L.}\ \bibnamefont {Watts}}, \bibinfo {author} {\bibfnamefont {J.~M.}\ \bibnamefont {Lattimer}}, \ and\ \bibinfo {author} {\bibfnamefont {W.~C.~G.}\ \bibnamefont {Ho}},\ }\href {\doibase 10.3847/2041-8213/ac089a} {\bibfield  {journal} {\bibinfo  {journal} {Astrophys. J. Lett.}\ }\textbf {\bibinfo {volume} {918}},\ \bibinfo {pages} {L29} (\bibinfo {year} {2021})},\ \Eprint {http://arxiv.org/abs/2105.06981} {arXiv:2105.06981 [astro-ph.HE]} \BibitemShut {NoStop}%
\bibitem [{\citenamefont {Rutherford}\ \emph {et~al.}(2024)\citenamefont {Rutherford} \emph {et~al.}}]{Rutherford:2024}%
  \BibitemOpen
  \bibfield  {author} {\bibinfo {author} {\bibfnamefont {N.}~\bibnamefont {Rutherford}} \emph {et~al.},\ }\href {\doibase 10.3847/2041-8213/ad5f02} {\bibfield  {journal} {\bibinfo  {journal} {{Accepted at Astrophysical Journal Letters}}\ } (\bibinfo {year} {2024}),\ 10.3847/2041-8213/ad5f02},\ \Eprint {http://arxiv.org/abs/2407.06790} {arXiv:2407.06790 [astro-ph.HE]} \BibitemShut {NoStop}%
\bibitem [{\citenamefont {Paschalidis}\ \emph {et~al.}(2018)\citenamefont {Paschalidis}, \citenamefont {Yagi}, \citenamefont {Alvarez-Castillo}, \citenamefont {Blaschke},\ and\ \citenamefont {Sedrakian}}]{Paschalidis:2017qmb}%
  \BibitemOpen
  \bibfield  {author} {\bibinfo {author} {\bibfnamefont {V.}~\bibnamefont {Paschalidis}}, \bibinfo {author} {\bibfnamefont {K.}~\bibnamefont {Yagi}}, \bibinfo {author} {\bibfnamefont {D.}~\bibnamefont {Alvarez-Castillo}}, \bibinfo {author} {\bibfnamefont {D.~B.}\ \bibnamefont {Blaschke}}, \ and\ \bibinfo {author} {\bibfnamefont {A.}~\bibnamefont {Sedrakian}},\ }\href {\doibase 10.1103/PhysRevD.97.084038} {\bibfield  {journal} {\bibinfo  {journal} {Phys. Rev.}\ }\textbf {\bibinfo {volume} {D97}},\ \bibinfo {pages} {084038} (\bibinfo {year} {2018})},\ \Eprint {http://arxiv.org/abs/1712.00451} {arXiv:1712.00451 [astro-ph.HE]} \BibitemShut {NoStop}%
\bibitem [{\citenamefont {Christian}\ \emph {et~al.}(2019)\citenamefont {Christian}, \citenamefont {Zacchi},\ and\ \citenamefont {Schaffner-Bielich}}]{Christian:2018jyd}%
  \BibitemOpen
  \bibfield  {author} {\bibinfo {author} {\bibfnamefont {J.-E.}\ \bibnamefont {Christian}}, \bibinfo {author} {\bibfnamefont {A.}~\bibnamefont {Zacchi}}, \ and\ \bibinfo {author} {\bibfnamefont {J.}~\bibnamefont {Schaffner-Bielich}},\ }\href {\doibase 10.1103/PhysRevD.99.023009} {\bibfield  {journal} {\bibinfo  {journal} {Phys. Rev.}\ }\textbf {\bibinfo {volume} {D99}},\ \bibinfo {pages} {023009} (\bibinfo {year} {2019})},\ \Eprint {http://arxiv.org/abs/1809.03333} {arXiv:1809.03333 [astro-ph.HE]} \BibitemShut {NoStop}%
\bibitem [{\citenamefont {Montana}\ \emph {et~al.}(2019)\citenamefont {Montana}, \citenamefont {Tolos}, \citenamefont {Hanauske},\ and\ \citenamefont {Rezzolla}}]{Montana:2018bkb}%
  \BibitemOpen
  \bibfield  {author} {\bibinfo {author} {\bibfnamefont {G.}~\bibnamefont {Montana}}, \bibinfo {author} {\bibfnamefont {L.}~\bibnamefont {Tolos}}, \bibinfo {author} {\bibfnamefont {M.}~\bibnamefont {Hanauske}}, \ and\ \bibinfo {author} {\bibfnamefont {L.}~\bibnamefont {Rezzolla}},\ }\href {\doibase 10.1103/PhysRevD.99.103009} {\bibfield  {journal} {\bibinfo  {journal} {Phys. Rev. D}\ }\textbf {\bibinfo {volume} {99}},\ \bibinfo {pages} {103009} (\bibinfo {year} {2019})},\ \Eprint {http://arxiv.org/abs/1811.10929} {arXiv:1811.10929 [astro-ph.HE]} \BibitemShut {NoStop}%
\bibitem [{\citenamefont {Sieniawska}\ \emph {et~al.}(2019)\citenamefont {Sieniawska}, \citenamefont {Turczanski}, \citenamefont {Bejger},\ and\ \citenamefont {Zdunik}}]{Sieniawska:2018zzj}%
  \BibitemOpen
  \bibfield  {author} {\bibinfo {author} {\bibfnamefont {M.}~\bibnamefont {Sieniawska}}, \bibinfo {author} {\bibfnamefont {W.}~\bibnamefont {Turczanski}}, \bibinfo {author} {\bibfnamefont {M.}~\bibnamefont {Bejger}}, \ and\ \bibinfo {author} {\bibfnamefont {J.~L.}\ \bibnamefont {Zdunik}},\ }\href {\doibase 10.1051/0004-6361/201833969} {\bibfield  {journal} {\bibinfo  {journal} {Astron. Astrophys.}\ }\textbf {\bibinfo {volume} {622}},\ \bibinfo {pages} {A174} (\bibinfo {year} {2019})},\ \Eprint {http://arxiv.org/abs/1807.11581} {arXiv:1807.11581 [astro-ph.HE]} \BibitemShut {NoStop}%
\bibitem [{\citenamefont {Christian}\ and\ \citenamefont {Schaffner-Bielich}(2020)}]{Christian:2019}%
  \BibitemOpen
  \bibfield  {author} {\bibinfo {author} {\bibfnamefont {J.-E.}\ \bibnamefont {Christian}}\ and\ \bibinfo {author} {\bibfnamefont {J.}~\bibnamefont {Schaffner-Bielich}},\ }\href {\doibase 10.3847/2041-8213/ab8af4} {\bibfield  {journal} {\bibinfo  {journal} {Astrophys. J. Lett.}\ }\textbf {\bibinfo {volume} {894}},\ \bibinfo {pages} {L8} (\bibinfo {year} {2020})},\ \Eprint {http://arxiv.org/abs/1912.09809} {arXiv:1912.09809 [astro-ph.HE]} \BibitemShut {NoStop}%
\bibitem [{\citenamefont {Ivanenko}\ and\ \citenamefont {Kurdgelaidze}(1965)}]{Ivanenko:1965dg}%
  \BibitemOpen
  \bibfield  {author} {\bibinfo {author} {\bibfnamefont {D.~D.}\ \bibnamefont {Ivanenko}}\ and\ \bibinfo {author} {\bibfnamefont {D.~F.}\ \bibnamefont {Kurdgelaidze}},\ }\href@noop {} {\bibfield  {journal} {\bibinfo  {journal} {Astrophys.}\ }\textbf {\bibinfo {volume} {1}},\ \bibinfo {pages} {251} (\bibinfo {year} {1965})}\BibitemShut {NoStop}%
\bibitem [{\citenamefont {Itoh}(1970)}]{Itoh:1970uw}%
  \BibitemOpen
  \bibfield  {author} {\bibinfo {author} {\bibfnamefont {N.}~\bibnamefont {Itoh}},\ }\href {\doibase 10.1143/PTP.44.291} {\bibfield  {journal} {\bibinfo  {journal} {Prog.Theor.Phys.}\ }\textbf {\bibinfo {volume} {44}},\ \bibinfo {pages} {291} (\bibinfo {year} {1970})}\BibitemShut {NoStop}%
\bibitem [{\citenamefont {Alford}\ \emph {et~al.}(2005)\citenamefont {Alford}, \citenamefont {Braby}, \citenamefont {Paris},\ and\ \citenamefont {Reddy}}]{Alford:2004pf}%
  \BibitemOpen
  \bibfield  {author} {\bibinfo {author} {\bibfnamefont {M.}~\bibnamefont {Alford}}, \bibinfo {author} {\bibfnamefont {M.}~\bibnamefont {Braby}}, \bibinfo {author} {\bibfnamefont {M.}~\bibnamefont {Paris}}, \ and\ \bibinfo {author} {\bibfnamefont {S.}~\bibnamefont {Reddy}},\ }\href {\doibase 10.1086/430902} {\bibfield  {journal} {\bibinfo  {journal} {Astrophys.J.}\ }\textbf {\bibinfo {volume} {629}},\ \bibinfo {pages} {969} (\bibinfo {year} {2005})},\ \Eprint {http://arxiv.org/abs/nucl-th/0411016} {arXiv:nucl-th/0411016 [nucl-th]} \BibitemShut {NoStop}%
\bibitem [{\citenamefont {Coelho}\ \emph {et~al.}(2010)\citenamefont {Coelho}, \citenamefont {Lenzi}, \citenamefont {Malheiro}, \citenamefont {Marinho},\ and\ \citenamefont {Fiolhais}}]{Coelho:2010fv}%
  \BibitemOpen
  \bibfield  {author} {\bibinfo {author} {\bibfnamefont {J.}~\bibnamefont {Coelho}}, \bibinfo {author} {\bibfnamefont {C.}~\bibnamefont {Lenzi}}, \bibinfo {author} {\bibfnamefont {M.}~\bibnamefont {Malheiro}}, \bibinfo {author} {\bibfnamefont {J.}~\bibnamefont {Marinho}, \bibfnamefont {R.M.}}, \ and\ \bibinfo {author} {\bibfnamefont {M.}~\bibnamefont {Fiolhais}},\ }\href {\doibase 10.1142/S0218271810017597} {\bibfield  {journal} {\bibinfo  {journal} {Int.J.Mod.Phys.}\ }\textbf {\bibinfo {volume} {D19}},\ \bibinfo {pages} {1521} (\bibinfo {year} {2010})},\ \Eprint {http://arxiv.org/abs/1001.1661} {arXiv:1001.1661 [nucl-th]} \BibitemShut {NoStop}%
\bibitem [{\citenamefont {Chen}\ \emph {et~al.}(2011)\citenamefont {Chen}, \citenamefont {Baldo}, \citenamefont {Burgio},\ and\ \citenamefont {Schulze}}]{Chen:2011my}%
  \BibitemOpen
  \bibfield  {author} {\bibinfo {author} {\bibfnamefont {H.}~\bibnamefont {Chen}}, \bibinfo {author} {\bibfnamefont {M.}~\bibnamefont {Baldo}}, \bibinfo {author} {\bibfnamefont {G.}~\bibnamefont {Burgio}}, \ and\ \bibinfo {author} {\bibfnamefont {H.-J.}\ \bibnamefont {Schulze}},\ }\href {\doibase 10.1103/PhysRevD.84.105023} {\bibfield  {journal} {\bibinfo  {journal} {Phys.Rev.}\ }\textbf {\bibinfo {volume} {D84}},\ \bibinfo {pages} {105023} (\bibinfo {year} {2011})},\ \Eprint {http://arxiv.org/abs/1107.2497} {arXiv:1107.2497 [nucl-th]} \BibitemShut {NoStop}%
\bibitem [{\citenamefont {Masuda}\ \emph {et~al.}(2013)\citenamefont {Masuda}, \citenamefont {Hatsuda},\ and\ \citenamefont {Takatsuka}}]{Masuda:2012kf}%
  \BibitemOpen
  \bibfield  {author} {\bibinfo {author} {\bibfnamefont {K.}~\bibnamefont {Masuda}}, \bibinfo {author} {\bibfnamefont {T.}~\bibnamefont {Hatsuda}}, \ and\ \bibinfo {author} {\bibfnamefont {T.}~\bibnamefont {Takatsuka}},\ }\href {\doibase 10.1088/0004-637X/764/1/12} {\bibfield  {journal} {\bibinfo  {journal} {Astrophys. J.}\ }\textbf {\bibinfo {volume} {764}},\ \bibinfo {pages} {12} (\bibinfo {year} {2013})},\ \Eprint {http://arxiv.org/abs/1205.3621} {arXiv:1205.3621 [nucl-th]} \BibitemShut {NoStop}%
\bibitem [{\citenamefont {Yasutake}\ \emph {et~al.}(2014)\citenamefont {Yasutake}, \citenamefont {Lastowiecki}, \citenamefont {Benic}, \citenamefont {Blaschke}, \citenamefont {Maruyama},\ and\ \citenamefont {Tatsumi}}]{Yasutake:2014oxa}%
  \BibitemOpen
  \bibfield  {author} {\bibinfo {author} {\bibfnamefont {N.}~\bibnamefont {Yasutake}}, \bibinfo {author} {\bibfnamefont {R.}~\bibnamefont {Lastowiecki}}, \bibinfo {author} {\bibfnamefont {S.}~\bibnamefont {Benic}}, \bibinfo {author} {\bibfnamefont {D.}~\bibnamefont {Blaschke}}, \bibinfo {author} {\bibfnamefont {T.}~\bibnamefont {Maruyama}}, \ and\ \bibinfo {author} {\bibfnamefont {T.}~\bibnamefont {Tatsumi}},\ }\href {\doibase 10.1103/PhysRevC.89.065803} {\bibfield  {journal} {\bibinfo  {journal} {Phys.Rev.}\ }\textbf {\bibinfo {volume} {C89}},\ \bibinfo {pages} {065803} (\bibinfo {year} {2014})},\ \Eprint {http://arxiv.org/abs/1403.7492} {arXiv:1403.7492 [astro-ph.HE]} \BibitemShut {NoStop}%
\bibitem [{\citenamefont {Zacchi}\ \emph {et~al.}(2016)\citenamefont {Zacchi}, \citenamefont {Hanauske},\ and\ \citenamefont {Schaffner-Bielich}}]{Zacchi:2015oma}%
  \BibitemOpen
  \bibfield  {author} {\bibinfo {author} {\bibfnamefont {A.}~\bibnamefont {Zacchi}}, \bibinfo {author} {\bibfnamefont {M.}~\bibnamefont {Hanauske}}, \ and\ \bibinfo {author} {\bibfnamefont {J.}~\bibnamefont {Schaffner-Bielich}},\ }\href {\doibase 10.1103/PhysRevD.93.065011} {\bibfield  {journal} {\bibinfo  {journal} {Phys. Rev.}\ }\textbf {\bibinfo {volume} {D93}},\ \bibinfo {pages} {065011} (\bibinfo {year} {2016})},\ \Eprint {http://arxiv.org/abs/1510.00180} {arXiv:1510.00180 [nucl-th]} \BibitemShut {NoStop}%
\bibitem [{\citenamefont {K\"ampfer}(1981)}]{Kampfer:1981yr}%
  \BibitemOpen
  \bibfield  {author} {\bibinfo {author} {\bibfnamefont {B.}~\bibnamefont {K\"ampfer}},\ }\href {\doibase 10.1088/0305-4470/14/11/009} {\bibfield  {journal} {\bibinfo  {journal} {J.Phys.}\ }\textbf {\bibinfo {volume} {A14}},\ \bibinfo {pages} {L471} (\bibinfo {year} {1981})}\BibitemShut {NoStop}%
\bibitem [{\citenamefont {{Glendenning}}\ and\ \citenamefont {{Kettner}}(2000)}]{Glendenning:1998ag}%
  \BibitemOpen
  \bibfield  {author} {\bibinfo {author} {\bibfnamefont {N.~K.}\ \bibnamefont {{Glendenning}}}\ and\ \bibinfo {author} {\bibfnamefont {C.}~\bibnamefont {{Kettner}}},\ }\href {\doibase 10.48550/arXiv.astro-ph/9807155} {\bibfield  {journal} {\bibinfo  {journal} {Astronomy and Astrophysics}\ }\textbf {\bibinfo {volume} {353}},\ \bibinfo {pages} {L9} (\bibinfo {year} {2000})},\ \Eprint {http://arxiv.org/abs/astro-ph/9807155} {arXiv:astro-ph/9807155 [astro-ph]} \BibitemShut {NoStop}%
\bibitem [{\citenamefont {Schertler}\ \emph {et~al.}(2000)\citenamefont {Schertler}, \citenamefont {Greiner}, \citenamefont {Schaffner-Bielich},\ and\ \citenamefont {Thoma}}]{Schertler:2000xq}%
  \BibitemOpen
  \bibfield  {author} {\bibinfo {author} {\bibfnamefont {K.}~\bibnamefont {Schertler}}, \bibinfo {author} {\bibfnamefont {C.}~\bibnamefont {Greiner}}, \bibinfo {author} {\bibfnamefont {J.}~\bibnamefont {Schaffner-Bielich}}, \ and\ \bibinfo {author} {\bibfnamefont {M.~H.}\ \bibnamefont {Thoma}},\ }\href@noop {} {\bibfield  {journal} {\bibinfo  {journal} {NP}\ }\textbf {\bibinfo {volume} {A677}},\ \bibinfo {pages} {463} (\bibinfo {year} {2000})},\ \Eprint {http://arxiv.org/abs/astro-ph/0001467} {astro-ph/0001467} \BibitemShut {NoStop}%
\bibitem [{\citenamefont {Schaffner-Bielich}\ \emph {et~al.}(2002)\citenamefont {Schaffner-Bielich}, \citenamefont {Hanauske}, \citenamefont {St{\"o}cker},\ and\ \citenamefont {Greiner}}]{SchaffnerBielich:2002ki}%
  \BibitemOpen
  \bibfield  {author} {\bibinfo {author} {\bibfnamefont {J.}~\bibnamefont {Schaffner-Bielich}}, \bibinfo {author} {\bibfnamefont {M.}~\bibnamefont {Hanauske}}, \bibinfo {author} {\bibfnamefont {H.}~\bibnamefont {St{\"o}cker}}, \ and\ \bibinfo {author} {\bibfnamefont {W.}~\bibnamefont {Greiner}},\ }\href@noop {} {\bibfield  {journal} {\bibinfo  {journal} {Physical Review Letters}\ }\textbf {\bibinfo {volume} {89}},\ \bibinfo {pages} {171101} (\bibinfo {year} {2002})},\ \Eprint {http://arxiv.org/abs/astro-ph/0005490} {astro-ph/0005490} \BibitemShut {NoStop}%
\bibitem [{\citenamefont {Zdunik}\ and\ \citenamefont {Haensel}(2013)}]{Zdunik:2012dj}%
  \BibitemOpen
  \bibfield  {author} {\bibinfo {author} {\bibfnamefont {J.}~\bibnamefont {Zdunik}}\ and\ \bibinfo {author} {\bibfnamefont {P.}~\bibnamefont {Haensel}},\ }\href {\doibase 10.1051/0004-6361/201220697} {\bibfield  {journal} {\bibinfo  {journal} {Astron.Astrophys.}\ }\textbf {\bibinfo {volume} {551}},\ \bibinfo {pages} {A61} (\bibinfo {year} {2013})},\ \Eprint {http://arxiv.org/abs/1211.1231} {arXiv:1211.1231 [astro-ph.SR]} \BibitemShut {NoStop}%
\bibitem [{\citenamefont {Alford}\ \emph {et~al.}(2015)\citenamefont {Alford}, \citenamefont {Burgio}, \citenamefont {Han}, \citenamefont {Taranto},\ and\ \citenamefont {Zappal\`a}}]{Alford:2015dpa}%
  \BibitemOpen
  \bibfield  {author} {\bibinfo {author} {\bibfnamefont {M.~G.}\ \bibnamefont {Alford}}, \bibinfo {author} {\bibfnamefont {G.}~\bibnamefont {Burgio}}, \bibinfo {author} {\bibfnamefont {S.}~\bibnamefont {Han}}, \bibinfo {author} {\bibfnamefont {G.}~\bibnamefont {Taranto}}, \ and\ \bibinfo {author} {\bibfnamefont {D.}~\bibnamefont {Zappal\`a}},\ }\href {\doibase 10.1103/PhysRevD.92.083002} {\bibfield  {journal} {\bibinfo  {journal} {Phys. Rev.}\ }\textbf {\bibinfo {volume} {D92}},\ \bibinfo {pages} {083002} (\bibinfo {year} {2015})},\ \Eprint {http://arxiv.org/abs/1501.07902} {arXiv:1501.07902 [nucl-th]} \BibitemShut {NoStop}%
\bibitem [{\citenamefont {Blaschke}\ and\ \citenamefont {Alvarez-Castillo}(2016)}]{Blaschke:2015uva}%
  \BibitemOpen
  \bibfield  {author} {\bibinfo {author} {\bibfnamefont {D.}~\bibnamefont {Blaschke}}\ and\ \bibinfo {author} {\bibfnamefont {D.~E.}\ \bibnamefont {Alvarez-Castillo}},\ }\bibfield  {booktitle} {\emph {\bibinfo {booktitle} {{Proceedings, 11th Conference on Quark Confinement and the Hadron Spectrum (Confinement XI): St. Petersburg, Russia, September 8-12, 2014}}},\ }\href {\doibase 10.1063/1.4938602} {\bibfield  {journal} {\bibinfo  {journal} {AIP Conf. Proc.}\ }\textbf {\bibinfo {volume} {1701}},\ \bibinfo {pages} {020013} (\bibinfo {year} {2016})},\ \Eprint {http://arxiv.org/abs/1503.03834} {arXiv:1503.03834 [astro-ph.HE]} \BibitemShut {NoStop}%
\bibitem [{\citenamefont {Alford}\ and\ \citenamefont {Sedrakian}(2017)}]{Alford:2017qgh}%
  \BibitemOpen
  \bibfield  {author} {\bibinfo {author} {\bibfnamefont {M.~G.}\ \bibnamefont {Alford}}\ and\ \bibinfo {author} {\bibfnamefont {A.}~\bibnamefont {Sedrakian}},\ }\href {\doibase 10.1103/PhysRevLett.119.161104} {\bibfield  {journal} {\bibinfo  {journal} {Phys. Rev. Lett.}\ }\textbf {\bibinfo {volume} {119}},\ \bibinfo {pages} {161104} (\bibinfo {year} {2017})},\ \Eprint {http://arxiv.org/abs/1706.01592} {arXiv:1706.01592 [astro-ph.HE]} \BibitemShut {NoStop}%
\bibitem [{\citenamefont {Christian}\ \emph {et~al.}(2018)\citenamefont {Christian}, \citenamefont {Zacchi},\ and\ \citenamefont {Schaffner-Bielich}}]{Christian:2017jni}%
  \BibitemOpen
  \bibfield  {author} {\bibinfo {author} {\bibfnamefont {J.-E.}\ \bibnamefont {Christian}}, \bibinfo {author} {\bibfnamefont {A.}~\bibnamefont {Zacchi}}, \ and\ \bibinfo {author} {\bibfnamefont {J.}~\bibnamefont {Schaffner-Bielich}},\ }\href {\doibase 10.1140/epja/i2018-12472-y} {\bibfield  {journal} {\bibinfo  {journal} {Eur. Phys. J.}\ }\textbf {\bibinfo {volume} {A54}},\ \bibinfo {pages} {28} (\bibinfo {year} {2018})},\ \Eprint {http://arxiv.org/abs/1707.07524} {arXiv:1707.07524 [astro-ph.HE]} \BibitemShut {NoStop}%
\bibitem [{\citenamefont {de~Carvalho}\ \emph {et~al.}(2015)\citenamefont {de~Carvalho}, \citenamefont {Negreiros}, \citenamefont {Orsaria}, \citenamefont {Contrera}, \citenamefont {Weber},\ and\ \citenamefont {Spinella}}]{deCarvalho:2015}%
  \BibitemOpen
  \bibfield  {author} {\bibinfo {author} {\bibfnamefont {S.~M.}\ \bibnamefont {de~Carvalho}}, \bibinfo {author} {\bibfnamefont {R.}~\bibnamefont {Negreiros}}, \bibinfo {author} {\bibfnamefont {M.}~\bibnamefont {Orsaria}}, \bibinfo {author} {\bibfnamefont {G.~A.}\ \bibnamefont {Contrera}}, \bibinfo {author} {\bibfnamefont {F.}~\bibnamefont {Weber}}, \ and\ \bibinfo {author} {\bibfnamefont {W.}~\bibnamefont {Spinella}},\ }\href {\doibase 10.1103/PhysRevC.92.035810} {\bibfield  {journal} {\bibinfo  {journal} {Phys. Rev. C}\ }\textbf {\bibinfo {volume} {92}},\ \bibinfo {pages} {035810} (\bibinfo {year} {2015})}\BibitemShut {NoStop}%
\bibitem [{\citenamefont {Lyra}\ \emph {et~al.}(2023)\citenamefont {Lyra}, \citenamefont {Moreira}, \citenamefont {Negreiros}, \citenamefont {Gomes},\ and\ \citenamefont {Dexheimer}}]{Lyra:2023}%
  \BibitemOpen
  \bibfield  {author} {\bibinfo {author} {\bibfnamefont {F.}~\bibnamefont {Lyra}}, \bibinfo {author} {\bibfnamefont {L.}~\bibnamefont {Moreira}}, \bibinfo {author} {\bibfnamefont {R.}~\bibnamefont {Negreiros}}, \bibinfo {author} {\bibfnamefont {R.~O.}\ \bibnamefont {Gomes}}, \ and\ \bibinfo {author} {\bibfnamefont {V.}~\bibnamefont {Dexheimer}},\ }\href {\doibase 10.1103/PhysRevC.107.025806} {\bibfield  {journal} {\bibinfo  {journal} {Phys. Rev. C}\ }\textbf {\bibinfo {volume} {107}},\ \bibinfo {pages} {025806} (\bibinfo {year} {2023})},\ \Eprint {http://arxiv.org/abs/2206.01679} {arXiv:2206.01679 [astro-ph.HE]} \BibitemShut {NoStop}%
\bibitem [{\citenamefont {Mendes}\ \emph {et~al.}(2022)\citenamefont {Mendes}, \citenamefont {Fattoyev}, \citenamefont {Cumming},\ and\ \citenamefont {Gale}}]{Mendes:2022}%
  \BibitemOpen
  \bibfield  {author} {\bibinfo {author} {\bibfnamefont {M.}~\bibnamefont {Mendes}}, \bibinfo {author} {\bibfnamefont {F.~J.}\ \bibnamefont {Fattoyev}}, \bibinfo {author} {\bibfnamefont {A.}~\bibnamefont {Cumming}}, \ and\ \bibinfo {author} {\bibfnamefont {C.}~\bibnamefont {Gale}},\ }\href {\doibase 10.3847/1538-4357/ac9138} {\bibfield  {journal} {\bibinfo  {journal} {The Astrophysical Journal}\ }\textbf {\bibinfo {volume} {938}},\ \bibinfo {pages} {119} (\bibinfo {year} {2022})}\BibitemShut {NoStop}%
\bibitem [{\citenamefont {Potekhin}\ \emph {et~al.}(2023)\citenamefont {Potekhin}, \citenamefont {Gusakov},\ and\ \citenamefont {Chugunov}}]{Potekhin:2023}%
  \BibitemOpen
  \bibfield  {author} {\bibinfo {author} {\bibfnamefont {A.~Y.}\ \bibnamefont {Potekhin}}, \bibinfo {author} {\bibfnamefont {M.~E.}\ \bibnamefont {Gusakov}}, \ and\ \bibinfo {author} {\bibfnamefont {A.~I.}\ \bibnamefont {Chugunov}},\ }\href {\doibase https://doi.org/10.48550/arXiv.2303.08716} {\enquote {\bibinfo {title} {Thermal evolution of neutron stars in soft x-ray transients with thermodynamically consistent models of the accreted crust},}\ }\bibinfo {howpublished} {Pre-print} (\bibinfo {year} {2023}),\ \Eprint {http://arxiv.org/abs/2303.08716} {arXiv:2303.08716 [astro-ph.HE]} \BibitemShut {NoStop}%
\bibitem [{\citenamefont {Reed}\ \emph {et~al.}(2024)\citenamefont {Reed}, \citenamefont {Fattoyev}, \citenamefont {Horowitz},\ and\ \citenamefont {Piekarewicz}}]{Reed:2023cap}%
  \BibitemOpen
  \bibfield  {author} {\bibinfo {author} {\bibfnamefont {B.~T.}\ \bibnamefont {Reed}}, \bibinfo {author} {\bibfnamefont {F.~J.}\ \bibnamefont {Fattoyev}}, \bibinfo {author} {\bibfnamefont {C.~J.}\ \bibnamefont {Horowitz}}, \ and\ \bibinfo {author} {\bibfnamefont {J.}~\bibnamefont {Piekarewicz}},\ }\href {\doibase 10.1103/PhysRevC.109.035803} {\bibfield  {journal} {\bibinfo  {journal} {Phys. Rev. C}\ }\textbf {\bibinfo {volume} {109}},\ \bibinfo {pages} {035803} (\bibinfo {year} {2024})},\ \Eprint {http://arxiv.org/abs/2305.19376} {arXiv:2305.19376 [nucl-th]} \BibitemShut {NoStop}%
\bibitem [{\citenamefont {Zhang}\ and\ \citenamefont {Chen}(2014)}]{Zhang:2014yfa}%
  \BibitemOpen
  \bibfield  {author} {\bibinfo {author} {\bibfnamefont {Z.}~\bibnamefont {Zhang}}\ and\ \bibinfo {author} {\bibfnamefont {L.-W.}\ \bibnamefont {Chen}},\ }\href {\doibase 10.1103/PhysRevC.90.064317} {\bibfield  {journal} {\bibinfo  {journal} {Phys. Rev. C}\ }\textbf {\bibinfo {volume} {90}},\ \bibinfo {pages} {064317} (\bibinfo {year} {2014})},\ \Eprint {http://arxiv.org/abs/1407.8054} {arXiv:1407.8054 [nucl-th]} \BibitemShut {NoStop}%
\bibitem [{\citenamefont {Adhikari}\ \emph {et~al.}(2022)\citenamefont {Adhikari} \emph {et~al.}}]{CREX:2022kgg}%
  \BibitemOpen
  \bibfield  {author} {\bibinfo {author} {\bibfnamefont {D.}~\bibnamefont {Adhikari}} \emph {et~al.} (\bibinfo {collaboration} {CREX}),\ }\href {\doibase 10.1103/PhysRevLett.129.042501} {\bibfield  {journal} {\bibinfo  {journal} {Phys. Rev. Lett.}\ }\textbf {\bibinfo {volume} {129}},\ \bibinfo {pages} {042501} (\bibinfo {year} {2022})},\ \Eprint {http://arxiv.org/abs/2205.11593} {arXiv:2205.11593 [nucl-ex]} \BibitemShut {NoStop}%
\bibitem [{\citenamefont {Reed}\ \emph {et~al.}(2021)\citenamefont {Reed}, \citenamefont {Fattoyev}, \citenamefont {Horowitz},\ and\ \citenamefont {Piekarewicz}}]{Reed:2021nqk}%
  \BibitemOpen
  \bibfield  {author} {\bibinfo {author} {\bibfnamefont {B.~T.}\ \bibnamefont {Reed}}, \bibinfo {author} {\bibfnamefont {F.~J.}\ \bibnamefont {Fattoyev}}, \bibinfo {author} {\bibfnamefont {C.~J.}\ \bibnamefont {Horowitz}}, \ and\ \bibinfo {author} {\bibfnamefont {J.}~\bibnamefont {Piekarewicz}},\ }\href {\doibase 10.1103/PhysRevLett.126.172503} {\bibfield  {journal} {\bibinfo  {journal} {Phys. Rev. Lett.}\ }\textbf {\bibinfo {volume} {126}},\ \bibinfo {pages} {172503} (\bibinfo {year} {2021})},\ \Eprint {http://arxiv.org/abs/2101.03193} {arXiv:2101.03193 [nucl-th]} \BibitemShut {NoStop}%
\bibitem [{\citenamefont {Zhang}\ and\ \citenamefont {Chen}(2023)}]{Zhang:2022bni}%
  \BibitemOpen
  \bibfield  {author} {\bibinfo {author} {\bibfnamefont {Z.}~\bibnamefont {Zhang}}\ and\ \bibinfo {author} {\bibfnamefont {L.-W.}\ \bibnamefont {Chen}},\ }\href {\doibase 10.1103/PhysRevC.108.024317} {\bibfield  {journal} {\bibinfo  {journal} {Phys. Rev. C}\ }\textbf {\bibinfo {volume} {108}},\ \bibinfo {pages} {024317} (\bibinfo {year} {2023})},\ \Eprint {http://arxiv.org/abs/2207.03328} {arXiv:2207.03328 [nucl-th]} \BibitemShut {NoStop}%
\bibitem [{\citenamefont {Adhikari}\ \emph {et~al.}(2021)\citenamefont {Adhikari} \emph {et~al.}}]{PREX:2021umo}%
  \BibitemOpen
  \bibfield  {author} {\bibinfo {author} {\bibfnamefont {D.}~\bibnamefont {Adhikari}} \emph {et~al.} (\bibinfo {collaboration} {PREX}),\ }\href {\doibase 10.1103/PhysRevLett.126.172502} {\bibfield  {journal} {\bibinfo  {journal} {Phys. Rev. Lett.}\ }\textbf {\bibinfo {volume} {126}},\ \bibinfo {pages} {172502} (\bibinfo {year} {2021})},\ \Eprint {http://arxiv.org/abs/2102.10767} {arXiv:2102.10767 [nucl-ex]} \BibitemShut {NoStop}%
\bibitem [{\citenamefont {Cai}\ and\ \citenamefont {Chen}(2017)}]{Cai:2014kya}%
  \BibitemOpen
  \bibfield  {author} {\bibinfo {author} {\bibfnamefont {B.-J.}\ \bibnamefont {Cai}}\ and\ \bibinfo {author} {\bibfnamefont {L.-W.}\ \bibnamefont {Chen}},\ }\href {\doibase 10.1007/s41365-017-0329-1} {\bibfield  {journal} {\bibinfo  {journal} {Nucl. Sci. Tech.}\ }\textbf {\bibinfo {volume} {28}},\ \bibinfo {pages} {185} (\bibinfo {year} {2017})},\ \Eprint {http://arxiv.org/abs/1402.4242} {arXiv:1402.4242 [nucl-th]} \BibitemShut {NoStop}%
\bibitem [{\citenamefont {Rufa}\ \emph {et~al.}(1988)\citenamefont {Rufa}, \citenamefont {Reinhard}, \citenamefont {Maruhn}, \citenamefont {Greiner},\ and\ \citenamefont {Strayer}}]{Rufa:1988zz}%
  \BibitemOpen
  \bibfield  {author} {\bibinfo {author} {\bibfnamefont {M.}~\bibnamefont {Rufa}}, \bibinfo {author} {\bibfnamefont {P.~G.}\ \bibnamefont {Reinhard}}, \bibinfo {author} {\bibfnamefont {J.~A.}\ \bibnamefont {Maruhn}}, \bibinfo {author} {\bibfnamefont {W.}~\bibnamefont {Greiner}}, \ and\ \bibinfo {author} {\bibfnamefont {M.~R.}\ \bibnamefont {Strayer}},\ }\href {\doibase 10.1103/PhysRevC.38.390} {\bibfield  {journal} {\bibinfo  {journal} {Phys. Rev. C}\ }\textbf {\bibinfo {volume} {38}},\ \bibinfo {pages} {390} (\bibinfo {year} {1988})}\BibitemShut {NoStop}%
\bibitem [{\citenamefont {Fattoyev}\ \emph {et~al.}(2013)\citenamefont {Fattoyev}, \citenamefont {Carvajal}, \citenamefont {Newton},\ and\ \citenamefont {Li}}]{Fattoyev:2012uu}%
  \BibitemOpen
  \bibfield  {author} {\bibinfo {author} {\bibfnamefont {F.~J.}\ \bibnamefont {Fattoyev}}, \bibinfo {author} {\bibfnamefont {J.}~\bibnamefont {Carvajal}}, \bibinfo {author} {\bibfnamefont {W.~G.}\ \bibnamefont {Newton}}, \ and\ \bibinfo {author} {\bibfnamefont {B.-A.}\ \bibnamefont {Li}},\ }\href {\doibase 10.1103/PhysRevC.87.015806} {\bibfield  {journal} {\bibinfo  {journal} {Phys. Rev. C}\ }\textbf {\bibinfo {volume} {87}},\ \bibinfo {pages} {015806} (\bibinfo {year} {2013})},\ \Eprint {http://arxiv.org/abs/1210.3402} {arXiv:1210.3402 [nucl-th]} \BibitemShut {NoStop}%
\bibitem [{\citenamefont {Ghosh}\ \emph {et~al.}(2022)\citenamefont {Ghosh}, \citenamefont {Pradhan}, \citenamefont {Chatterjee},\ and\ \citenamefont {Schaffner-Bielich}}]{Ghosh:2022lam}%
  \BibitemOpen
  \bibfield  {author} {\bibinfo {author} {\bibfnamefont {S.}~\bibnamefont {Ghosh}}, \bibinfo {author} {\bibfnamefont {B.~K.}\ \bibnamefont {Pradhan}}, \bibinfo {author} {\bibfnamefont {D.}~\bibnamefont {Chatterjee}}, \ and\ \bibinfo {author} {\bibfnamefont {J.}~\bibnamefont {Schaffner-Bielich}},\ }\href {\doibase 10.3389/fspas.2022.864294} {\bibfield  {journal} {\bibinfo  {journal} {Front. Astron. Space Sci.}\ }\textbf {\bibinfo {volume} {9}},\ \bibinfo {pages} {864294} (\bibinfo {year} {2022})},\ \Eprint {http://arxiv.org/abs/2203.03156} {arXiv:2203.03156 [astro-ph.HE]} \BibitemShut {NoStop}%
\bibitem [{\citenamefont {Piekarewicz}\ and\ \citenamefont {Centelles}(2009)}]{Piekarewicz:2008nh}%
  \BibitemOpen
  \bibfield  {author} {\bibinfo {author} {\bibfnamefont {J.}~\bibnamefont {Piekarewicz}}\ and\ \bibinfo {author} {\bibfnamefont {M.}~\bibnamefont {Centelles}},\ }\href@noop {} {\bibfield  {journal} {\bibinfo  {journal} {Phys. Rev.}\ }\textbf {\bibinfo {volume} {C79}},\ \bibinfo {pages} {054311} (\bibinfo {year} {2009})}\BibitemShut {NoStop}%
\bibitem [{\citenamefont {Chen}\ and\ \citenamefont {Piekarewicz}(2014{\natexlab{b}})}]{Wang:2012}%
  \BibitemOpen
  \bibfield  {author} {\bibinfo {author} {\bibfnamefont {W.-C.}\ \bibnamefont {Chen}}\ and\ \bibinfo {author} {\bibfnamefont {J.}~\bibnamefont {Piekarewicz}},\ }\href@noop {} {\bibfield  {journal} {\bibinfo  {journal} {Phys. Rev.}\ }\textbf {\bibinfo {volume} {C90}},\ \bibinfo {pages} {044305} (\bibinfo {year} {2014}{\natexlab{b}})}\BibitemShut {NoStop}%
\bibitem [{\citenamefont {Angeli}\ and\ \citenamefont {Marinova}(2013)}]{Angeli:2013}%
  \BibitemOpen
  \bibfield  {author} {\bibinfo {author} {\bibfnamefont {I.}~\bibnamefont {Angeli}}\ and\ \bibinfo {author} {\bibfnamefont {K.}~\bibnamefont {Marinova}},\ }\href@noop {} {\bibfield  {journal} {\bibinfo  {journal} {At. Data Nucl. Data Tables}\ }\textbf {\bibinfo {volume} {99}},\ \bibinfo {pages} {69 } (\bibinfo {year} {2013})}\BibitemShut {NoStop}%
\bibitem [{\citenamefont {Fattoyev}\ \emph {et~al.}(2010)\citenamefont {Fattoyev}, \citenamefont {Horowitz}, \citenamefont {Piekarewicz},\ and\ \citenamefont {Shen}}]{Fattoyev:2010mx}%
  \BibitemOpen
  \bibfield  {author} {\bibinfo {author} {\bibfnamefont {F.~J.}\ \bibnamefont {Fattoyev}}, \bibinfo {author} {\bibfnamefont {C.~J.}\ \bibnamefont {Horowitz}}, \bibinfo {author} {\bibfnamefont {J.}~\bibnamefont {Piekarewicz}}, \ and\ \bibinfo {author} {\bibfnamefont {G.}~\bibnamefont {Shen}},\ }\href {\doibase 10.1103/PhysRevC.82.055803} {\bibfield  {journal} {\bibinfo  {journal} {Phys. Rev.}\ }\textbf {\bibinfo {volume} {C82}},\ \bibinfo {pages} {055803} (\bibinfo {year} {2010})}\BibitemShut {NoStop}%
\bibitem [{\citenamefont {Chen}\ and\ \citenamefont {Piekarewicz}(2014{\natexlab{c}})}]{Chen:2014sca}%
  \BibitemOpen
  \bibfield  {author} {\bibinfo {author} {\bibfnamefont {W.-C.}\ \bibnamefont {Chen}}\ and\ \bibinfo {author} {\bibfnamefont {J.}~\bibnamefont {Piekarewicz}},\ }\href@noop {} {\bibfield  {journal} {\bibinfo  {journal} {Phys. Rev.}\ }\textbf {\bibinfo {volume} {C90}},\ \bibinfo {pages} {044305} (\bibinfo {year} {2014}{\natexlab{c}})}\BibitemShut {NoStop}%
\bibitem [{\citenamefont {Lattimer}\ \emph {et~al.}(1991)\citenamefont {Lattimer}, \citenamefont {Pethick}, \citenamefont {Prakash},\ and\ \citenamefont {Haensel}}]{Lattimer:1991}%
  \BibitemOpen
  \bibfield  {author} {\bibinfo {author} {\bibfnamefont {J.~M.}\ \bibnamefont {Lattimer}}, \bibinfo {author} {\bibfnamefont {C.~J.}\ \bibnamefont {Pethick}}, \bibinfo {author} {\bibfnamefont {M.}~\bibnamefont {Prakash}}, \ and\ \bibinfo {author} {\bibfnamefont {P.}~\bibnamefont {Haensel}},\ }\href {\doibase 10.1103/PhysRevLett.66.2701} {\bibfield  {journal} {\bibinfo  {journal} {Phys. Rev. Lett.}\ }\textbf {\bibinfo {volume} {66}},\ \bibinfo {pages} {2701} (\bibinfo {year} {1991})}\BibitemShut {NoStop}%
\bibitem [{\citenamefont {Kl\"ahn}\ \emph {et~al.}(2006)\citenamefont {Kl\"ahn}, \citenamefont {Blaschke}, \citenamefont {Typel}, \citenamefont {van Dalen}, \citenamefont {Faessler}, \citenamefont {Fuchs}, \citenamefont {Gaitanos}, \citenamefont {Grigorian}, \citenamefont {Ho}, \citenamefont {Kolomeitsev}, \citenamefont {Miller}, \citenamefont {R\"opke}, \citenamefont {Tr\"umper}, \citenamefont {Voskresensky}, \citenamefont {Weber},\ and\ \citenamefont {Wolter}}]{Klahn:2006}%
  \BibitemOpen
  \bibfield  {author} {\bibinfo {author} {\bibfnamefont {T.}~\bibnamefont {Kl\"ahn}}, \bibinfo {author} {\bibfnamefont {D.}~\bibnamefont {Blaschke}}, \bibinfo {author} {\bibfnamefont {S.}~\bibnamefont {Typel}}, \bibinfo {author} {\bibfnamefont {E.~N.~E.}\ \bibnamefont {van Dalen}}, \bibinfo {author} {\bibfnamefont {A.}~\bibnamefont {Faessler}}, \bibinfo {author} {\bibfnamefont {C.}~\bibnamefont {Fuchs}}, \bibinfo {author} {\bibfnamefont {T.}~\bibnamefont {Gaitanos}}, \bibinfo {author} {\bibfnamefont {H.}~\bibnamefont {Grigorian}}, \bibinfo {author} {\bibfnamefont {A.}~\bibnamefont {Ho}}, \bibinfo {author} {\bibfnamefont {E.~E.}\ \bibnamefont {Kolomeitsev}}, \bibinfo {author} {\bibfnamefont {M.~C.}\ \bibnamefont {Miller}}, \bibinfo {author} {\bibfnamefont {G.}~\bibnamefont {R\"opke}}, \bibinfo {author} {\bibfnamefont {J.}~\bibnamefont {Tr\"umper}}, \bibinfo {author} {\bibfnamefont {D.~N.}\ \bibnamefont {Voskresensky}}, \bibinfo {author} {\bibfnamefont {F.}~\bibnamefont {Weber}}, \ and\ \bibinfo {author}
  {\bibfnamefont {H.~H.}\ \bibnamefont {Wolter}},\ }\href {\doibase 10.1103/PhysRevC.74.035802} {\bibfield  {journal} {\bibinfo  {journal} {Phys. Rev. C}\ }\textbf {\bibinfo {volume} {74}},\ \bibinfo {pages} {035802} (\bibinfo {year} {2006})}\BibitemShut {NoStop}%
\bibitem [{\citenamefont {Fattoyev}(2011)}]{Fattoyev:2011}%
  \BibitemOpen
  \bibfield  {author} {\bibinfo {author} {\bibfnamefont {F.}~\bibnamefont {Fattoyev}},\ }\emph {\bibinfo {title} {Sensitivity of neutron star properties to the equation of state}},\ \href@noop {} {\bibinfo {type} {phdthesis}},\ \bibinfo  {school} {The Florida State University} (\bibinfo {year} {2011})\BibitemShut {NoStop}%
\bibitem [{\citenamefont {{Lattimer}}\ and\ \citenamefont {{Steiner}}(2014)}]{Lattimer:2014}%
  \BibitemOpen
  \bibfield  {author} {\bibinfo {author} {\bibfnamefont {J.~M.}\ \bibnamefont {{Lattimer}}}\ and\ \bibinfo {author} {\bibfnamefont {A.~W.}\ \bibnamefont {{Steiner}}},\ }\href {\doibase 10.1140/epja/i2014-14040-y} {\bibfield  {journal} {\bibinfo  {journal} {Eur. Phys. J. A}\ }\textbf {\bibinfo {volume} {50}},\ \bibinfo {eid} {40} (\bibinfo {year} {2014})},\ \Eprint {http://arxiv.org/abs/1403.1186} {arXiv:1403.1186 [nucl-th]} \BibitemShut {NoStop}%
\bibitem [{\citenamefont {Mendes}\ \emph {et~al.}(2021)\citenamefont {Mendes}, \citenamefont {Cumming}, \citenamefont {Gale},\ and\ \citenamefont {Fattoyev}}]{Mendes:2021}%
  \BibitemOpen
  \bibfield  {author} {\bibinfo {author} {\bibfnamefont {M.}~\bibnamefont {Mendes}}, \bibinfo {author} {\bibfnamefont {A.}~\bibnamefont {Cumming}}, \bibinfo {author} {\bibfnamefont {C.}~\bibnamefont {Gale}}, \ and\ \bibinfo {author} {\bibfnamefont {F.~J.}\ \bibnamefont {Fattoyev}},\ }in\ \href {\doibase 10.1142/9789811269776_0309} {\emph {\bibinfo {booktitle} {{16th Marcel Grossmann Meeting on~Recent Developments in Theoretical and Experimental General Relativity, Astrophysics and Relativistic Field Theories}}}}\ (\bibinfo {year} {2021})\ \Eprint {http://arxiv.org/abs/2110.11077} {arXiv:2110.11077 [nucl-th]} \BibitemShut {NoStop}%
\bibitem [{\citenamefont {Alford}\ \emph {et~al.}(2013)\citenamefont {Alford}, \citenamefont {Han},\ and\ \citenamefont {Prakash}}]{Alford:2013}%
  \BibitemOpen
  \bibfield  {author} {\bibinfo {author} {\bibfnamefont {M.~G.}\ \bibnamefont {Alford}}, \bibinfo {author} {\bibfnamefont {S.}~\bibnamefont {Han}}, \ and\ \bibinfo {author} {\bibfnamefont {M.}~\bibnamefont {Prakash}},\ }\href {\doibase 10.1103/PhysRevD.88.083013} {\bibfield  {journal} {\bibinfo  {journal} {Phys. Rev. D}\ }\textbf {\bibinfo {volume} {88}},\ \bibinfo {pages} {083013} (\bibinfo {year} {2013})}\BibitemShut {NoStop}%
\bibitem [{\citenamefont {Christian}\ \emph {et~al.}(2024)\citenamefont {Christian}, \citenamefont {Schaffner-Bielich},\ and\ \citenamefont {Rosswog}}]{Christian:2023hez}%
  \BibitemOpen
  \bibfield  {author} {\bibinfo {author} {\bibfnamefont {J.-E.}\ \bibnamefont {Christian}}, \bibinfo {author} {\bibfnamefont {J.}~\bibnamefont {Schaffner-Bielich}}, \ and\ \bibinfo {author} {\bibfnamefont {S.}~\bibnamefont {Rosswog}},\ }\href {\doibase 10.1103/PhysRevD.109.063035} {\bibfield  {journal} {\bibinfo  {journal} {Phys. Rev. D}\ }\textbf {\bibinfo {volume} {109}},\ \bibinfo {pages} {063035} (\bibinfo {year} {2024})},\ \Eprint {http://arxiv.org/abs/2312.10148} {arXiv:2312.10148 [nucl-th]} \BibitemShut {NoStop}%
\bibitem [{\citenamefont {{Baym}}\ \emph {et~al.}(1971)\citenamefont {{Baym}}, \citenamefont {{Pethick}},\ and\ \citenamefont {{Sutherland}}}]{Baym:1971}%
  \BibitemOpen
  \bibfield  {author} {\bibinfo {author} {\bibfnamefont {G.}~\bibnamefont {{Baym}}}, \bibinfo {author} {\bibfnamefont {C.}~\bibnamefont {{Pethick}}}, \ and\ \bibinfo {author} {\bibfnamefont {P.}~\bibnamefont {{Sutherland}}},\ }\href {\doibase https://doi.org/10.1086/151216} {\bibfield  {journal} {\bibinfo  {journal} {Astrophys. J.}\ }\textbf {\bibinfo {volume} {170}},\ \bibinfo {pages} {299} (\bibinfo {year} {1971})}\BibitemShut {NoStop}%
\bibitem [{\citenamefont {Negele}\ and\ \citenamefont {Vautherin}(1973)}]{Negele:1973}%
  \BibitemOpen
  \bibfield  {author} {\bibinfo {author} {\bibfnamefont {J.}~\bibnamefont {Negele}}\ and\ \bibinfo {author} {\bibfnamefont {D.}~\bibnamefont {Vautherin}},\ }\href {\doibase https://doi.org/10.1016/0375-9474(73)90349-7} {\bibfield  {journal} {\bibinfo  {journal} {Nucl. Phys. A}\ }\textbf {\bibinfo {volume} {207}},\ \bibinfo {pages} {298} (\bibinfo {year} {1973})}\BibitemShut {NoStop}%
\bibitem [{\citenamefont {Piekarewicz}\ and\ \citenamefont {Fattoyev}(2019)}]{Piekarewicz:2018sgy}%
  \BibitemOpen
  \bibfield  {author} {\bibinfo {author} {\bibfnamefont {J.}~\bibnamefont {Piekarewicz}}\ and\ \bibinfo {author} {\bibfnamefont {F.~J.}\ \bibnamefont {Fattoyev}},\ }\href {\doibase 10.1103/PhysRevC.99.045802} {\bibfield  {journal} {\bibinfo  {journal} {Phys. Rev. C}\ }\textbf {\bibinfo {volume} {99}},\ \bibinfo {pages} {045802} (\bibinfo {year} {2019})},\ \Eprint {http://arxiv.org/abs/1812.09974} {arXiv:1812.09974 [nucl-th]} \BibitemShut {NoStop}%
\bibitem [{\citenamefont {Kalaitzis}\ \emph {et~al.}(2019)\citenamefont {Kalaitzis}, \citenamefont {Motta},\ and\ \citenamefont {Thomas}}]{Kalaitzis:2019dqc}%
  \BibitemOpen
  \bibfield  {author} {\bibinfo {author} {\bibfnamefont {A.~M.}\ \bibnamefont {Kalaitzis}}, \bibinfo {author} {\bibfnamefont {T.~F.}\ \bibnamefont {Motta}}, \ and\ \bibinfo {author} {\bibfnamefont {A.~W.}\ \bibnamefont {Thomas}},\ }\href {\doibase 10.1142/S0218301319500812} {\bibfield  {journal} {\bibinfo  {journal} {Int. J. Mod. Phys. E}\ }\textbf {\bibinfo {volume} {28}},\ \bibinfo {pages} {1950081} (\bibinfo {year} {2019})},\ \Eprint {http://arxiv.org/abs/1905.05907} {arXiv:1905.05907 [astro-ph.HE]} \BibitemShut {NoStop}%
\bibitem [{\citenamefont {Biswas}\ \emph {et~al.}(2019)\citenamefont {Biswas}, \citenamefont {Nandi}, \citenamefont {Char},\ and\ \citenamefont {Bose}}]{Biswas:2019}%
  \BibitemOpen
  \bibfield  {author} {\bibinfo {author} {\bibfnamefont {B.}~\bibnamefont {Biswas}}, \bibinfo {author} {\bibfnamefont {R.}~\bibnamefont {Nandi}}, \bibinfo {author} {\bibfnamefont {P.}~\bibnamefont {Char}}, \ and\ \bibinfo {author} {\bibfnamefont {S.}~\bibnamefont {Bose}},\ }\href {\doibase 10.1103/PhysRevD.100.044056} {\bibfield  {journal} {\bibinfo  {journal} {Phys. Rev. D}\ }\textbf {\bibinfo {volume} {100}},\ \bibinfo {pages} {044056} (\bibinfo {year} {2019})}\BibitemShut {NoStop}%
\bibitem [{\citenamefont {Ji}\ \emph {et~al.}(2019)\citenamefont {Ji}, \citenamefont {Hu}, \citenamefont {Bao},\ and\ \citenamefont {Shen}}]{Ji:2019hxe}%
  \BibitemOpen
  \bibfield  {author} {\bibinfo {author} {\bibfnamefont {F.}~\bibnamefont {Ji}}, \bibinfo {author} {\bibfnamefont {J.}~\bibnamefont {Hu}}, \bibinfo {author} {\bibfnamefont {S.}~\bibnamefont {Bao}}, \ and\ \bibinfo {author} {\bibfnamefont {H.}~\bibnamefont {Shen}},\ }\href {\doibase 10.1103/PhysRevC.100.045801} {\bibfield  {journal} {\bibinfo  {journal} {Phys. Rev. C}\ }\textbf {\bibinfo {volume} {100}},\ \bibinfo {pages} {045801} (\bibinfo {year} {2019})},\ \Eprint {http://arxiv.org/abs/1910.02266} {arXiv:1910.02266 [nucl-th]} \BibitemShut {NoStop}%
\bibitem [{\citenamefont {Perot}\ \emph {et~al.}(2020)\citenamefont {Perot}, \citenamefont {Chamel},\ and\ \citenamefont {Sourie}}]{Perot:2020gux}%
  \BibitemOpen
  \bibfield  {author} {\bibinfo {author} {\bibfnamefont {L.}~\bibnamefont {Perot}}, \bibinfo {author} {\bibfnamefont {N.}~\bibnamefont {Chamel}}, \ and\ \bibinfo {author} {\bibfnamefont {A.}~\bibnamefont {Sourie}},\ }\href {\doibase 10.1103/PhysRevC.101.015806} {\bibfield  {journal} {\bibinfo  {journal} {Phys. Rev. C}\ }\textbf {\bibinfo {volume} {101}},\ \bibinfo {pages} {015806} (\bibinfo {year} {2020})},\ \Eprint {http://arxiv.org/abs/2001.11068} {arXiv:2001.11068 [astro-ph.HE]} \BibitemShut {NoStop}%
\bibitem [{\citenamefont {Gittins}\ \emph {et~al.}(2020)\citenamefont {Gittins}, \citenamefont {Andersson},\ and\ \citenamefont {Pereira}}]{Gittins:2020mll}%
  \BibitemOpen
  \bibfield  {author} {\bibinfo {author} {\bibfnamefont {F.}~\bibnamefont {Gittins}}, \bibinfo {author} {\bibfnamefont {N.}~\bibnamefont {Andersson}}, \ and\ \bibinfo {author} {\bibfnamefont {J.~P.}\ \bibnamefont {Pereira}},\ }\href {\doibase 10.1103/PhysRevD.101.103025} {\bibfield  {journal} {\bibinfo  {journal} {Phys. Rev. D}\ }\textbf {\bibinfo {volume} {101}},\ \bibinfo {pages} {103025} (\bibinfo {year} {2020})},\ \Eprint {http://arxiv.org/abs/2003.05449} {arXiv:2003.05449 [astro-ph.HE]} \BibitemShut {NoStop}%
\bibitem [{\citenamefont {Piekarewicz}\ \emph {et~al.}(2014)\citenamefont {Piekarewicz}, \citenamefont {Fattoyev},\ and\ \citenamefont {Horowitz}}]{Piekarewicz:2014}%
  \BibitemOpen
  \bibfield  {author} {\bibinfo {author} {\bibfnamefont {J.}~\bibnamefont {Piekarewicz}}, \bibinfo {author} {\bibfnamefont {F.~J.}\ \bibnamefont {Fattoyev}}, \ and\ \bibinfo {author} {\bibfnamefont {C.~J.}\ \bibnamefont {Horowitz}},\ }\href {\doibase 10.1103/PhysRevC.90.015803} {\bibfield  {journal} {\bibinfo  {journal} {Phys. Rev. C}\ }\textbf {\bibinfo {volume} {90}},\ \bibinfo {pages} {015803} (\bibinfo {year} {2014})}\BibitemShut {NoStop}%
\bibitem [{\citenamefont {Carriere}\ \emph {et~al.}(2003)\citenamefont {Carriere}, \citenamefont {Horowitz},\ and\ \citenamefont {Piekarewicz}}]{Carriere:2002bx}%
  \BibitemOpen
  \bibfield  {author} {\bibinfo {author} {\bibfnamefont {J.}~\bibnamefont {Carriere}}, \bibinfo {author} {\bibfnamefont {C.~J.}\ \bibnamefont {Horowitz}}, \ and\ \bibinfo {author} {\bibfnamefont {J.}~\bibnamefont {Piekarewicz}},\ }\href {\doibase 10.1086/376515} {\bibfield  {journal} {\bibinfo  {journal} {Astrophys. J.}\ }\textbf {\bibinfo {volume} {593}},\ \bibinfo {pages} {463} (\bibinfo {year} {2003})},\ \Eprint {http://arxiv.org/abs/nucl-th/0211015} {arXiv:nucl-th/0211015} \BibitemShut {NoStop}%
\bibitem [{\citenamefont {Dittmann}\ \emph {et~al.}(2024)\citenamefont {Dittmann} \emph {et~al.}}]{Dittmann:2024mbo}%
  \BibitemOpen
  \bibfield  {author} {\bibinfo {author} {\bibfnamefont {A.~J.}\ \bibnamefont {Dittmann}} \emph {et~al.},\ }\href@noop {} {\  (\bibinfo {year} {2024})},\ \Eprint {http://arxiv.org/abs/2406.14467} {arXiv:2406.14467 [astro-ph.HE]} \BibitemShut {NoStop}%
\bibitem [{\citenamefont {Salmi}\ \emph {et~al.}(2024)\citenamefont {Salmi} \emph {et~al.}}]{Salmi:2024}%
  \BibitemOpen
  \bibfield  {author} {\bibinfo {author} {\bibfnamefont {T.}~\bibnamefont {Salmi}} \emph {et~al.},\ }\href {\doibase 10.3847/1538-4357/ad5f1f} {\bibfield  {journal} {\bibinfo  {journal} {Astrophys. J.}\ }\textbf {\bibinfo {volume} {974}},\ \bibinfo {pages} {294} (\bibinfo {year} {2024})},\ \Eprint {http://arxiv.org/abs/2406.14466} {arXiv:2406.14466 [astro-ph.HE]} \BibitemShut {NoStop}%
\bibitem [{\citenamefont {Abbott}(2018)}]{LIGO:2018}%
  \BibitemOpen
  \bibfield  {author} {\bibinfo {author} {\bibfnamefont {B.~P. e.~a.}\ \bibnamefont {Abbott}},\ }\href {\doibase 10.1103/PhysRevLett.121.161101} {\bibfield  {journal} {\bibinfo  {journal} {Phys. Rev. Lett.}\ }\textbf {\bibinfo {volume} {121}},\ \bibinfo {pages} {161101} (\bibinfo {year} {2018})}\BibitemShut {NoStop}%
\bibitem [{Note1()}]{Note1}%
  \BibitemOpen
  \bibinfo {note} {Mass-radius posteriors for all models of \cite {Vinciguerra:2023qxq} can be found at the Zenodo repository: https://zenodo.org/records/7646352}\BibitemShut {NoStop}%
\bibitem [{\citenamefont {{Seidov}}(1971)}]{Seidov:1971}%
  \BibitemOpen
  \bibfield  {author} {\bibinfo {author} {\bibfnamefont {Z.~F.}\ \bibnamefont {{Seidov}}},\ }\href@noop {} {\bibfield  {journal} {\bibinfo  {journal} {Soviet Astronomy}\ }\textbf {\bibinfo {volume} {15}},\ \bibinfo {pages} {347} (\bibinfo {year} {1971})}\BibitemShut {NoStop}%
\bibitem [{\citenamefont {Alford}(2024)}]{Alford:2024}%
  \BibitemOpen
  \bibfield  {author} {\bibinfo {author} {\bibfnamefont {M.}~\bibnamefont {Alford}},\ }\href@noop {} {}\bibinfo {howpublished} {private communication} (\bibinfo {year} {2024})\BibitemShut {NoStop}%
\bibitem [{\citenamefont {Yakovlev}\ \emph {et~al.}(2001)\citenamefont {Yakovlev}, \citenamefont {Kaminker}, \citenamefont {Gnedin},\ and\ \citenamefont {Haensel}}]{Yakovlev:2001}%
  \BibitemOpen
  \bibfield  {author} {\bibinfo {author} {\bibfnamefont {D.}~\bibnamefont {Yakovlev}}, \bibinfo {author} {\bibfnamefont {A.}~\bibnamefont {Kaminker}}, \bibinfo {author} {\bibfnamefont {O.}~\bibnamefont {Gnedin}}, \ and\ \bibinfo {author} {\bibfnamefont {P.}~\bibnamefont {Haensel}},\ }\href {\doibase https://doi.org/10.1016/S0370-1573(00)00131-9} {\bibfield  {journal} {\bibinfo  {journal} {Physics Reports}\ }\textbf {\bibinfo {volume} {354}},\ \bibinfo {pages} {1} (\bibinfo {year} {2001})}\BibitemShut {NoStop}%
\bibitem [{\citenamefont {Iwamoto}(1980)}]{Iwamoto:1980}%
  \BibitemOpen
  \bibfield  {author} {\bibinfo {author} {\bibfnamefont {N.}~\bibnamefont {Iwamoto}},\ }\href {\doibase 10.1103/PhysRevLett.44.1637} {\bibfield  {journal} {\bibinfo  {journal} {Phys. Rev. Lett.}\ }\textbf {\bibinfo {volume} {44}},\ \bibinfo {pages} {1637} (\bibinfo {year} {1980})}\BibitemShut {NoStop}%
\bibitem [{\citenamefont {Carter}\ and\ \citenamefont {Prakash}(2002)}]{Carter:2002}%
  \BibitemOpen
  \bibfield  {author} {\bibinfo {author} {\bibfnamefont {G.~W.}\ \bibnamefont {Carter}}\ and\ \bibinfo {author} {\bibfnamefont {M.}~\bibnamefont {Prakash}},\ }\href {\doibase 10.1016/s0370-2693(01)01452-6} {\bibfield  {journal} {\bibinfo  {journal} {Physics Letters B}\ }\textbf {\bibinfo {volume} {525}},\ \bibinfo {pages} {249} (\bibinfo {year} {2002})}\BibitemShut {NoStop}%
\bibitem [{\citenamefont {Ho}\ \emph {et~al.}(2015)\citenamefont {Ho}, \citenamefont {Elshamouty}, \citenamefont {Heinke},\ and\ \citenamefont {Potekhin}}]{Ho:2015}%
  \BibitemOpen
  \bibfield  {author} {\bibinfo {author} {\bibfnamefont {W.~C.~G.}\ \bibnamefont {Ho}}, \bibinfo {author} {\bibfnamefont {K.~G.}\ \bibnamefont {Elshamouty}}, \bibinfo {author} {\bibfnamefont {C.~O.}\ \bibnamefont {Heinke}}, \ and\ \bibinfo {author} {\bibfnamefont {A.~Y.}\ \bibnamefont {Potekhin}},\ }\href {\doibase 10.1103/PhysRevC.91.015806} {\bibfield  {journal} {\bibinfo  {journal} {Phys. Rev. C}\ }\textbf {\bibinfo {volume} {91}},\ \bibinfo {pages} {015806} (\bibinfo {year} {2015})}\BibitemShut {NoStop}%
\bibitem [{\citenamefont {Shternin}\ \emph {et~al.}(2011)\citenamefont {Shternin}, \citenamefont {Yakovlev}, \citenamefont {Heinke}, \citenamefont {Ho},\ and\ \citenamefont {Patnaude}}]{Shternin:2011}%
  \BibitemOpen
  \bibfield  {author} {\bibinfo {author} {\bibfnamefont {P.~S.}\ \bibnamefont {Shternin}}, \bibinfo {author} {\bibfnamefont {D.~G.}\ \bibnamefont {Yakovlev}}, \bibinfo {author} {\bibfnamefont {C.~O.}\ \bibnamefont {Heinke}}, \bibinfo {author} {\bibfnamefont {W.~C.~G.}\ \bibnamefont {Ho}}, \ and\ \bibinfo {author} {\bibfnamefont {D.~J.}\ \bibnamefont {Patnaude}},\ }\href {\doibase 10.1111/j.1745-3933.2011.01015.x} {\bibfield  {journal} {\bibinfo  {journal} {Monthly Notices of the Royal Astronomical Society: Letters}\ }\textbf {\bibinfo {volume} {412}},\ \bibinfo {pages} {L108} (\bibinfo {year} {2011})},\ \Eprint {http://arxiv.org/abs/https://academic.oup.com/mnrasl/article-pdf/412/1/L108/3262203/412-1-L108.pdf} {https://academic.oup.com/mnrasl/article-pdf/412/1/L108/3262203/412-1-L108.pdf} \BibitemShut {NoStop}%
\bibitem [{\citenamefont {Amundsen}\ and\ \citenamefont {Østgaard}(1985)}]{Amundsen_2:1985}%
  \BibitemOpen
  \bibfield  {author} {\bibinfo {author} {\bibfnamefont {L.}~\bibnamefont {Amundsen}}\ and\ \bibinfo {author} {\bibfnamefont {E.}~\bibnamefont {Østgaard}},\ }\href {\doibase https://doi.org/10.1016/0375-9474(85)90140-X} {\bibfield  {journal} {\bibinfo  {journal} {Nuclear Physics A}\ }\textbf {\bibinfo {volume} {442}},\ \bibinfo {pages} {163} (\bibinfo {year} {1985})}\BibitemShut {NoStop}%
\bibitem [{\citenamefont {Elgar\o{}y}\ \emph {et~al.}(1996)\citenamefont {Elgar\o{}y}, \citenamefont {Engvik}, \citenamefont {Hjorth-Jensen},\ and\ \citenamefont {Osnes}}]{Elgaroy_3:1996}%
  \BibitemOpen
  \bibfield  {author} {\bibinfo {author} {\bibfnamefont {O.}~\bibnamefont {Elgar\o{}y}}, \bibinfo {author} {\bibfnamefont {L.}~\bibnamefont {Engvik}}, \bibinfo {author} {\bibfnamefont {M.}~\bibnamefont {Hjorth-Jensen}}, \ and\ \bibinfo {author} {\bibfnamefont {E.}~\bibnamefont {Osnes}},\ }\href {\doibase https://doi.org/10.1016/0375-9474(96)00217-5} {\bibfield  {journal} {\bibinfo  {journal} {Nuclear Physics A}\ }\textbf {\bibinfo {volume} {607}},\ \bibinfo {pages} {425} (\bibinfo {year} {1996})}\BibitemShut {NoStop}%
\bibitem [{\citenamefont {Chen}\ \emph {et~al.}(1993)\citenamefont {Chen}, \citenamefont {Clark}, \citenamefont {Davé},\ and\ \citenamefont {Khodel}}]{Chen:1993}%
  \BibitemOpen
  \bibfield  {author} {\bibinfo {author} {\bibfnamefont {J.}~\bibnamefont {Chen}}, \bibinfo {author} {\bibfnamefont {J.}~\bibnamefont {Clark}}, \bibinfo {author} {\bibfnamefont {R.}~\bibnamefont {Davé}}, \ and\ \bibinfo {author} {\bibfnamefont {V.}~\bibnamefont {Khodel}},\ }\href {\doibase https://doi.org/10.1016/0375-9474(93)90314-N} {\bibfield  {journal} {\bibinfo  {journal} {Nuclear Physics A}\ }\textbf {\bibinfo {volume} {555}},\ \bibinfo {pages} {59} (\bibinfo {year} {1993})}\BibitemShut {NoStop}%
\bibitem [{\citenamefont {Baldo}\ \emph {et~al.}(1992)\citenamefont {Baldo}, \citenamefont {Cugnon}, \citenamefont {Lejeune},\ and\ \citenamefont {Lombardo}}]{Baldo:1992}%
  \BibitemOpen
  \bibfield  {author} {\bibinfo {author} {\bibfnamefont {M.}~\bibnamefont {Baldo}}, \bibinfo {author} {\bibfnamefont {J.}~\bibnamefont {Cugnon}}, \bibinfo {author} {\bibfnamefont {A.}~\bibnamefont {Lejeune}}, \ and\ \bibinfo {author} {\bibfnamefont {U.}~\bibnamefont {Lombardo}},\ }\href {\doibase https://doi.org/10.1016/0375-9474(92)90387-Y} {\bibfield  {journal} {\bibinfo  {journal} {Nuclear Physics A}\ }\textbf {\bibinfo {volume} {536}},\ \bibinfo {pages} {349} (\bibinfo {year} {1992})}\BibitemShut {NoStop}%
\bibitem [{\citenamefont {Chao}\ \emph {et~al.}(1972)\citenamefont {Chao}, \citenamefont {Clark},\ and\ \citenamefont {Yang}}]{Chao:1972}%
  \BibitemOpen
  \bibfield  {author} {\bibinfo {author} {\bibfnamefont {N.-C.}\ \bibnamefont {Chao}}, \bibinfo {author} {\bibfnamefont {J.}~\bibnamefont {Clark}}, \ and\ \bibinfo {author} {\bibfnamefont {C.-H.}\ \bibnamefont {Yang}},\ }\href {\doibase https://doi.org/10.1016/0375-9474(72)90373-9} {\bibfield  {journal} {\bibinfo  {journal} {Nuclear Physics A}\ }\textbf {\bibinfo {volume} {179}},\ \bibinfo {pages} {320} (\bibinfo {year} {1972})}\BibitemShut {NoStop}%
\bibitem [{\citenamefont {{Levenfish}}\ and\ \citenamefont {{Yakovlev}}(1994)}]{Yakovlev:1994}%
  \BibitemOpen
  \bibfield  {author} {\bibinfo {author} {\bibfnamefont {K.~P.}\ \bibnamefont {{Levenfish}}}\ and\ \bibinfo {author} {\bibfnamefont {D.~G.}\ \bibnamefont {{Yakovlev}}},\ }\href@noop {} {\bibfield  {journal} {\bibinfo  {journal} {Astronomy Letters}\ }\textbf {\bibinfo {volume} {20}},\ \bibinfo {pages} {43} (\bibinfo {year} {1994})}\BibitemShut {NoStop}%
\bibitem [{\citenamefont {Brown}\ \emph {et~al.}(2018)\citenamefont {Brown}, \citenamefont {Cumming}, \citenamefont {Fattoyev}, \citenamefont {Horowitz}, \citenamefont {Page},\ and\ \citenamefont {Reddy}}]{Brown:2018}%
  \BibitemOpen
  \bibfield  {author} {\bibinfo {author} {\bibfnamefont {E.~F.}\ \bibnamefont {Brown}}, \bibinfo {author} {\bibfnamefont {A.}~\bibnamefont {Cumming}}, \bibinfo {author} {\bibfnamefont {F.~J.}\ \bibnamefont {Fattoyev}}, \bibinfo {author} {\bibfnamefont {C.}~\bibnamefont {Horowitz}}, \bibinfo {author} {\bibfnamefont {D.}~\bibnamefont {Page}}, \ and\ \bibinfo {author} {\bibfnamefont {S.}~\bibnamefont {Reddy}},\ }\href {\doibase 10.1103/physrevlett.120.182701} {\bibfield  {journal} {\bibinfo  {journal} {Physical Review Letters}\ }\textbf {\bibinfo {volume} {120}} (\bibinfo {year} {2018}),\ 10.1103/physrevlett.120.182701}\BibitemShut {NoStop}%
\bibitem [{\citenamefont {Cumming}\ \emph {et~al.}(2017{\natexlab{a}})\citenamefont {Cumming}, \citenamefont {Brown}, \citenamefont {Fattoyev}, \citenamefont {Horowitz}, \citenamefont {Page},\ and\ \citenamefont {Reddy}}]{Cumming:2017}%
  \BibitemOpen
  \bibfield  {author} {\bibinfo {author} {\bibfnamefont {A.}~\bibnamefont {Cumming}}, \bibinfo {author} {\bibfnamefont {E.~F.}\ \bibnamefont {Brown}}, \bibinfo {author} {\bibfnamefont {F.~J.}\ \bibnamefont {Fattoyev}}, \bibinfo {author} {\bibfnamefont {C.~J.}\ \bibnamefont {Horowitz}}, \bibinfo {author} {\bibfnamefont {D.}~\bibnamefont {Page}}, \ and\ \bibinfo {author} {\bibfnamefont {S.}~\bibnamefont {Reddy}},\ }\href {\doibase 10.1103/PhysRevC.95.025806} {\bibfield  {journal} {\bibinfo  {journal} {Phys. Rev. C}\ }\textbf {\bibinfo {volume} {95}},\ \bibinfo {pages} {025806} (\bibinfo {year} {2017}{\natexlab{a}})}\BibitemShut {NoStop}%
\bibitem [{\citenamefont {{Parikh}}\ \emph {et~al.}(2019)\citenamefont {{Parikh}}, \citenamefont {{Wijnands}}, \citenamefont {{Ootes}}, \citenamefont {{Page}}, \citenamefont {{Degenaar}}, \citenamefont {{Bahramian}}, \citenamefont {{Brown}}, \citenamefont {{Cackett}}, \citenamefont {{Cumming}}, \citenamefont {{Heinke}}, \citenamefont {{Homan}}, \citenamefont {{Rouco Escorial}},\ and\ \citenamefont {{Wijngaarden}}}]{Parikh:2019}%
  \BibitemOpen
  \bibfield  {author} {\bibinfo {author} {\bibfnamefont {A.~S.}\ \bibnamefont {{Parikh}}}, \bibinfo {author} {\bibfnamefont {R.}~\bibnamefont {{Wijnands}}}, \bibinfo {author} {\bibfnamefont {L.~S.}\ \bibnamefont {{Ootes}}}, \bibinfo {author} {\bibfnamefont {D.}~\bibnamefont {{Page}}}, \bibinfo {author} {\bibfnamefont {N.}~\bibnamefont {{Degenaar}}}, \bibinfo {author} {\bibfnamefont {A.}~\bibnamefont {{Bahramian}}}, \bibinfo {author} {\bibfnamefont {E.~F.}\ \bibnamefont {{Brown}}}, \bibinfo {author} {\bibfnamefont {E.~M.}\ \bibnamefont {{Cackett}}}, \bibinfo {author} {\bibfnamefont {A.}~\bibnamefont {{Cumming}}}, \bibinfo {author} {\bibfnamefont {C.}~\bibnamefont {{Heinke}}}, \bibinfo {author} {\bibfnamefont {J.}~\bibnamefont {{Homan}}}, \bibinfo {author} {\bibfnamefont {A.}~\bibnamefont {{Rouco Escorial}}}, \ and\ \bibinfo {author} {\bibfnamefont {M.~J.~P.}\ \bibnamefont {{Wijngaarden}}},\ }\href {\doibase 10.1051/0004-6361/201834412} {\bibfield  {journal} {\bibinfo  {journal} {Astronomy \& Astrophysics}\
  }\textbf {\bibinfo {volume} {624}},\ \bibinfo {eid} {A84} (\bibinfo {year} {2019})},\ \Eprint {http://arxiv.org/abs/1810.05626} {arXiv:1810.05626 [astro-ph.HE]} \BibitemShut {NoStop}%
\bibitem [{\citenamefont {{Gudmundsson}}\ \emph {et~al.}(1983)\citenamefont {{Gudmundsson}}, \citenamefont {{Pethick}},\ and\ \citenamefont {{Epstein}}}]{Gudmundsson:1983}%
  \BibitemOpen
  \bibfield  {author} {\bibinfo {author} {\bibfnamefont {E.~H.}\ \bibnamefont {{Gudmundsson}}}, \bibinfo {author} {\bibfnamefont {C.~J.}\ \bibnamefont {{Pethick}}}, \ and\ \bibinfo {author} {\bibfnamefont {R.~I.}\ \bibnamefont {{Epstein}}},\ }\href {\doibase 10.1086/161292} {\bibfield  {journal} {\bibinfo  {journal} {\apj}\ }\textbf {\bibinfo {volume} {272}},\ \bibinfo {pages} {286} (\bibinfo {year} {1983})}\BibitemShut {NoStop}%
\bibitem [{\citenamefont {Cumming}\ \emph {et~al.}(2017{\natexlab{b}})\citenamefont {Cumming}, \citenamefont {Brown}, \citenamefont {Fattoyev}, \citenamefont {Horowitz}, \citenamefont {Page},\ and\ \citenamefont {Reddy}}]{Cumming:2016}%
  \BibitemOpen
  \bibfield  {author} {\bibinfo {author} {\bibfnamefont {A.}~\bibnamefont {Cumming}}, \bibinfo {author} {\bibfnamefont {E.~F.}\ \bibnamefont {Brown}}, \bibinfo {author} {\bibfnamefont {F.~J.}\ \bibnamefont {Fattoyev}}, \bibinfo {author} {\bibfnamefont {C.~J.}\ \bibnamefont {Horowitz}}, \bibinfo {author} {\bibfnamefont {D.}~\bibnamefont {Page}}, \ and\ \bibinfo {author} {\bibfnamefont {S.}~\bibnamefont {Reddy}},\ }\href {\doibase 10.1103/PhysRevC.95.025806} {\bibfield  {journal} {\bibinfo  {journal} {Phys. Rev. C}\ }\textbf {\bibinfo {volume} {95}},\ \bibinfo {pages} {025806} (\bibinfo {year} {2017}{\natexlab{b}})},\ \Eprint {http://arxiv.org/abs/1608.07532} {arXiv:1608.07532 [astro-ph.HE]} \BibitemShut {NoStop}%
\bibitem [{\citenamefont {{Potekhin, A. Y.}}\ \emph {et~al.}(2019)\citenamefont {{Potekhin, A. Y.}}, \citenamefont {{Chugunov, A. I.}},\ and\ \citenamefont {{Chabrier, G.}}}]{Potekhin:2019}%
  \BibitemOpen
  \bibfield  {author} {\bibinfo {author} {\bibnamefont {{Potekhin, A. Y.}}}, \bibinfo {author} {\bibnamefont {{Chugunov, A. I.}}}, \ and\ \bibinfo {author} {\bibnamefont {{Chabrier, G.}}},\ }\href {\doibase 10.1051/0004-6361/201936003} {\bibfield  {journal} {\bibinfo  {journal} {A\&A}\ }\textbf {\bibinfo {volume} {629}},\ \bibinfo {pages} {A88} (\bibinfo {year} {2019})}\BibitemShut {NoStop}%
\bibitem [{\citenamefont {Heinke}\ \emph {et~al.}(2009)\citenamefont {Heinke}, \citenamefont {Jonker}, \citenamefont {Wijnands}, \citenamefont {Deloye},\ and\ \citenamefont {Taam}}]{Heinke:2009}%
  \BibitemOpen
  \bibfield  {author} {\bibinfo {author} {\bibfnamefont {C.~O.}\ \bibnamefont {Heinke}}, \bibinfo {author} {\bibfnamefont {P.~G.}\ \bibnamefont {Jonker}}, \bibinfo {author} {\bibfnamefont {R.}~\bibnamefont {Wijnands}}, \bibinfo {author} {\bibfnamefont {C.~J.}\ \bibnamefont {Deloye}}, \ and\ \bibinfo {author} {\bibfnamefont {R.~E.}\ \bibnamefont {Taam}},\ }\href {\doibase 10.1088/0004-637x/691/2/1035} {\bibfield  {journal} {\bibinfo  {journal} {The Astrophysical Journal}\ }\textbf {\bibinfo {volume} {691}},\ \bibinfo {pages} {1035} (\bibinfo {year} {2009})}\BibitemShut {NoStop}%
\bibitem [{\citenamefont {{Potekhin}}\ \emph {et~al.}(1997)\citenamefont {{Potekhin}}, \citenamefont {{Chabrier}},\ and\ \citenamefont {{Yakovlev}}}]{Potekhin:1997}%
  \BibitemOpen
  \bibfield  {author} {\bibinfo {author} {\bibfnamefont {A.~Y.}\ \bibnamefont {{Potekhin}}}, \bibinfo {author} {\bibfnamefont {G.}~\bibnamefont {{Chabrier}}}, \ and\ \bibinfo {author} {\bibfnamefont {D.~G.}\ \bibnamefont {{Yakovlev}}},\ }\href {\doibase 10.48550/arXiv.astro-ph/9706148} {\bibfield  {journal} {\bibinfo  {journal} {Astronomy and Astrophysics}\ }\textbf {\bibinfo {volume} {323}},\ \bibinfo {pages} {415} (\bibinfo {year} {1997})},\ \Eprint {http://arxiv.org/abs/astro-ph/9706148} {arXiv:astro-ph/9706148 [astro-ph]} \BibitemShut {NoStop}%
\bibitem [{\citenamefont {Burrows}\ \emph {et~al.}(2019)\citenamefont {Burrows}, \citenamefont {Radice}, \citenamefont {Vartanyan}, \citenamefont {Nagakura}, \citenamefont {Skinner},\ and\ \citenamefont {Dolence}}]{Burrows:2019}%
  \BibitemOpen
  \bibfield  {author} {\bibinfo {author} {\bibfnamefont {A.}~\bibnamefont {Burrows}}, \bibinfo {author} {\bibfnamefont {D.}~\bibnamefont {Radice}}, \bibinfo {author} {\bibfnamefont {D.}~\bibnamefont {Vartanyan}}, \bibinfo {author} {\bibfnamefont {H.}~\bibnamefont {Nagakura}}, \bibinfo {author} {\bibfnamefont {M.~A.}\ \bibnamefont {Skinner}}, \ and\ \bibinfo {author} {\bibfnamefont {J.~C.}\ \bibnamefont {Dolence}},\ }\href {\doibase 10.1093/mnras/stz3223} {\bibfield  {journal} {\bibinfo  {journal} {Monthly Notices of the Royal Astronomical Society}\ }\textbf {\bibinfo {volume} {491}},\ \bibinfo {pages} {2715} (\bibinfo {year} {2019})},\ \Eprint {http://arxiv.org/abs/https://academic.oup.com/mnras/article-pdf/491/2/2715/31221715/stz3223.pdf} {https://academic.oup.com/mnras/article-pdf/491/2/2715/31221715/stz3223.pdf} \BibitemShut {NoStop}%
\bibitem [{\citenamefont {Suwa}\ \emph {et~al.}(2018)\citenamefont {Suwa}, \citenamefont {Yoshida}, \citenamefont {Shibata}, \citenamefont {Umeda},\ and\ \citenamefont {Takahashi}}]{Suwa2018}%
  \BibitemOpen
  \bibfield  {author} {\bibinfo {author} {\bibfnamefont {Y.}~\bibnamefont {Suwa}}, \bibinfo {author} {\bibfnamefont {T.}~\bibnamefont {Yoshida}}, \bibinfo {author} {\bibfnamefont {M.}~\bibnamefont {Shibata}}, \bibinfo {author} {\bibfnamefont {H.}~\bibnamefont {Umeda}}, \ and\ \bibinfo {author} {\bibfnamefont {K.}~\bibnamefont {Takahashi}},\ }\href {\doibase 10.1093/mnras/sty2460} {\bibfield  {journal} {\bibinfo  {journal} {Monthly Notices of the Royal Astronomical Society}\ }\textbf {\bibinfo {volume} {481}},\ \bibinfo {pages} {3305} (\bibinfo {year} {2018})},\ \Eprint {http://arxiv.org/abs/https://academic.oup.com/mnras/article-pdf/481/3/3305/25834769/sty2460.pdf} {https://academic.oup.com/mnras/article-pdf/481/3/3305/25834769/sty2460.pdf} \BibitemShut {NoStop}%
\bibitem [{\citenamefont {{Radice}}\ \emph {et~al.}(2017)\citenamefont {{Radice}}, \citenamefont {{Burrows}}, \citenamefont {{Vartanyan}}, \citenamefont {{Skinner}},\ and\ \citenamefont {{Dolence}}}]{Radice2017}%
  \BibitemOpen
  \bibfield  {author} {\bibinfo {author} {\bibfnamefont {D.}~\bibnamefont {{Radice}}}, \bibinfo {author} {\bibfnamefont {A.}~\bibnamefont {{Burrows}}}, \bibinfo {author} {\bibfnamefont {D.}~\bibnamefont {{Vartanyan}}}, \bibinfo {author} {\bibfnamefont {M.~A.}\ \bibnamefont {{Skinner}}}, \ and\ \bibinfo {author} {\bibfnamefont {J.~C.}\ \bibnamefont {{Dolence}}},\ }\href {\doibase 10.3847/1538-4357/aa92c5} {\bibfield  {journal} {\bibinfo  {journal} {\apj}\ }\textbf {\bibinfo {volume} {850}},\ \bibinfo {eid} {43} (\bibinfo {year} {2017})},\ \Eprint {http://arxiv.org/abs/1702.03927} {arXiv:1702.03927 [astro-ph.HE]} \BibitemShut {NoStop}%
\bibitem [{\citenamefont {{Janka}}\ \emph {et~al.}(2008)\citenamefont {{Janka}}, \citenamefont {{M{\"u}ller}}, \citenamefont {{Kitaura}},\ and\ \citenamefont {{Buras}}}]{Janka2008}%
  \BibitemOpen
  \bibfield  {author} {\bibinfo {author} {\bibfnamefont {H.~T.}\ \bibnamefont {{Janka}}}, \bibinfo {author} {\bibfnamefont {B.}~\bibnamefont {{M{\"u}ller}}}, \bibinfo {author} {\bibfnamefont {F.~S.}\ \bibnamefont {{Kitaura}}}, \ and\ \bibinfo {author} {\bibfnamefont {R.}~\bibnamefont {{Buras}}},\ }\href {\doibase 10.1051/0004-6361:20079334} {\bibfield  {journal} {\bibinfo  {journal} {Astronomy \& Astrophysics}\ }\textbf {\bibinfo {volume} {485}},\ \bibinfo {pages} {199} (\bibinfo {year} {2008})},\ \Eprint {http://arxiv.org/abs/0712.4237} {arXiv:0712.4237 [astro-ph]} \BibitemShut {NoStop}%
\bibitem [{\citenamefont {{Fischer}}\ \emph {et~al.}(2010)\citenamefont {{Fischer}}, \citenamefont {{Whitehouse}}, \citenamefont {{Mezzacappa}}, \citenamefont {{Thielemann}},\ and\ \citenamefont {{Liebend{\"o}rfer}}}]{Fischer2010}%
  \BibitemOpen
  \bibfield  {author} {\bibinfo {author} {\bibfnamefont {T.}~\bibnamefont {{Fischer}}}, \bibinfo {author} {\bibfnamefont {S.~C.}\ \bibnamefont {{Whitehouse}}}, \bibinfo {author} {\bibfnamefont {A.}~\bibnamefont {{Mezzacappa}}}, \bibinfo {author} {\bibfnamefont {F.~K.}\ \bibnamefont {{Thielemann}}}, \ and\ \bibinfo {author} {\bibfnamefont {M.}~\bibnamefont {{Liebend{\"o}rfer}}},\ }\href {\doibase 10.1051/0004-6361/200913106} {\bibfield  {journal} {\bibinfo  {journal} {Astronomy \& Astrophysics}\ }\textbf {\bibinfo {volume} {517}},\ \bibinfo {eid} {A80} (\bibinfo {year} {2010})},\ \Eprint {http://arxiv.org/abs/0908.1871} {arXiv:0908.1871 [astro-ph.HE]} \BibitemShut {NoStop}%
\bibitem [{\citenamefont {Christian}\ and\ \citenamefont {Schaffner-Bielich}(2022)}]{Christian:2021}%
  \BibitemOpen
  \bibfield  {author} {\bibinfo {author} {\bibfnamefont {J.-E.}\ \bibnamefont {Christian}}\ and\ \bibinfo {author} {\bibfnamefont {J.}~\bibnamefont {Schaffner-Bielich}},\ }\href {\doibase 10.3847/1538-4357/ac75cf} {\bibfield  {journal} {\bibinfo  {journal} {Astrophys. J.}\ }\textbf {\bibinfo {volume} {935}},\ \bibinfo {pages} {122} (\bibinfo {year} {2022})},\ \Eprint {http://arxiv.org/abs/2109.04191} {arXiv:2109.04191 [astro-ph.HE]} \BibitemShut {NoStop}%
\bibitem [{\citenamefont {Schaffner-Bielich}(2020)}]{Schaffner-Bielich:2020}%
  \BibitemOpen
  \bibfield  {author} {\bibinfo {author} {\bibfnamefont {J.}~\bibnamefont {Schaffner-Bielich}},\ }\href@noop {} {\emph {\bibinfo {title} {Compact Star Physics}}}\ (\bibinfo  {publisher} {Cambridge University Press},\ \bibinfo {year} {2020})\BibitemShut {NoStop}%
\bibitem [{\citenamefont {Buballa}(2005)}]{Buballa:2003}%
  \BibitemOpen
  \bibfield  {author} {\bibinfo {author} {\bibfnamefont {M.}~\bibnamefont {Buballa}},\ }\href {\doibase 10.1016/j.physrep.2004.11.004} {\bibfield  {journal} {\bibinfo  {journal} {Phys. Rept.}\ }\textbf {\bibinfo {volume} {407}},\ \bibinfo {pages} {205} (\bibinfo {year} {2005})},\ \Eprint {http://arxiv.org/abs/hep-ph/0402234} {arXiv:hep-ph/0402234} \BibitemShut {NoStop}%
\end{thebibliography}%
\bibliographystyle{apsrev4-1}

\end{document}